\documentclass[twocolumn]{aastex631}

\usepackage{natbib}
\usepackage{epsfig}
\usepackage{url}
\usepackage[maxfloats=256]{morefloats}
\newcommand{\eal}[2]{\ifmmode{\mathrm{#1\,#2}}\else{#1\textsc{$\,$\lowercase{#2}}}\fi\xspace}
\newcommand{\feal}[2]{\ifmmode{\mathrm{#1\,#2}}\else{[#1\textsc{$\,$\lowercase{#2}}]}\fi\xspace}
\newcommand{\hfeal}[2]{\ifmmode{\mathrm{#1\,#2}}\else{#1\textsc{$\,$\lowercase{#2}}]}\fi\xspace}

\graphicspath{{./}{figures/}}

\begin{document}

\title[]{X-ray and gamma-ray study for 2023 nova eruption of V1716 Sco}

\author{H.-H. Wang}
\affiliation{School of Physics and Engineering, Henan University of Science and Technology, Luoyang 471023, China}
\email{wanghh33@mail.sysu.edu.cn}
\author{H.-D. Yan}
\affiliation{Department of Astronomy, School of Physics, Huazhong University of Science and Technology, Wuhan 430074, China}
\email{yanhd125@alumni.hust.edu.cn}
\author{J. Takata}
\affiliation{Department of Astronomy, School of Physics, Huazhong University of Science and Technology, Wuhan 430074, China}
\email{takata@hust.edu.cn}
\author{L.C.-C., Lin}
\affiliation{Department of Physics, National Cheng Kung University, Tainan 701401, Taiwan}


\begin{abstract}
We report the results of X-ray and gamma-ray analyses of the classical-nova V1716 Sco using data taken by \verb|Swift|, \verb|NICER|, \verb|NuSTAR| and \verb|Fermi|-LAT. We confirm gamma-ray emission at a significant level exceeding 8~$\sigma$ in the one-day bin immediately following  the optical eruption.  The gamma-ray emission, with a Test Statistic value more than four, persists for approximately 40 days and its total emitted energy reaches of the order of $\sim 10^{42}~{\rm erg}$.
The X-ray emission  being concurrent with the gamma-ray emission is  described by the optically thin thermal plasma emission and it is likely dominated by the emission from the gas heated up by the shock. This X-ray component acquires the flux peak at approximately 20 days after the eruption and  the observed X-ray emission enters super soft  state (SSS) about 45 days after the eruption.  The spectrum in the SSS phase is explained by the white dwarf's atmosphere model with a temperature of $\sim 50$~keV and a luminosity of $\sim 10^{37-38}~{\rm erg~s^{-1}}$. The gamma-ray and X-ray emission properties of V1716 Sco are similar to those of other classical novae. Unlike other classical nova, the X-ray emission initially resolved by the \verb|Swift| occurs earlier, during a period when the gamma-ray emission is still at a detectable flux level by \verb|Fermi|-LAT observations. Using the X-ray emission properties observed  before the SSS phase, we interpret that  the nova produces initial slow and less dense outflow, which is eventually overtaken by the fast and dense outflow that causes the main outburst, and the X-ray emission is powered by the forward shock that propagates in the slow outflow. During the SSS, the \verb|NICER| data reveal a quasi-periodic oscillation with a observed period of $\sim 79$~seconds with a possible temporal variation, and indicates the temporal variation of the emission region on the white dwarf's surface.

\end{abstract}

\section{Introduction}
Novae are thermonuclear eruptions that occur in binary systems, where a white dwarf (hereafter WD) accretes matter from its companion. The energy released from the thermonuclear eruption causes a dramatic expansion and ejection of the accreted envelope. Observations have shown that the ejected matter expands into the surrounding environment at speeds ranging from hundreds to thousands of ${\rm km~s^{-1}}$  \citep{Gallagher1978} and have confirmed multi-wavelength emission from radio to TeV gamma-ray bands~\citep{Chomiuk2021,Acciari2022NatAs,HESS2022}. Hard X-ray and gamma-ray emissions are thought to be evidence of the formation of the shock due to the novae outflows \citep{Metzger2014,Li2017}. The \verb|Fermi| Large Area Telescope (hereafter \verb|Fermi|-LAT) has confirmed GeV emissions from  20~novae and potential emissions from 6 sources, since its launch in 2008\footnote{\url{https://asd.gsfc.nasa.gov/Koji.Mukai/novae/latnovae.html}}. 
It is argued that the collisions of the multiple ejecta (internal shock) or the interaction between the ejecta and preexisting medium surrounding the binary can cause the shock~\citep{Della2020A&AR,Aydi2020ApJ,Chomiuk2021}, resulting in the production of gamma-rays through leptonic and/or hadronic processes~\citep[e.g.][]{vurm2018, Chomiuk2021}. Most novae detected in the GeV range are classified as classical novae, typically having a main-sequence star as the companion. These shocks are thought to be internal, resulting from collisions of the multiple ejecta \citep{Metzger2014}.

The observed X-ray emission from novae is typically characterized by thermal radiation from the hot WD and/or the shocked matter~\citep{Orio2001A&A,Mukai2001,Mukai2008,Chomiuk2014Natur}. The soft X-ray emission with an effective temperature of $<0.1$~keV can reach a luminosity of $L_X>10^{36}~{\rm erg~s^{-1}}$. The 
soft X-ray emission is thought to originate from a hot WD sustained by residual nuclear burning. As the ejected material spreads out, the surrounding environment becomes optically thin, allowing the soft X-ray emission from the hot WD becomes visible~\citep{Page2020AdSpR}. The emission in the soft X-ray band defines the Supersoft Source (SSS) phase, when the ejecta became transparent to X-rays from the central source~\citep{Bode2008book}. Observations during SSS phase of some novae have confirmed  quasi-periodic oscillations (QPOs) with a period in the range of 10 to 100~s~\citep{page2020-mnras,beardmore2019,ness2015,orio2022ApJ}.
Several possibilities for  the origin of the QPOs  have been suggested: for example, 
the spin modulation of the WD with a strong magnetic field is the most likely explanation for the QPOs~\citep{Drake2021,Li2022ApJ}.  
Another possibility is the g-mode (buoyancy) pulsations driven by an ionisation-opacity instability, which is expected to produce a  period of the order of 10~s or less \citep{osborne2011-apj,drake2003-apj,wolf2018-apj}. Consequently, the exact origin of QPO in novae remains unclear.

The nova V1716 Sco (also known as PNV J17224490-4137160; Nova Sco 2023) was discovered by Andrew Pearce on 2023 April 20.678 UT and visually confirmed on April 20.705 UT at magnitude
8.0~\footnote{\url{http://www.cbat.eps.harvard.edu/unconf/followups/J17224490-4137160.html}}. Data from the All-Sky Automated Survey for Supernovae (ASAS-SN) revealed a pre-discovery detection on 2023 April 20.410 UT \citep{Sokolovsky2023ATel16018}, and spectroscopically confirmed as a classical (Fe II) nova \citep{Walter2023ATel16003,Shore2023ATel16004}. In this paper we adopt the date of the first ASAS-SN detection as the eruption start time $t_0$ = UT 2023-04-20.410 = JD 2460054.910 = MJD 60054.410.

\cite{Cheung2023ATel16002} reported the detection of gamma-ray emission ($>5 \sigma$ significance level) using \verb|Fermi|-LAT data taken from 2023-04-21 00:00:00 to 24:00:00 UTC.  The $>$100~MeV flux averaged over that  period was $F_{\gamma}=({6.5}\pm{2.1})\times 10^{-7}~{\rm ph~cm^{-2}s^{-1}}$ and the photon index=1.9$\pm$0.2.
Hard X-rays were detected by \verb|NuSTAR| on 2023 April 21.89 UT, with the X-ray spectrum being consistent with a heavily absorbed thermal plasma \citep{Sokolovsky2023ATel16018}. 
\verb|Swift| detected the X-ray emission 
on 2023 May 01 and confirmed an additional soft component appeared after 2023 May 31 \citep{Page2023ATel16069}. \cite{Dethero2023ATel16167} reported the power spectrum of V1716 Sco using \verb|NICER| data and confirmed  a strong pulsation around a period of $\sim 80$~s, which
may be a SSS QPO as this period appears to be slightly varying.

In this study, we report the results of detailed GeV and X-ray analyses of nova~V1716 Sco. In Section 2, we describe the data analysis conducted by \verb|Swift|, \verb|NICER|, \verb|NuSTAR| and \verb|Fermi|-LAT observations. Section 3 shows the results of data reduction. The discussion and summary are found in Sections 4 and 5, respectively.

\section{data reduction}
\begin{figure*}
    \centering
    \includegraphics[scale=0.56]{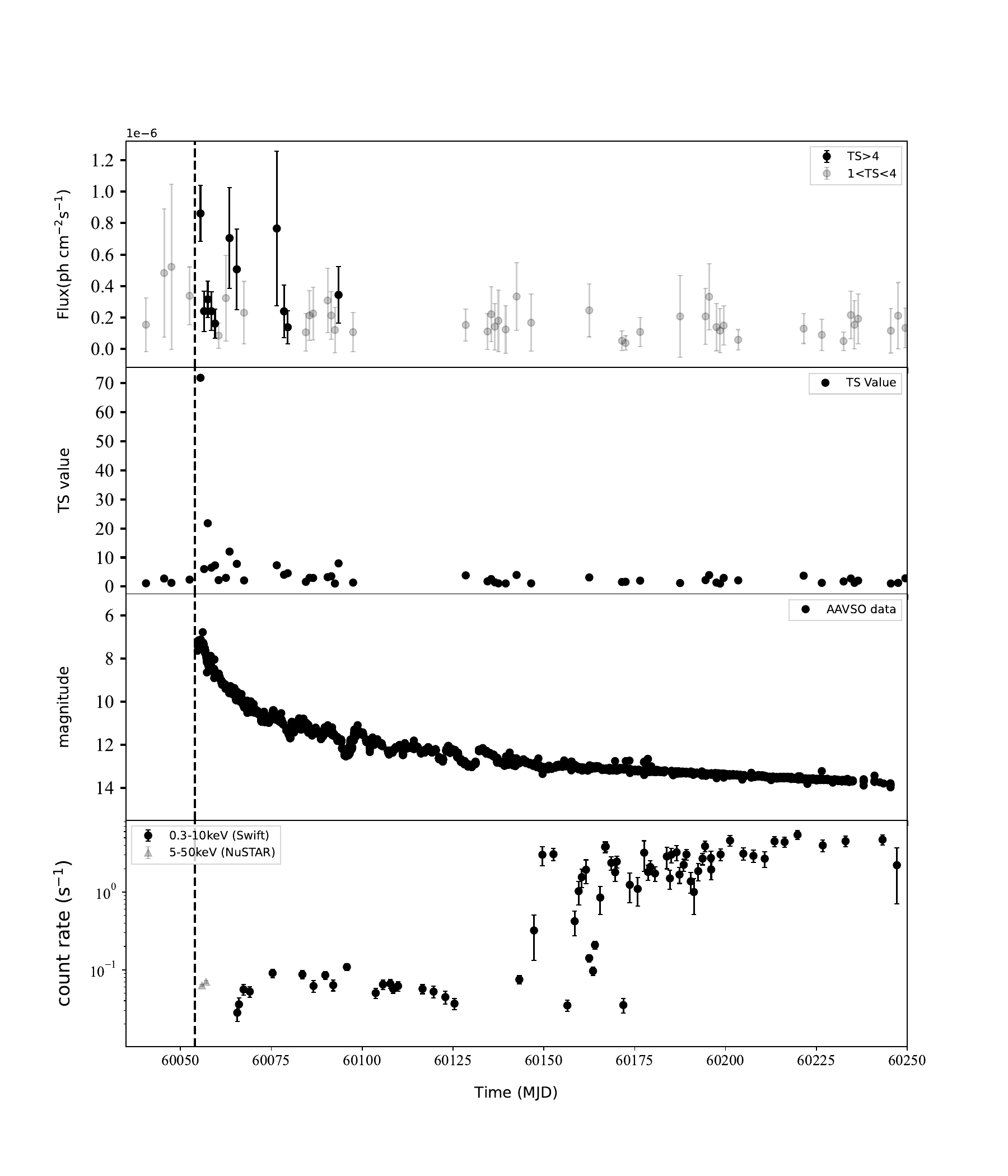}
    \caption{Top panel: The photon flux with 1-day bin using Fermi-LAT data. The black dots and the light black dots  are the flux with the TS value large than 4~(corresponding to a detection significance of $\sim$ 2~$\sigma$) and 1~($\sim$1~$\sigma$), respectively. Second panel: The evolution of the  TS values. Third panel: The AAVSO V-band light curve from optical data in site of \url{http://aavso.org/lcg}. Bottom panel: The count rate of X-ray data taken by the  Swift observations  (filled circles)
      and NuSTAR observations (triangles). The vertical dash line shows the epoch of nova eruption in optical bands.}
    \label{lighcurve-1d}
\end{figure*}
\subsection{Fermi-LAT data}
We performed a binned analysis using the standard \verb|Fermi|-LAT ScienceTool package, which is available from the \verb|Fermi|-LAT Science Support Center\footnote{\url{https://fermi.gsfc.nasa.gov/ssc/data/access/lat/}}.
We selected Pass 8 data in the energy band of 0.1-300\, GeV. The data for the fourth \verb|Fermi|-LAT catalog (4FGL DR4) were taken during the period August 2008 to August 2022 covering 14 years~\citep{4fgl-dr4,4fgl-dr3}.
We conducted a binned analysis using a gamma-ray emission model file based on the 4FGL DR4 catalog. To avoid contamination from Earth's limb, we included only events with zenith angles less than 90 degrees. Our analysis limited the  events from the point source or Galactic diffuse class (\verb|event class = 128|) and utilized data from both the front and back sections of the tracker (\verb|evttype = 3|).

For the analysis of the GeV emission from the target, 
we selected the data  with the energy above 100~MeV and the time epoch to cover from MJD~60040, which is $\sim$ 15 days before the detection of its optical eruption~(at 2023 April 20.410= MJD~60054.410), to MJD~60250. 
 We constructed a background emission model that incorporates both the Galactic diffuse emission (\verb|gll_iem_v07|) and the isotropic diffuse emission (\verb|iso_P8R3_SOURCE_V3_v1|) provided by the \verb|Fermi|-LAT Science Support Center. A gamma-ray emission model for the whole ROI was built using all sources in the fourth \verb|Fermi|-LAT catalog~\citep{FERMI2020} located within $20^{o}$ of the nova V1716 Sco, and the target is included in the model at the nova position of (R.A., decl.)=($17^{o}22^{'}44.88^{''},
-41^{o}37^{'}16.0^{''}$).

\subsection{NuSTAR data}
\verb|NuSTAR| observed V1716 Sco between 2023-04-21 21:36:56($t_0$ + 1.5d) to 2023-04-23 09:54:11 ($t_0$ + 3.0d) (ObsID:80801335002) after a day of optical eruption with a total exposure time 70~ks. For the analysis, we used the tasks of \verb|nupipeline| and \verb|nuproducts| to extract source and background spectra and light curves  from the focal plane modules A (FPMA) and B (FPMB). We generated the source and background extraction region files using \verb|DS9| by choosing a circular region of $\sim$ 50" radius centered on the source and the background region close to the source, respectively. We grouped the channels at least 30 counts per bin for \verb|NuSTAR| FPM A/B data.

\subsection{Swift-XRT data}
\verb|Swift| had continuously monitored V1716 Sco since the discovery of the nova eruption. We create the light curve and hardness ratio of the X-ray emission using the XRT web tool\footnote{\url{https://www.swift.ac.uk/user\_objects/}}~\citep{Evans2007A&A,Evans2009MNRAS}.  We use only grade 0 events in the analysis to minimize the optical loading.  To investigate the spectral properties, we downloaded the archival data from HEASARC Browse\footnote{\url{https://heasarc.gsfc.nasa.gov/cgi-bin/W3Browse/w3browse.pl}} and performed  the analysis with the HEASOFT version 6.31.1 and its SWIFTDAS package with the updated calibration files. The clean event lists were obtained using the task  \verb|xrtpipeline| of the HEASOFT and extract the spectrum using \verb|Xselect|. We grouped the source spectra to ensure at least 1 count per spectral bin and fit the spectra using \verb|Xspec|.

\cite{woodward2024} reported the analysis of the Swift data and fitted  the spectra  with 
the black body  emission model (hereafter BB model). 
Because of its large count rate in SSS phase,  the Swift data could be affected by pile-up. We evaluated the pile-up level by refereeing the analysis thread\footnote{\url{https://www.swift.ac.uk/analysis/xrt/pileup.php}}, and we found that the Swift data taken in SSS phase were indeed affected by the pile-up.  We removed the pile-up region when extracting the spectra file in \verb|Xselect|, and run the command \verb|xrtarf| to create an ARF corrected for the loss of counts caused by this annular exclusion. 

\subsection{NICER data}
Neutron Star Interior Composition Explorer (\verb|NICER|) observed the target in SSS phase and covered  from three month to  fourth month after the  optical eruption. We apply the standard task \verb|nicerl12| to extract the cleaned event file and  perform the barycentric time correction using the \verb|barycorr| task of \verb|HEASoft|. \cite{Dethero2023ATel16167} found the QPO with a period of $\sim 80$~s in \verb|NICER| data. For further studying of the QPO,  we use Lomb-Scargle (LS) periodogram~\citep{lomb1976,Scargle1982} and
search for the periodic modulation in the light curve (10~s time bin) in each data set. We confirmed the significant signal from 14 data sets, as shown in Table~\ref{table-lsp}. To perform the phase-resolved spectroscopy for each data set,  we extract the events of the on-/off-pulse phase using \verb|xselect| and the phase information
determined from the command \verb|efold| of \verb|Heasoft|. We apply \verb|niextspec| to extract
the spectra of on- and off-pulse phases and obtain the spectrum of the pulsed component by extract spectrum in off-pulse phase from that in on-pulse phase.  We use \verb|nicerl3-spect| task to create the response matrix file  and ancillary response file, and  utilize  \verb|grppha| task to group the generated spectrum such that the new grouping contains a minimum of 30 counts in each bin.

\section{results}
\subsection{GeV emission properities}
To describe the TS value and flux time evolution, we conducted a refit of the gamma-ray data in each bin using binned likelihood analysis ($gtlike$). We created a daily light curve to enable us  a more precise measurement of the epoch of the gamma-ray emission, as shown in Figure~\ref{lighcurve-1d}.
In the daily light curve of Figure~\ref{lighcurve-1d}, the TS value reached to the maximum value ($\sim 70$), which corresponds to a detection significance level larger  $8~\sigma$ (i.e., $\sqrt{TS}$ is about detection significance in $\sigma$). After the peak, the TS value decayed rapidly and the emissions with TS $>4$ ($\sigma>2$) were confirmed until $\sim 40$~days after the nova eruption.

\begin{figure}
    \centering
    \includegraphics[scale=0.35]{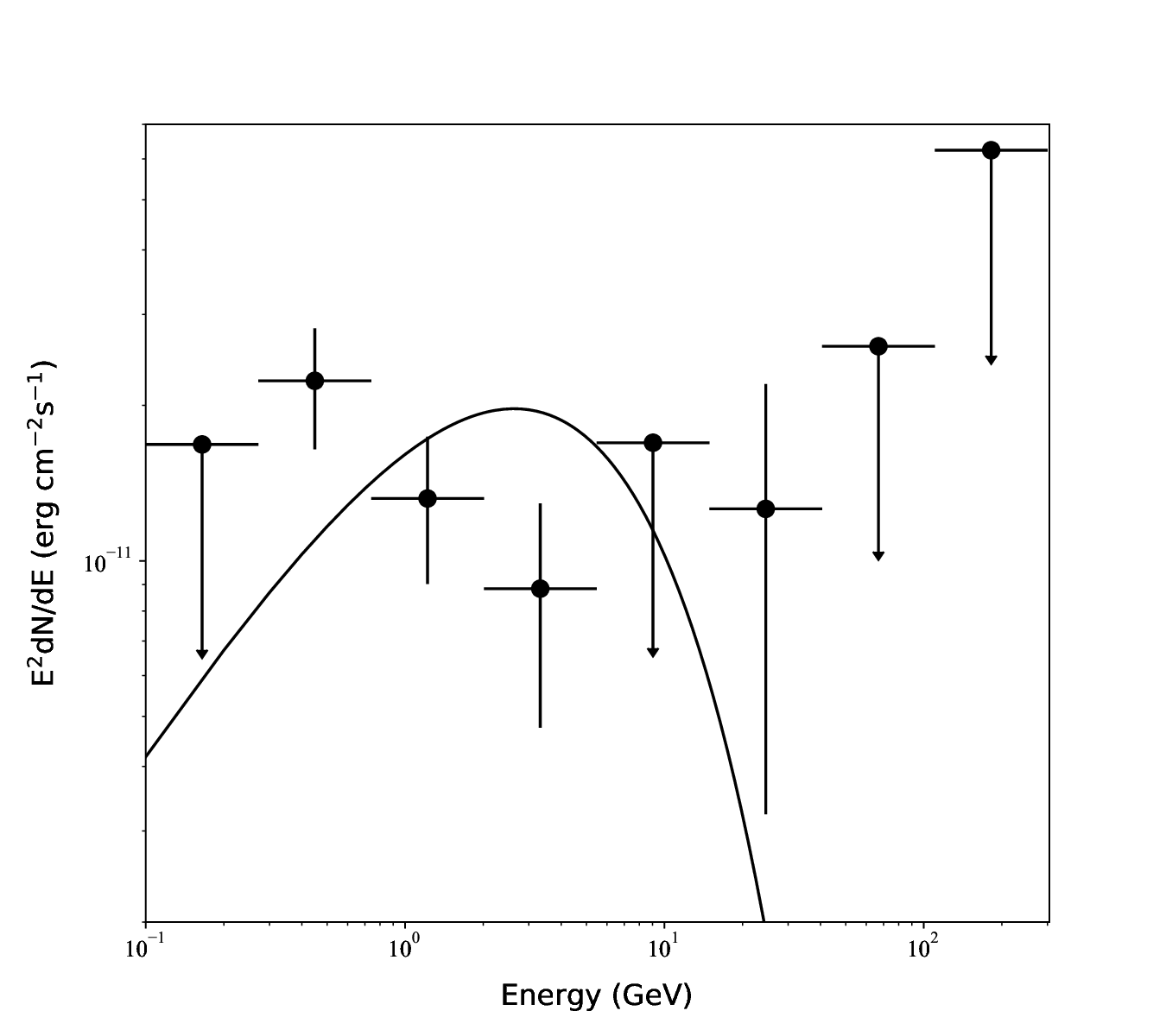}
    \caption{The gamma-ray spectrum of V1716 Sco observed by  Fermi-LAT. The black line shows the best-fitting models with the  equation~(\ref{eq2}).}
    \label{fig:spec-gamma}
\end{figure}

To generate the spectrum, we performed the likelihood analysis using the data obtained from MJD~60055 to MJD~60094, during which the emission with $TS>4$ were confirmed.  The gamma-ray spectrum can be well describe by a power-law function with an exponential cut-off, as describe:
\begin{equation}
\centering
\label{eq2}
    \frac{dN}{d{E}} \propto E^{-\gamma_{1}}{\rm exp}\left[-\left(\frac{E}{E_c}\right)^{\gamma_{2}}\right],
\end{equation}
where we fixed to $\gamma_2=2/3$. We obtained a power-law index 
of $\gamma_1=1.98(7)$ and a cut-off energy of $E_{c}=22.1(1)$~GeV. We obtained an averaged energy flux of  
$F_{\gamma}=1.4(1)\times 10^{-11}~{\rm erg~cm^{-2}s^{-1}}$ in 0.1-300 GeV bands. Figure~\ref{fig:spec-gamma} represents the 
spectrum in GeV bands. We also extracted the energy flux for the time bin that has $TS\sim 70$ and obtained $F_\gamma\sim  1.4(3)\times 10^{-10}~{\rm erg~cm^{-2}~s^{-1}}$.

\begin{figure}
     \centering
     \includegraphics[scale=0.36]{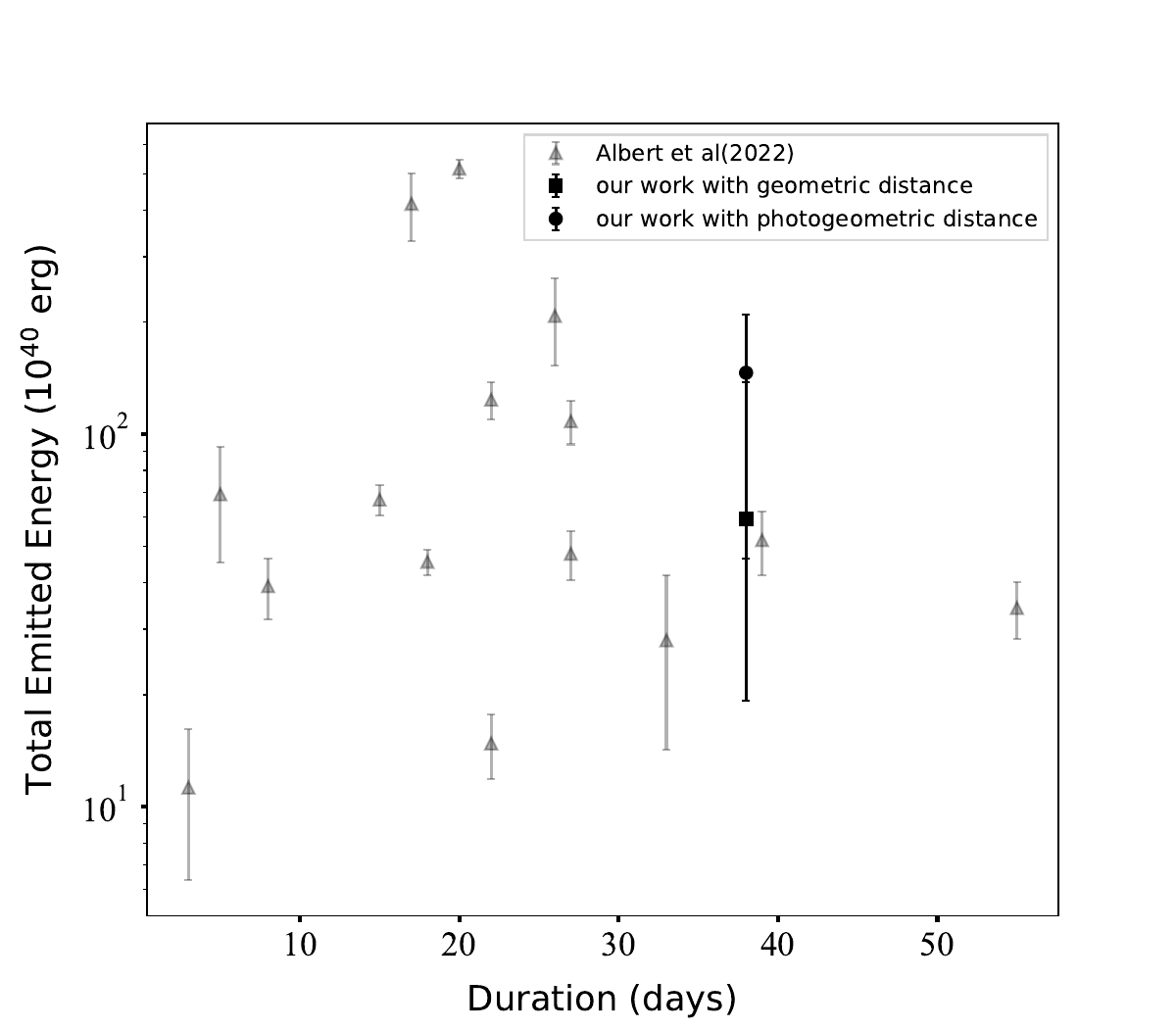}
     \caption{Total emitted GeV energy and  duration of GeV emission of the novae detected in the GeV range. The duration is defined by the epoch during which the emission with $TS>4$ lasted.  The data illustrated with triangles are taken from \protect\cite{Albert2022}. The symbols with the square and filled circles correspond to the emission energy of nova V1716 Sco estimated with the geometric distance and photogeometric distance, respectively.}
     \label{t-energy}
 \end{figure}

According to GAIA archive\footnote{\url{https://dc.g-vo.org/gedr3dist/q/cone/form}}, the distance of nova V1716 Sco is estimated as $3.16^{+2.13}_{-1.62}$~kpc or $4.96^{+1.75}_{-1.06}$~kpc in geometric or photogeometric measurements~\citep{Bailer-Jones2021AJ}. To estimate the total emitted energy in the gamma-ray bands, we integrated the daily flux detected with  $TS>4$ and obtained $0.59(1)\times 10^{42}~{\rm erg}$ for $d$=3.16~kpc and  $1.46(7)\times 10^{42}~{\rm erg}$ for d=4.96~kpc.  Figure~\ref{t-energy} shows the total emitted energy and duration of the  gamma-ray emission of GeV novae. It can be seen that the total emission gamma-ray of V1716 Sco is similar to those of other novae detected by \verb|Fermi|-LAT.

\subsection{X-ray light curve}

\subsubsection{Long term variability}
\begin{figure}
    \centering
    \includegraphics[scale=0.35]{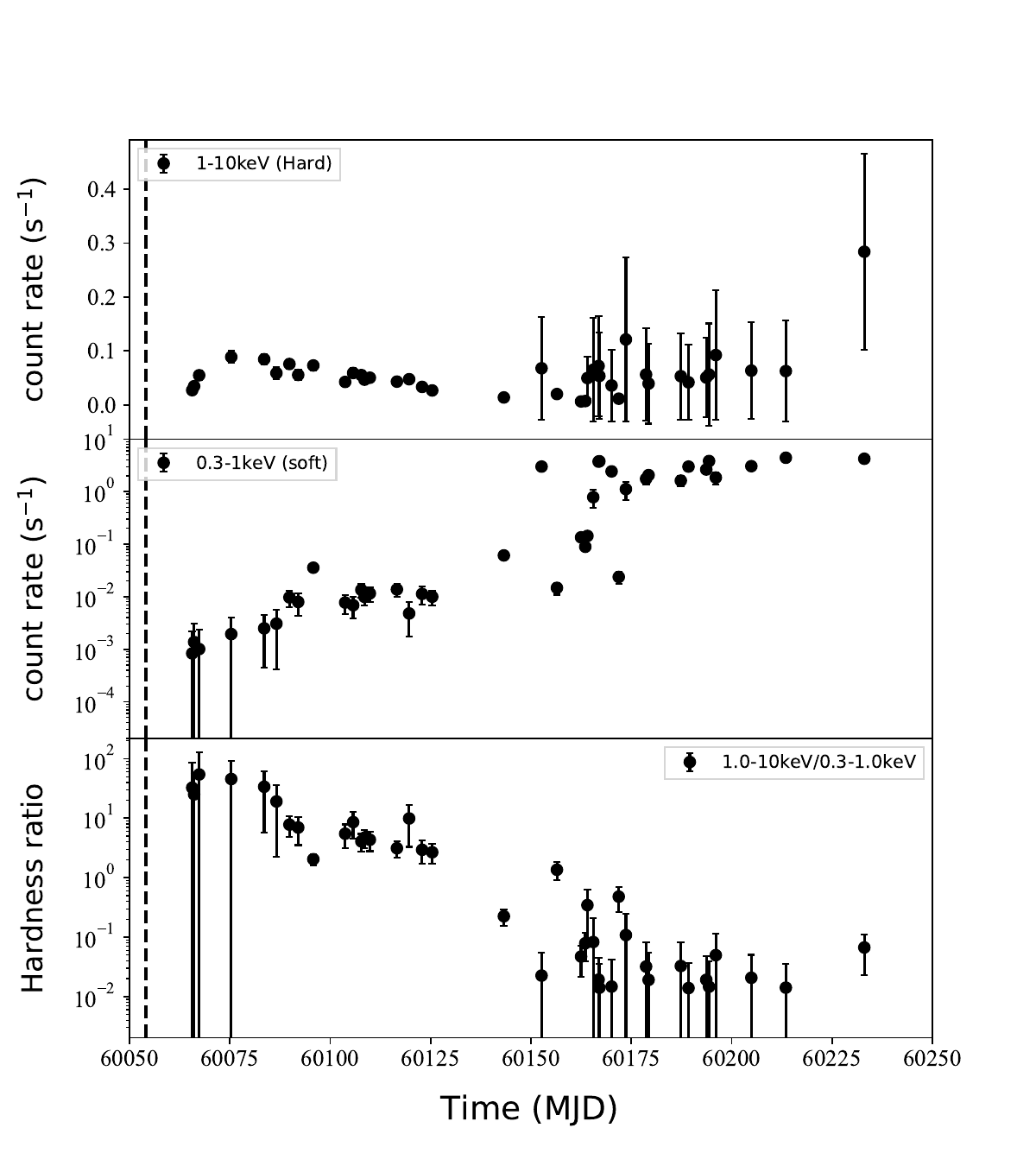}
    \caption{The light curves of hard X-rays (top panel) and soft X-rays (middle panel) taken by the  Swift observation. The bottom panel shows the hardness ratio (1.0-10.0~keV/0.3-1.0~keV)(bottom panel). The data points are after consideration of the pile-up effect. }
    \label{hardness}
\end{figure} 

The bottom panel of Figure~\ref{lighcurve-1d} shows 
the long term light curves taken by \verb|NuSTAR| and \verb|Swift|, and  Figure~\ref{hardness} represents the temporal evolution of the hardness of the X-ray emission measured by the \verb|Swift| observation.  As we can see in the figures, the counts rate measured by the Swift rapidly increases after its first detection around MJD~60060, and its X-ray emission at the initial stage is very hard with a hardness ratio of $>10$. The count rate measured by the \verb|Swift| reached to the local maximum value at about MJD~60075 (about 20 days eruption).

After the local peak at $\sim$MJD~60075, the X-ray count rate in the hard band (1.0-10~keV bands) and the hardness decreased, while
the count rate in soft bands (0.3-1.0~keV bands) rapidly increased. The hardness ratio in the third panel of Figure~\ref{hardness} decreased to $\sim 1$ at around MJD~60100, suggesting that the X-ray emission entered the SSS phase, as mentioned in \citet{Page2023ATel16069}. As the figure shows, the SSS phase was observed until the end of the monitoring by the Swift that was about 200~days after the eruption.

\begin{figure}
    \centering
\includegraphics[scale=0.4]{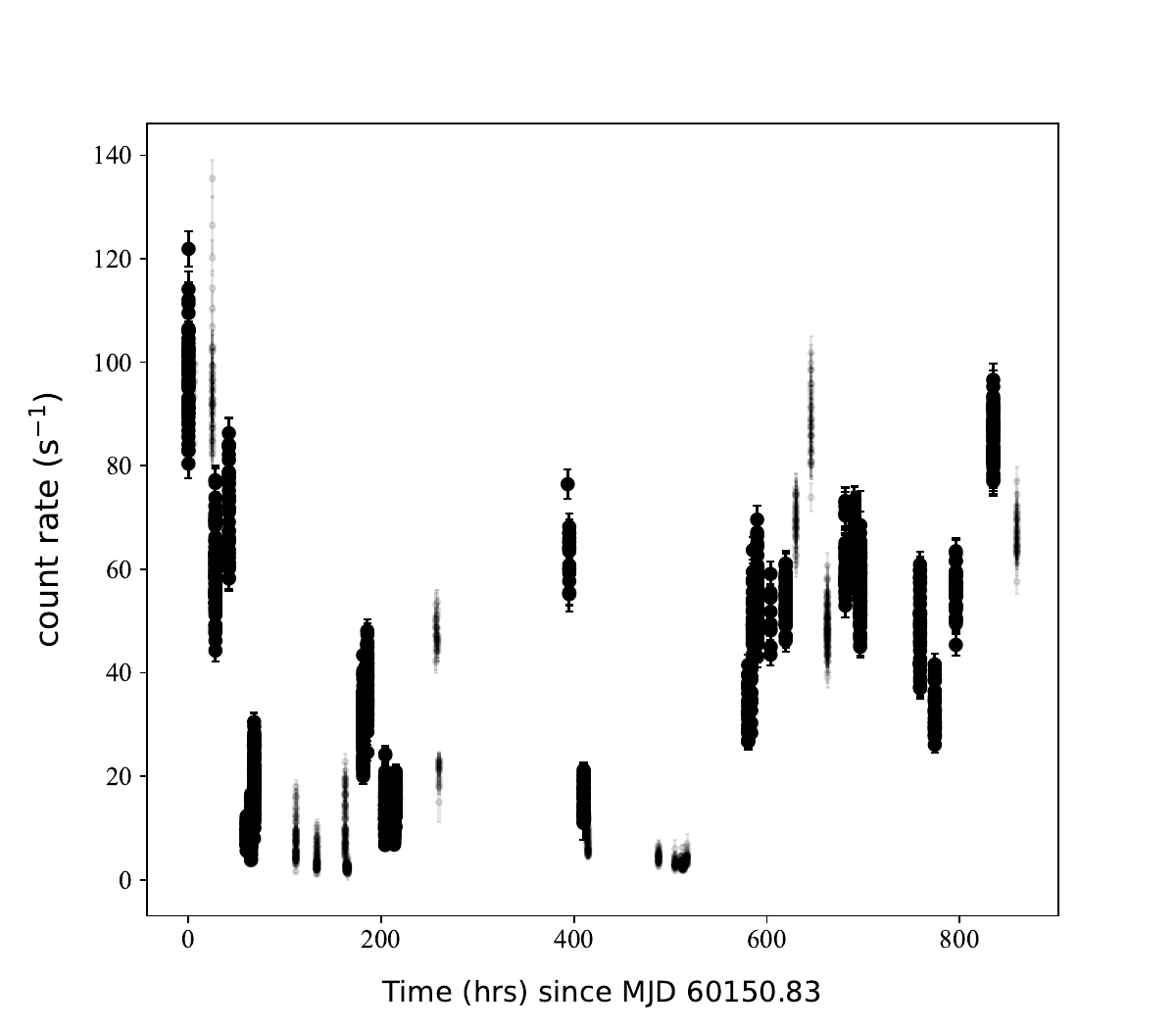}
    \caption{Light curve (10~s time bins) of nova V1716 Sco  measured by the  NICER observations carried out from  2023 July 25 to 2023 August 30.  The large and small circles indicate the dataset with and without the detection QPO signals, respectively. }
       \label{nicer-countsrate}
\end{figure}

\begin{figure}
    \centering
\includegraphics[scale=0.21]{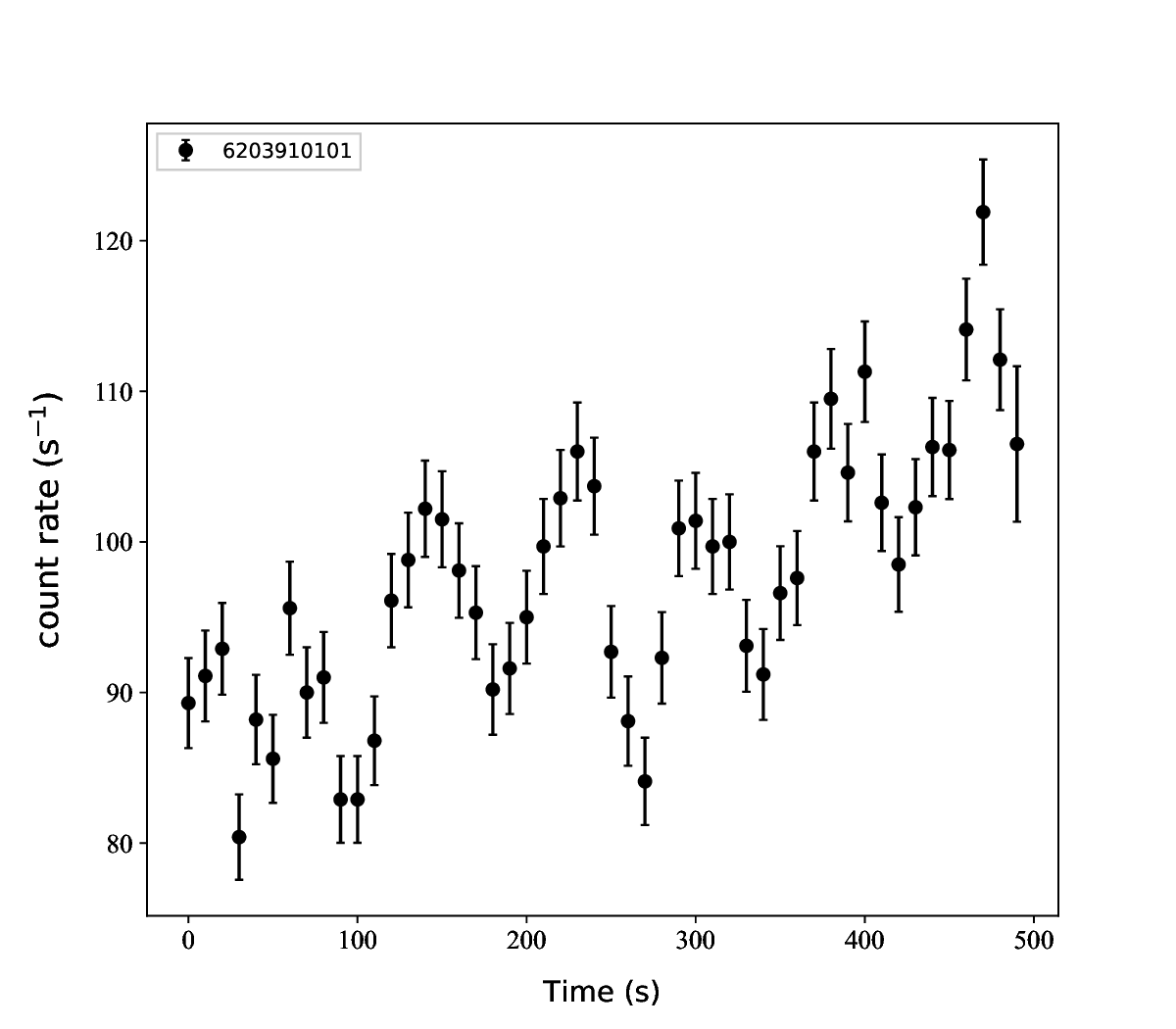}
\includegraphics[scale=0.21]{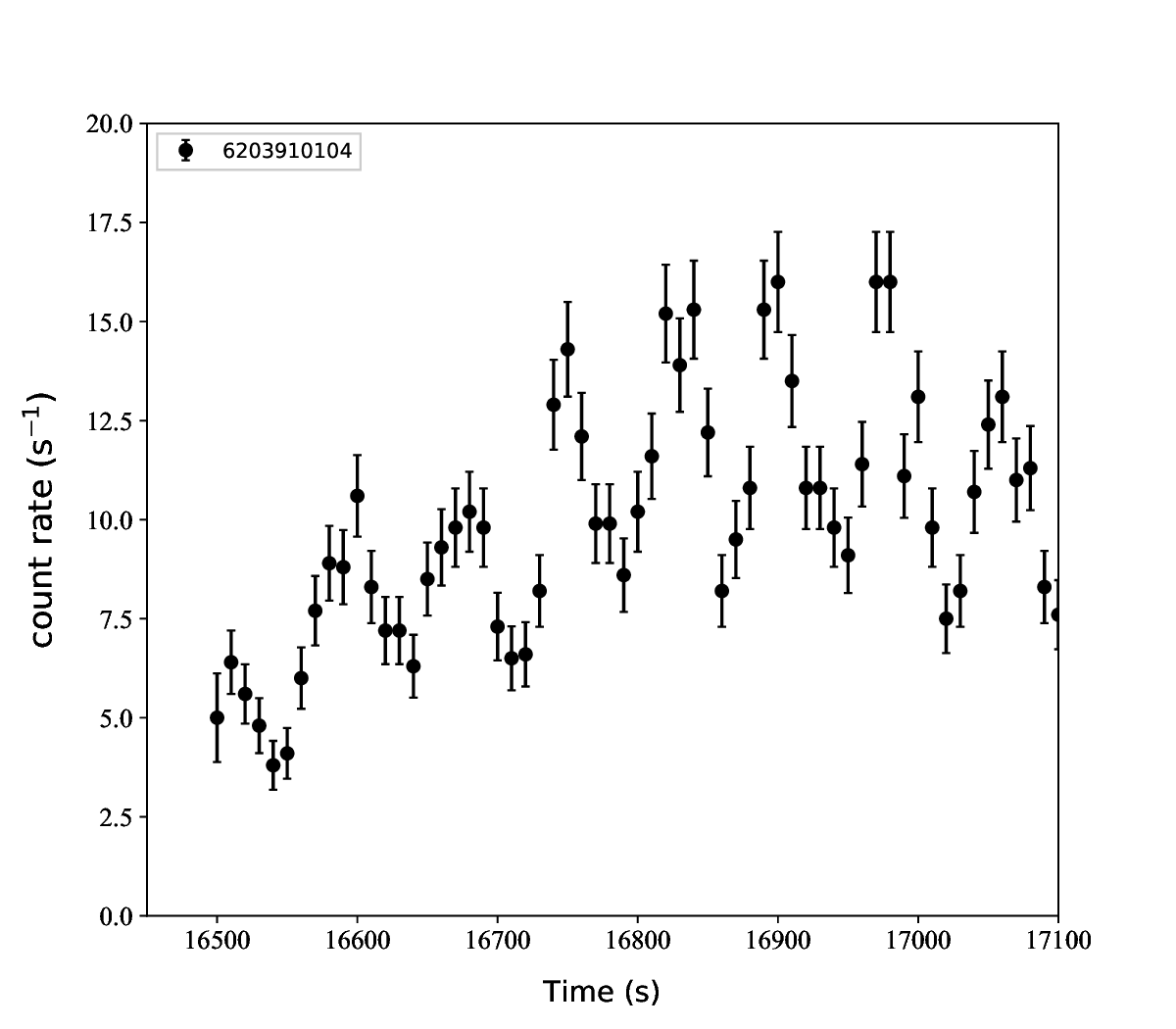} 
    \caption{Light curve in 0.3-10~keV energy bands of nova V1716 Sco using  the NICER data of obsID~6203910101~(left) and 6203910104~(right). These two examples illustrate  the  modulation with the period of $\sim 80$~s. }
    \label{nicer-rate1a4}
\end{figure}

\begin{table}[h]
\centering
\caption{The results from LS periodogram of each NICER data in full energy.}

\begin{tabular}{|c|c|c|c|}
    \hline
    \hline
ObsID&Start time(MJD)& Exposure(s)&Period(s)
  \\
         \hline
6203910101&60150.83&494&80.33$\pm$4.25\\
6203910103&60151.99&1077&75.69$\pm$7.23\\
6203910104&60153.34&2635&77.98$\pm$4.76\\
6203910109&60158.38&1619&77.94$\pm$6.09\\
6203910110&60159.33&1724&76.76$\pm$5.79\\
6203910112&60167.22&529&76.84$\pm$10.36\\
6203910113&60168.04&1130&77.88$\pm$7.30\\
6203910117&60175.02&1696&79.08$\pm$6.01\\
6203910118&60175.99&502&76.02$\pm$10.64\\
6203910121&60179.22&1683&78.57$\pm$8.52\\
6203910123&60182.45&435&78.41$\pm$11.72\\
6203910124&60183.09&326&78.79$\pm$13.71\\
6203910125&60184.00&402&77.50$\pm$12.10\\
6203910126&60185.61&536&80.40$\pm$10.94\\
  \hline
    \end{tabular}
    \label{table-lsp}
\end{table}

\begin{figure}
    \includegraphics[scale=0.4]{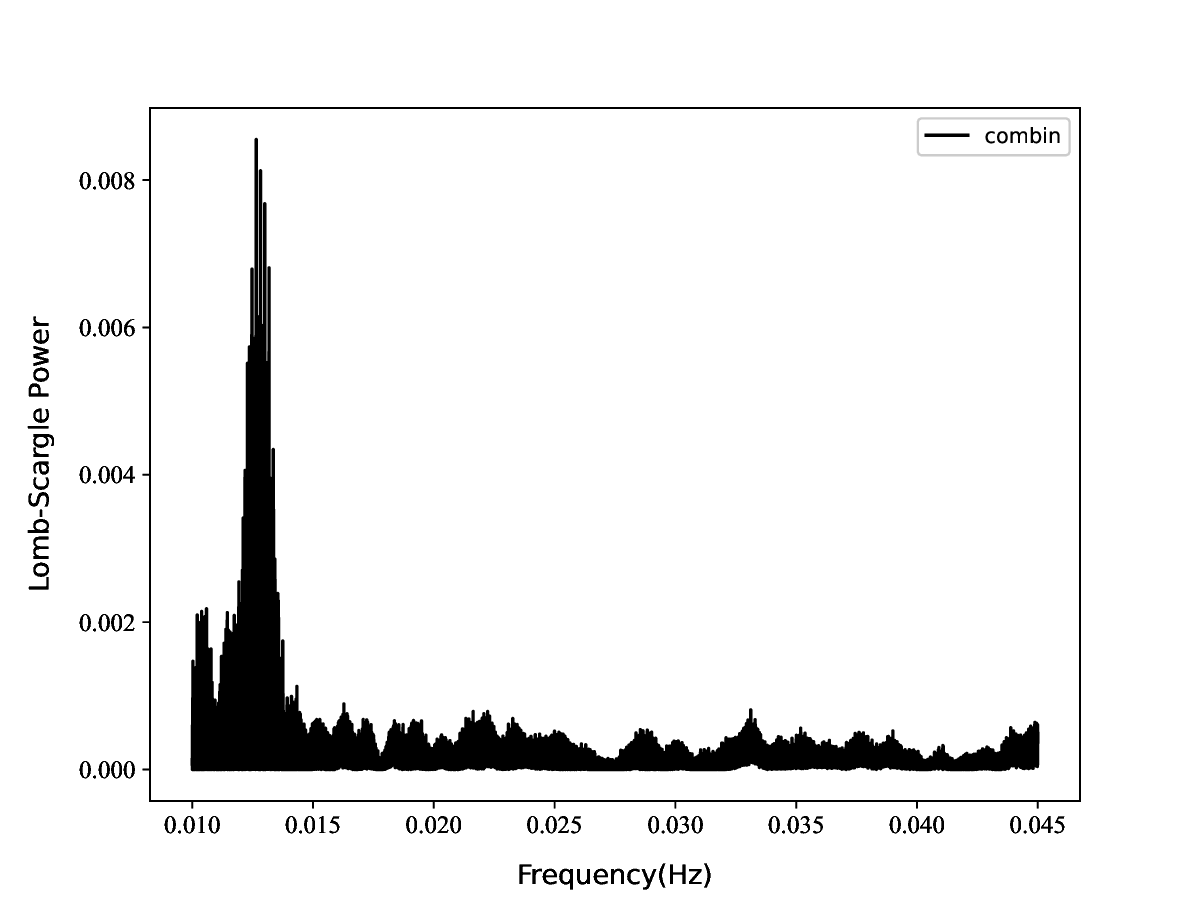}
    \includegraphics[scale=0.4]{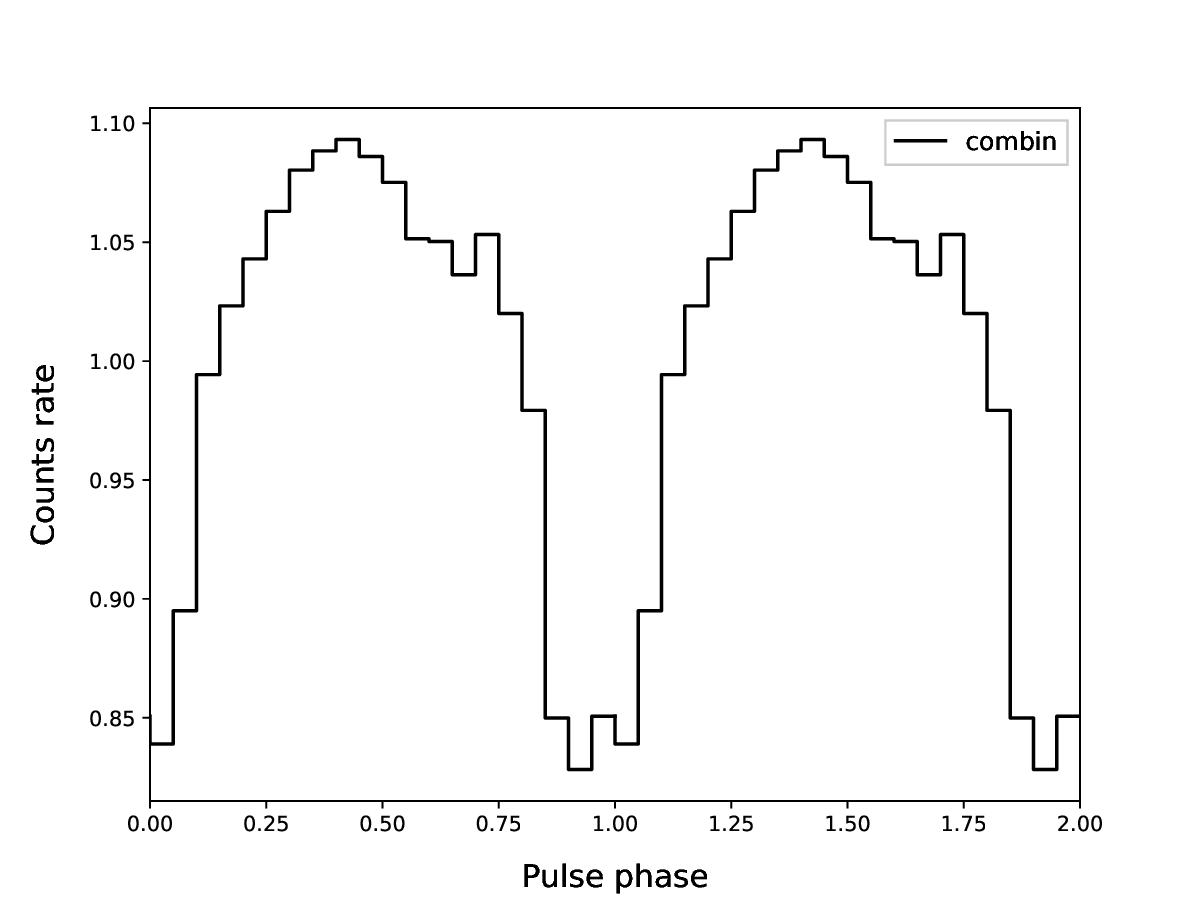}
    \caption{Upper panel: The LS periodorgram using the NICER data presented  in Table~\ref{table-lsp}. The peak frequency corresponds to a period of 79.10~s. Bottom panel:  Folded light curve with the period of  79.10~s. Two pulse phases are shown for clarity. }
    \label{fold-all-combin}
\end{figure}

\subsubsection{QPO}
Figure~\ref{nicer-countsrate} shows the long-term light curve taken by  \verb|NICER| in the SSS phases, and Figure~\ref{nicer-rate1a4}
 presents the light curves taken at MJD~60150.83 (left panel) and MJD~60153.34 (right panel).  In Figure~\ref{nicer-rate1a4}, we can clearly confirm a periodic modulation with a period of $\sim$ 80~s, as also reported by \cite{Dethero2023ATel16167}. 
Combining these 14 data sets presented in Table~\ref{table-lsp}, we obtain  the period of $79.10\pm 1.98$~s in LS periodogram, as the top panel of Figure~\ref{fold-all-combin} shows. In the bottom panel of   Figure~\ref{fold-all-combin}, we present the pulse profile folded with the period of 79.10~s.

\begin{figure}[h]
    \centering
    \includegraphics[scale=0.38]{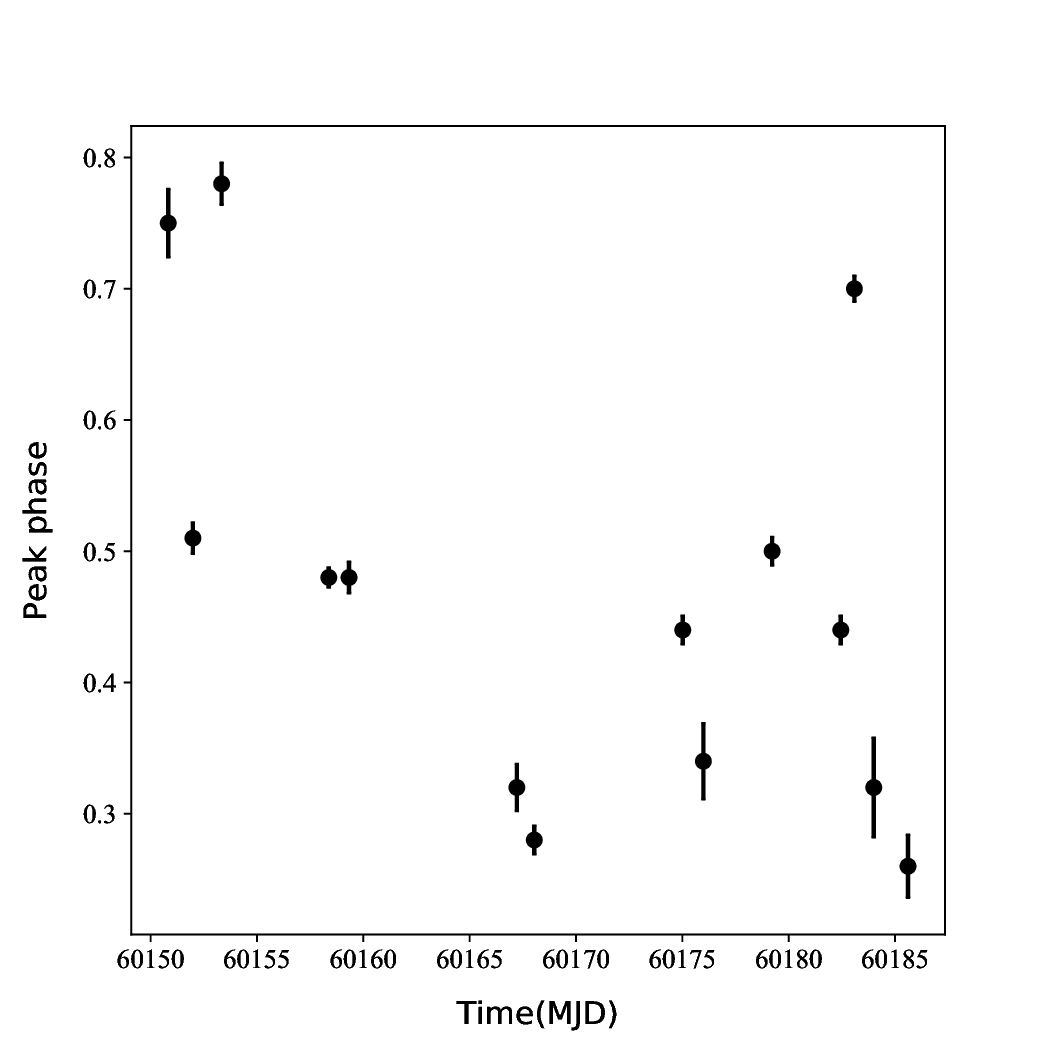}
    \caption{The position of the pulse peak peak phase against time of  NICER observations. }
    \label{fig:peak-phase}
\end{figure}

To investigate for any temporal evolution of the period signal, we created the LS-periodogram and obtain the period for each data set presented in Table~\ref{table-lsp}. 
As the fourth column in Table~\ref{table-lsp} shows, the periods obtained with different data sets are consistent within the error. Hence  we do not  confirm temporal evolution of the period using  the LS periodograms. We then investigate the temporal evolution of the pulse profiles. Figure~\ref{nicer-fold-every} presents 
the pulse profile each data set folded by $79.10$~s determined by whole 14 data sets and indicates that each pulse profile is described by a single broad peak.  We fit the pulse profile by a Gaussian function and obtain the fitting  parameters 
with the Monte Carlo method. As Figure~\ref{fig:peak-phase} shows, we observe that the position of the pulsed  peak indicates a rapid temporal variation. This variation of the peak position  makes difficult to create an ephemeris of the period evolution, suggesting that  the period is not stable over time. The unstabe periodic signal suggests that the emission region on the stellar surface shifts if the modulation is caused by the spin of the WD. Alternatively,  the  periodic modulation might be originated from the different mechanism (e.g. stellar  oscillation) rather than the spin of WD.

\subsection{X-ray spectrum}
\label{spectrum}
\begin{figure}
    \centering
    \includegraphics[scale=0.28,angle=270]{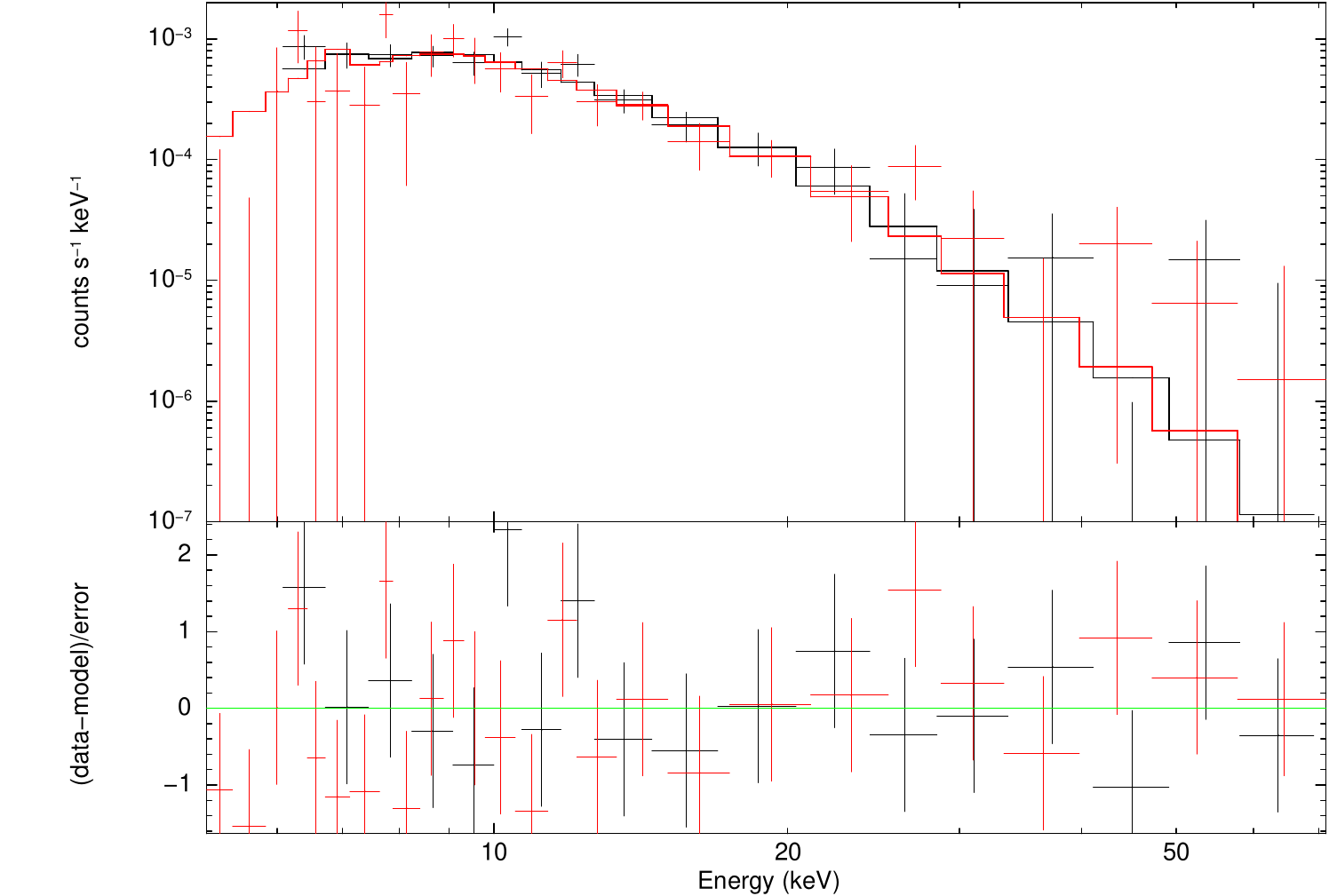}
    \caption{Phase-average spectrum in 5.0-70.0~keV taken by NuSTAR.  the observed spectra measured by  the  FPMA (black spectrum) and FPMB detectors (red spectrum) are simultaneously fitted by an absorbed optically thin thermal emission  model~(vphabs*vpaec).}
    \label{spec-nustar}
\end{figure}

Preliminary analysis  of \citet{Sokolovsky2023ATel16018} suggests that the X-ray spectrum taken by \verb|NuSTAR| data is consistent with that of a heavily absorbed thermal plasma with a temperature of $k_BT=31 \pm 13$ keV, where  $k_B$ is the Boltzmann constant, and the hydrogen column density of $N_H= (82 \pm 15) \times 10^{22} \text{cm}^{-2}$. In this study, we first fit the spectrum of \verb|NuSTAR| data  (Figure~\ref{spec-nustar})  with a power-law model (\verb|phabs*polwerlaw|) or  a thermal 
plasma emission (\verb|phabs*apec|). For the power-law model, 
we obtain a photon index of $2.04\pm 0.43$  and a hydrogen column density of $N_H =(58.45 \pm 24.06)\times10^{22}~\text{cm}^{-2}$, with $\chi^2 =46/44$. 
For the thermal 
plasma emission model, we obtain the temperature of $k_BT\sim 34.13 \pm 20.20~$keV, which is consistent with the result of \citet{Sokolovsky2023ATel16018},  and the column density of  $N_H =(42.28 \pm 17.06)\times10^{22}~\text{cm}^{-2}$ with the statistic of $\chi^2 =45/44$.

{The absorption of the X-ray at the initial stage will be mainly caused by the ejector of the nova eruption. Hence we fit the \textbf{NuSTAR} spectra with an absorbed thermal plasma model of \textbf{vphabs*vapec}, fixing the abundances to the values derived in \cite{woodward2024}.} The best fit for this
model provides  the column density of $N_{H}$(vphabs) = $(7.40\pm 2.33)\times 10^{22} ~\text{cm}^{-2}$ and temperature of $k_BT_{vapec} = 13.48 \pm 5.76~\text{keV}$ with $\chi^{2}=36/44$. These fitting parameters are slightly smaller than the values  using the interstellar absorption (\verb|phabs|).  It is reasonable to assume that the observed thermal X-ray emission at the initial stage
originates from the plasma heated by a shock produced by the ejector~\citep{Steinberg2018MNRAS}, and the observed temperature may be related with the shock speed as 
\begin{equation}
    k_{B}T \approx 1.2~{\rm keV}~\left(\frac{v_s}{10^3~{\text{km~s}^{-1}}}\right)^{2},
\end{equation}
where $v_s$ is the shock velocity.  The spectrum taken by the \verb|NuSTAR| implies that the shock speed can reach to $v_s>10^3~{\rm km~s^{-1}}$.

We obtain an unabsorbed flux of $(2-3)\times 10^{-12}~\text{erg}~\text{cm}^{-2}~\text{s}^{-1}$ in 5.0-70.0~keV energy bands which corresponds to a luminosity  of $\sim 10^{33}~{\rm erg~s^{-1}}$. During  \verb|NuSTAR| 
observation, the TS-value of the \verb|Fermi|-LAT observation reached to the maximum value of $\sim 70$, as Figure~\ref{lighcurve-1d}. It is therefore found that around the GeV peak, the gamma-ray flux ($F_\gamma\sim 1.4\times 10^{-10}~{\rm erg~cm^{-2}~s^{-1}}$) is about two orders of magnitude larger than the radiation luminosity in 5.0-70~keV bands, namely    $L_{\gamma}/L_X~\sim 100$, which is typical value for nova detected in GeV bands (see \citet{Gordon2021}).

The \verb|Swift| observation covered from the initial stage,  where the hardness ratio was much greater than unity,  to the SSS phase. For the initial stage, we employ
the thermal plasma emission model (\verb|phabs*(apec)|), as the observed  spectra (Figure~\ref{combin-swift-s}) did not indicated the thermal emission from the WD in the soft X-ray bands, and it was likely dominated by the emission from the gas heated by the shock.  This is the typical  situation for the initial stage of X-ray emissions from novae because the emission from the WD was heavily absorbed. The 
fitted parameters are summarized in the first four lines of the Table~\ref{table-model-1} before SSS and Table~\ref{table-model} in SSS.  We  see that the decrease tendency of the hydrogen column density, which is likely due to the expansion of the ejector. Within the error, on the other hand, we
do not confirm a  significant evolution of the temperature of the plasma.
The luminosity in 0.5-10.0 energy bands is estimated to be $L_{apec}\sim 10^{34}~\text{erg s}^{-1}$.

\begin{figure}[h]
    \centering
    \includegraphics[scale=0.35]{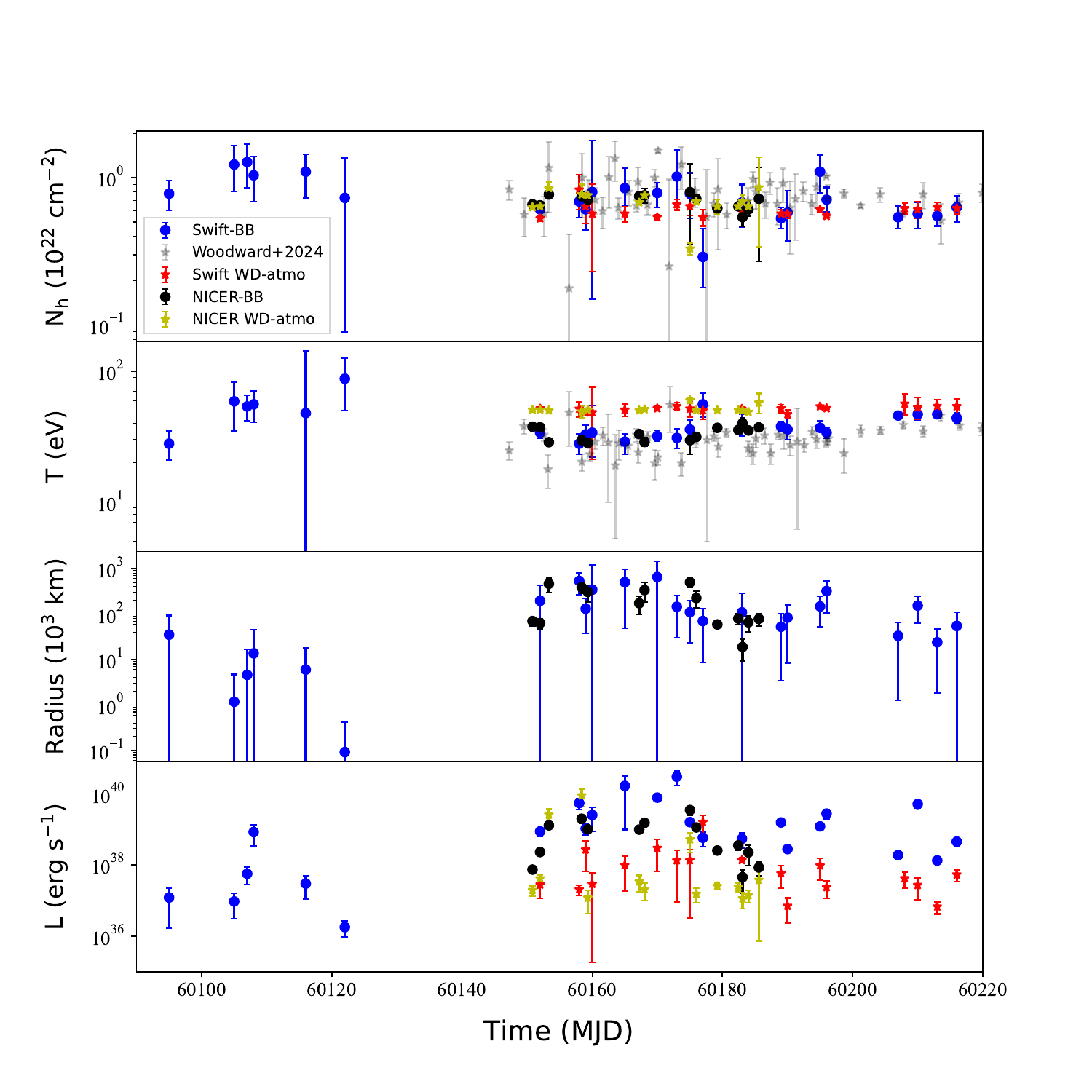}
    \caption{Fitting results for the spectra taken by Swift and Nicer observations. The symbols with the blue and black colors represent 
    the results of BB model for Swift ad Nicer data, respectively, while the symbols with the red and yellow colors 
    are the result of WD atmosphere models. For the WD atmosphere model, the radius is fixed to $10^9$~cm. The light dark star data are read from \citet{woodward2024}. }
    \label{fig:x-spec-para}
\end{figure}

For the SSS phase, we first fit the observed spectra with a BB model (\verb|phabs*(bbodyrad)|). The results of the fitting for \verb|Swift| data are summarized in Table~\ref{table-model} and Figure~\ref{fig:x-spec-para}.
Our results about Swift data are consistent with \citet{woodward2024}.
In the table, we see that the temperature is of the order of 30-50~eV, which is consistent with the typical values in the SSS phase of other novae~\citep{Chomiuk2021,page2022mnras,orio2022ApJ,Orio2023ApJ}. We see, however, that  the emission size ($R_{BB}>10^{10}$~cm) is larger than the radius of the WD, despite large error bars. Moreover, the luminosity ($L_{bb}$) in 0.3-1.0 energy bands can be the supper-Eddington value of $L_{bb}>10^{38}~{\rm erg~s^{-1}}$ with a distance of 4~kpc.

It has been argued that the BB model is inadequate for the emission in the SSS phase of the novae~\citep{Krautter1996,page2022mnras} and more appropriate model of  the emissions from the {hot WD's photosphere} has been investigated~\citep{Balman2001,suleimano2024}. We therefore fit the SSS component using the WD atmosphere emission model implemented in \verb|Xspec|~\citep{suleimano2024}.  During the fitting process, we find that the central value of the emission region is of the order of $R_{atom}\sim 10^9$~cm,  which is more consistent with the WD's radius, 
but it is not well constrained due to the large error bars. Hence, we fix the radius of the emission region to $R_{atmos}=10^9$~cm in this study. The results of the fitting are summarized in Table~\ref{table-model-swift-wd-atom} and in Figure~\ref{fig:x-spec-para}. 
As Figure~\ref{fig:x-spec-para} shows, the effective temperature of the atmosphere model is slightly larger than that of the pure BB model,  while its
expected X-ray luminosity becomes of the order of or smaller than the Eddington value.

{Due to its high timing resolution, the data collected by NICER will not be affected by  the pile-up with a current flux level of V1716~Sco, and it may be  more suitable for investigating the spectral properties during the SSS phase. Additionally, NICER data allows for phase-resolved spectroscopy, which can isolate the emission of the WD photosphere. As indicated in Figure~\ref{nicer-fold-every}, we obtain the pulsed spectrum by subtracting the off-pulse spectrum from the on-pulse spectrum. The results of the spectral fitting of the NICER data are  presented in  Tables~\ref{table-nicer-spec} for the BB model and~\ref{table-model-swift-wd-atom} for the  WD atmosphere model, respectively}. Figure~\ref{fig:x-spec-para}  also compares between the results of the \verb|Swift| and \verb|NICER| observations.  We can see in Figure~\ref{fig:x-spec-para} that the results of the fitting with \verb|NICER| data are consistent with those with \verb|Swift| data.  The X-ray luminosity in 0.2-1.0~keV bands with most of  datasets is well constrained and it is $L\sim (0.1-0.3)\times 10^{38}~{\rm erg~s^{-1}}$, although some datasets still predict the super-Eddington luminosity. This supper-Eddington luminosity may indicate a  flare-like activities.

The WD atmosphere model implemented in \verb|Xspec| fit the WD's gravity. We calculate the inferred WD's mass from the obtained gravity by fixing  $R_{WD}=10^9~{\rm cm}$ (fourth columns in Table~\ref{table-model-swift-wd-atom} or \ref{table-nicer-spec-wd-atom}). We find that with $R_{atmos}=10^{9}$~cm, the inferred mass is  consistent with the WD's mass ($M_{WD}<1.4M_{\odot}$, Chandrasekhar limit) within errors, but it is not well constrained due to  its large error sizes.

\begin{table*}
\centering
\caption{The parameters of X-ray data from Swift, using the model of apec or BB.}
\begin{tabular}{|c|c|c|c|c|c|c|c|}
    \hline
    \hline
ObsID&date(MJD)&N$_H$($10^{22}~cm^{-2}$)
&$k_{B}T$$_{bb}$~(eV)
&$R_{bb}(10^{3}~km)$
&$L_{bb}(10^{38}~\text{ergs}~\text{s}^{-1}$)
&$kT$$_{apec1}$~(keV)
&C-Stat/dof  \\
         \hline
00015990004&60066&4.42$\pm$1.64&-&-&-&4.56$\pm$5.15&34/34\\
00016007006&60067&6.95$\pm$2.03&-&-&-&2.33$\pm$0.80&46/42\\
00016007014&60075&2.33$\pm$0.56&-&-&-&2.41$\pm$0.81&57/64\\
00016023006&60089&1.57$\pm$0.42&-&-&-&1.90$\pm$0.47&62/55\\
00016023010&60095&0.78$\pm$0.18&28.85$\pm$7.10&35.32$\pm$57.96&
0.12$\pm$0.10&-&90/89\\
00016066006&60105&1.23$\pm$0.42&59.14$\pm$24.87&1.18$\pm$3.52&
0.094$\pm$0.063&-&41/48\\
00016066010&60107&1.28$\pm$0.43&54.02$\pm$12.53&4.60$\pm$12.13&
0.56$\pm$0.29&-&52/49\\
00016066012&60108&1.04$\pm$0.35&56.74$\pm$15.05&13.70$\pm$32.34&
8.51$\pm$5.01&-& 48/58\\
00015990006&60116&1.10$\pm$0.37&48.50$\pm$9.62&6.00$\pm$12.28&
0.29$\pm$0.18&-& 46/53\\
00015990008&60122&0.73$\pm$0.64&88.93$\pm$38.97&0.092$\pm$0.33&
0.017$\pm$0.0085&-& 34/44\\
 \hline
    \end{tabular}
    \label{table-model-1} \\
\end{table*}

\begin{table*}
\centering
\caption{The parameters of X-ray data from Swift in SSS, using the model BB.}
\begin{tabular}{|c|c|c|c|c|c|c|c|}
    \hline
    \hline
ObsID&date(MJD)&N$_H$($10^{22}~cm^{-2}$)
&$k_{B}T$$_{bb}$~(eV)
&$R_{bb}(10^{3}~km)$
&$L_{bb}(10^{38}~\text{ergs}~\text{s}^{-1}$)
&$kT$$_{apec1}$~(keV)
&C-Stat/dof  \\
         \hline
00015990016  & 60152 & 0.61$^{+0.071}_{-0.071}$& 34.85$^{+3.86}_{-3.10}$& 196.36$\pm$230.84 & 8.86$\pm$2.44&-&36/35 \\
00015990019&60158&0.69$^{+0.15}_{-0.15}$&28.78$^{+5.37}_{-4.93}$&536.65$\pm$271.49&307.31$\pm$104.27&-&40/36 \\
00015990020&60159&0.61$^{+0.16}_{-0.17}$&33.83$^{+5.61}_{-4.77}$&131.45$\pm$93.72&58.55$\pm$18.96&-&32/34 \\
00015990021  & 60160&  0.80$^{+0.99}_{-0.65}$ & 34.1$^{+12.29}_{-21.83}$& 345.26$\pm$880.98& 25.55$\pm$16.89 &-&8/11\\
00015990026 &  60165 & 0.85$^{+0.31}_{-0.21}$& 29.06$^{+4.44}_{-572}$& 502.79$\pm$454.31& 168.87$\pm$158.93&-&23/27\\
00015990030  & 60170 & 0.79$^{+0.14}_{-0.16}$ & 32.37$^{+3.65}_{-2.84}$& 654.64$\pm$820.17& 78.98$\pm$6.95&-&36/35\\
00015990034 &  60173 & 1.02$^{+0.53}_{-0.34}$&31.49$^{+5.22}_{-5.34}$&145.32$\pm$115.31&307.30$\pm$130.47&-&19/23\\
00015990036  & 60175 & 0.78$^{+0.30}_{-0.25}$ & 36.04$^{+6.66}_{-4.67}$&110.41$\pm$86.71& 16.12$\pm$1.75&-&16/25\\
00015990038  & 60177 & 0.29$^{+0.17}_{-0.10}$&42.71$^{+12.63}_{-10.92}$&70.42$\pm$61.84&5.85$\pm$2.56 &-&29/28\\
00015990042  & 60183 & 0.68$^{+0.22}_{-0.22}$& 37.91$^{+9.32}_{-6.44}$&109.51$\pm$173.86&5.46$\pm$2.46 &-&23/28\\
00015990048 &  60189&  0.53$^{+0.10}_{-0.08}$& 41.22$^{+3.55}_{-3.35}$&52.74$\pm$49.32&15.75$\pm$1.05&-&56/40\\
00015990049  & 60190&  0.58$^{+0.24}_{-0.21}$ & 36.62$^{+7.57}_{-5.87}$&83.50$\pm$75.04 &2.80$\pm$0.36&-&16/23\\
00015990054  & 60195 & 0.67$^{+0.32}_{-0.31}$&37.57$^{+4.87}_{-1.75}$&147.80$\pm$96.05& 12.23$\pm$1.08&-&35/34\\
00015990055  & 60196&  0.70$^{+0.16}_{-0.12}$&34.40$^{+3.22}_{-3.32}$&321.56$\pm$217.59&28.02$\pm$8.18&-&44/36\\

00015990059 &  60207  &0.54$^{+0.10}_{-0.08}$&45.72$^{+3.44}_{-3.22}$ &33.38$\pm$32.10&1.89$\pm$1.05&-&34/42\\
00015990060  & 60210  &0.57$^{+0.11}_{-0.12}$& 47.00$^{+5.36}_{-4.35}$&153.88$\pm$91.21 &52.18$\pm$11.68&-&47/37\\
00015990061 &  60213 & 0.55$^{+0.10}_{-0.08}$&47.95$^{+3.48}_{-3.21}$&24.03$\pm$22.18&1.34$\pm$0.17&-&51/47\\
00015990062  & 60216 &  0.63$^{+0.12}_{-0.13}$& 44.21$^{+4.36}_{-3.42}$&54.94$\pm$56.26&4.54$\pm$1.02&-&53/43 \\

  \hline
    \end{tabular}
    \label{table-model} \\
\end{table*}

\begin{table*}
\centering
\caption{The parameters of X-ray data from Swift, using the model of WD atmosphere in SSS phase. }
\begin{tabular}{|c|c|c|c|c|c|c|}
    \hline
    \hline
ObsID&Start time(MJD) 
&$N_H$($10^{22}~\text{cm}^{-2}$)
&$k_{B}T$$_{bb}$~(eV)&M$_{WD}$/M$_\sun$
&$L_{atmos}(10^{38}~\text{ergs}~\text{s}^{-1}$)
&C-Stat/dof  \\
\hline 
00015990016  & 60152 & 0.53$\pm$0.021& 52.05$\pm$1.33&1.63$\pm$0.43
&0.27$\pm$0.16&33/35 \\
00015990019&60158&0.83$\pm$0.22&69.16$\pm$6.11&2.63$\pm$5.26&1.14$\pm$0.36&39/36 \\
00015990020&60159&0.68$\pm$0.15&68.96$\pm$6.87&1.27$\pm$0.97&15.05$\pm$11.35&43/38 \\
00015990021 & 60160& 0.57$\pm$0.34 & 48.73$\pm$27.43&0.99$\pm$5.57&0.29$\pm$0.28 &9/11\\
00015990026 &  60165 & 0.57$\pm$0.068 & 50.87$\pm$5.70& 1.52$\pm$1.87&  0.99$\pm$0.81&22/27\\
00015990030  & 60170 &0.54$\pm$0.023 &  52.24$\pm$1.70& 1.69$\pm$0.60& 3.00$\pm$2.33&37/35\\
00015990034 &  60173 & 0.66$\pm$0.056&54.24$\pm$3.71&1.97$\pm$1.51&1.35$\pm$1.26&19/23\\
00015990036  & 60175 & 0.64$\pm$0.087 & 51.92$\pm$7.41&1.51$\pm$2.19& 1.35$\pm$1.32&15/25\\
0005990038  & 60177 & 0.54$\pm$0.07& 50.22$\pm$7.26&1.05$\pm$1.64&16.12$\pm$8.34 &26/28\\
00015990042&60183&0.59$\pm$0.029&51.81$\pm$1.95&1.60$\pm$0.59&1.37$\pm$0.21& 31/36\\
00015990048&60189&0.57$\pm$0.035&52.12$\pm$3.54&1.49$\pm$1.05&0.58$\pm$0.36& 48/45\\
00015990049&60190&0.57$\pm$0.039& 47.13$\pm$3.67&0.81$\pm$0.67&0.071$\pm$0.047& 53/44\\
00015990054&60195&0.61$\pm$0.018& 54.01$\pm$1.06&1.93$\pm$0.40&0.97$\pm$0.59& 41/44\\
00015990055&60196&0.55$\pm$0.015& 52.27$\pm$0.95&1.66$\pm$0.31&0.23$\pm$0.12& 52/44\\
00015990059&60207&0.62$\pm$0.053& 56.68$\pm$44.40&3.09$\pm$5.60&0.42$\pm$0.20& 50/45\\
00015990060&60210&0.61$\pm$0.077&53.00$\pm$10.07&1.39$\pm$2.79&0.27$\pm$0.16& 47/45\\
00015990061&60213&0.63$\pm$0.049&54.35$\pm$6.52&1.61$\pm$2.14&0.067$\pm$0.025& 43/45\\
00015990062&60216&0.62$\pm$0.052&54.07$\pm$7.26&1.58$\pm$2.33&0.53$\pm$0.19& 57/45\\
  \hline
    \end{tabular}
    \label{table-model-swift-wd-atom} \\
\end{table*}

\section{Discussion}

\subsection{X-ray emission region in early stage} 

As Figure~\ref{hardness} illustrates, the significant detection of the hard X-ray (1-10~keV) bands of the \verb|Swift| observation began at approximately 10~days after the optical eruption, during which gamma-ray emission was still detectable flux level for \verb|Fermi|-LAT observation. \citet{Gordon2021} summarized the X-ray behaviors observed  by \verb|Swift| for 13 Novae detected by \verb|Fermi|-LAT observations.  In their figure~3, we can see that the classical novae that have a main-sequence star do not show the X-ray detection by the \verb|Swift| until the gamma-ray have faded below the sensitivity of \verb|Fermi|-LAT observation. In the sample of \citet{Gordon2021}, only V407 Cyg is the case of a concurrent \verb|Swift| X-ray/gamma-ray detection, but it contains a red-giant companion star, which is  different from the typical classical novae.
Concurrent detection by  \verb|Swift| X-ray/gamma-ray observations are usually confirmed for the  recurrent novae\citep{cheung2022,buson2019,page2020-mnras}, in which the companion star is red-giant. Nova V1716 Sco, therefore, may be  rare  example of the  classical nova with a concurrent \verb|Swift| X-ray/gamma-ray detection.

The presence or absence of the X-ray emission detected by \verb|Swift| during the detection of the gamma-ray can help diagnose the environment of the emission regions. In cases where the companion star is a red giant, as seen in recurrent novae, the binary system is likely surrounded by a gas, due to a mass loss from the red giant. The nova ejecta collide with this surrounding gas, creating an expanding external shock~\citep{Bode2008book,HESS2022,zheng2022}. Due to the less dense environment in the external shock region, X-ray emissions from the heated gas may escape from the absorption while gamma-ray emissions are still detectable by \verb|Fermi|-LAT. 

For the case when the companion star is a main-sequence star, the nova emission will be dominated by the process occurred at the internal shock, as the circumbinary space and/or  an intrabinary space are not dense enough to form the external shock that produces an observable emission~\citep{Mukai2008,Chomiuk2014Natur,Steinberg2020}. The X-ray emission  from the shock-heated gas in the initial stage  will be significantly absorbed by the ejector, causing  a delay in the first detection  by \verb|Swift|  until  the gamma-ray emissions have faded below \verb|Fermi|-LAT's sensitivity threshold.

Our target, V1716 Sco, has been classified as the classical novae based on a spectroscopic information~\citep{Walter2023ATel16003,woodward2024}. Its X-ray detection by \verb|Swift| and \verb|NuSTAR| during the gamma-ray detection, however, indicate  the emission from the shock located in the less dense environment. \verb|NuSTAR| observation indicates the shock speed at approximately 2~days after the eruption is about $v_s\sim 3\times 10^8~{\rm cm~s^{-1}}$, indicating the radius of the expanding ejecta 
is of the order of $R_{ej}\sim v_s\times$ 2~days$\sim 5\times 10^{13}$~cm. \cite{woodward2024} estimate $M_{ej}\sim 4 \times 10^{-4}M_{\odot}$ as the ejected mass. If the mass ($M_{ej}$) is ejected in a scale of the days and the emission originates from a shock deeply embedded in the ejector, the expected column density at the epoch of \verb|NuSTAR| observation may  be of the order of 
\[
\Sigma\sim \frac{M_{ej}}{(4\pi R^3_{ej}m_p/3)}\times R_{ej}\sim 5\times 10^{25}~{\rm cm^{-2}},
\]
where $m_p$ is the proton mass.  The expected column density would be about two orders of magnitude  larger  than $\sim 10^{23}~{\rm cm^{-2}}$ of the the observed value. The first detection by the \verb|Swift| observatory was approximately 10 days after the eruption, and it measured that the speed of the shock is still of the order of $v_s\sim 2-3\times 10^{8}~{\rm cm~s^{-1}}$. Hence, 
if  the X-ray emission originated from the internal shock emission embedded in the ejector, the expected column density would be of the order of 
$\Sigma\sim 10^{24}~{\rm cm^{-2}}$, which still significantly larger than the observed value.  These estimations suggest that the observed X-ray emission in early stage is not originated from the internal shock deeply embedded in the ejector.

The observed column density  suggests the emission to be originated from the shock located in a less dense region and the X-ray emission in the initial stage is produced by the heated circumbinary matter.  To estimate the mass density of the unshocked materials in circumbinary space, we assume that the deceleration of the ejector motion had not yet begun  at the first  detection of the \verb|Swift| observations, since no significant change of the speed  was indicated by the \verb|NuSTAR| and the first \verb|Swift| observations. Using $v_s=3\times 10^8{\rm cm~s^{-1}}$ indicated by the observed temperature of the heated gas, we may estimate the swept mass by the external shock until the epoch of \verb|NuSTAR| observation as   
\[
\triangle M_1\sim 4\pi R_{ej}^2\Sigma m_p\sim  10^{-5}M_{\odot},
\]
where we used $R_{ej}=5\times 10^{13}~{\rm cm}$ and $\Sigma=5\times 10^{23}~{\rm cm^{-2}}$ from the observations.
The swept mass until the epoch of the first 
detection by \verb|Swift|, which is about 10~days after the eruption,  can be estimated as $\triangle M_2\sim 3\times 10^{-5}M_{\odot}$, 
where we used $R_{ej}=2.5\times 10^{14}~{\rm cm}$ and $\Sigma=5\times 10^{22}~{\rm cm^{-2}}$. From the values of the swept mass and the radius, we estimate the mass density of the unshocked gas  as $\rho_c\sim 6\times 10^{-16}~{\rm g~cm^{-3}}$.  This mass density will not be realized by the standard stellar wind from the main-sequence star, for which the mass density at $r\sim 10^{14}~{\rm cm}$ 
may be estimated as  $\rho_{wind}\sim 5\times 10^{-24}~{\rm g~cm^{-3}}$ with $\dot{M}_{wind}=10^{-13}M_{\odot}{\rm yr^{-1}}$ being  the typical mass-loss rate of a solar type star and the wind speed of $v_w=10^7{\rm cm~s^{-1}}$.

It has been argued that some novae in the initial stages generate a slow and less dense outflow, which is eventually overtaken by the fast dense outflow that causes  main outburst~\citep{Metzger2014,Chomiuk2021,Hachisu2022ApJ}. In this scenario, the X-ray emission produced in the forward shock region propagating into the  slow outflow would be escaped from the absorption.  Meanwhile the gamma-ray emission, which would be produced by the forward shock and/or reverse shock, is still detectable.  Our results of V1716 Sco is consistent with this scenario. 

As Figure~\ref{hardness} shows,  the count rate measured by the \verb|Swift| reached to a local peak at MJD~60075, which is about 20~days after the eruption. By assuming $v_s=3\times 10^8 {\rm cm~s^{-1}}$ and using the observed column density $N_H\sim (2-3)\times 10^{22}~{\rm cm^{-2}}$ (Table~2), 
the total mass swept by the shock is estimated to be $\triangle M\sim  10^{-4}M_{\odot}$.
This represents a significant fraction of the ejecta mass, $M_{ej} \approx 4 \times 10^{-4} M_{\odot}$. Consequently, we expect the ejecta to enter a deceleration phase around 20 days post-eruption. \cite{woodward2024} observed an ejecta speed to be $v_s \approx 10^8~{\rm cm~s^{-1}}$ about   $130$~days after the eruption, which aligns with our expectation, namely,  $3\times 10^8~{\rm cm~s^{-1}}(130{\rm days}/20~{\rm days})^{-1/3}\sim 1.6\times 10^8~{\rm cm~s^{-1}}$ if the mass density in the slow outflow
 has a radial distribution of $\rho\propto r^{-2}$, or $3\times 10^8~{\rm cm~s^{-1}}(130{\rm days}/20~{\rm days})^{-3/5}\sim 10^8~{\rm cm~s^{-1}}$ if $\rho\propto r^0$.

\subsection{Comparison of QPO of V1716 Sco with others}
The X-ray observations have confirmed the QPO signature in SSS phase for $\sim 10$ novae and the observed period is range  from 10~second to several thousand seconds~\citep{orio2022ApJ}. We reconfirmed that nova V1716 Sco is one of rare samples that show the QPO in the X-ray light curves and its period of $\sim 79.10$~s is within the range of the periods of the QPO novae. 
The folded light curve shows a single broad peak, which is also similar to those of other novae, such as, V407 Lupi~\citep{Aydi2018MNRAS}, V2491 Cyg~\citep{zemko2015apj} and  RS Oph~\cite{osborne2011-apj}.

As indicated in Figure~\ref{nicer-countsrate}, we did not confirm the QPO signals for some data sets; for example, some data sets have enough counts rate, such as OBS ID 6203910102, 6203910115, 620391016 with exposure more than 2ks, and counts rate around 100, but no QPO signals were confirmed with the LS-periodograms. The disappearance of the QPOs signal during the SSS phase was also confirmed for other novae 
 (e.g KT Eridan and Rs oph, ~\cite{ness2015};~\cite{Beardmore2010}). 

We could not find any periodic signal with \verb|NuSTAR| data.  It is expected that the X-ray emission detected by \verb|NuSTAR| originated from the shock emission, while the X-ray emission during SSS phase is produced by the WD.  The spin of the WD is more plausible explanation of the origin of the QPOs,  although  the confirmation of the periodic signal that is consistent with the QPO during SSS phase is necessary.  For example, QPO modulation of RS Oph  permanently disappeared  in the SSS, which may not be consistent with a simple WD rotation 
scenario \citep{beardmore2008}. We have demonstrated that the period of the modulations is not stable for nova V1716 Sco. Although this could be explained by the temporal variation of the geometry of the emission region on the WD surface, 
 the WD's spin scenario of the QPOs are still not conclusive.  
 The g-mode of the WD oscillation that is triggered by the $\varepsilon$-mechanism may be another possible reason to explain the QPOs of the novae.  \cite{wolf2018-apj} predicts, however, that non-radial g-mode pulsations model show tat stable pulsations with periods under $\sim$10 s, which is shorter than the typical period of QPOs of the novae.

\section{summary}

We conducted a joint analysis of \verb|NuSTAR|, \verb|Swift|, \verb|NICER|, and \verb|Fermi|-LAT observations of a classical nova V1716~Sco. 
We confirmed that the gamma-ray emissions emerged on a day after the optical eruption with a TS value of 70. The duration of the gamma-ray activity with a TS value above 4 lasted for 40 days and the estimated total gamma-ray output is  $\sim 10^{42}~{\rm erg}$ in total. 
 The X-ray emission of V1716 Sco was confirmed by \verb|NuSTAR| observations on a day after the optical eruption 
 and by the \verb|Swift| observations after 10 days  from the eruption.  The spectra taken within $\sim 20$ days from the eruption were fitted by the optically thin thermal plasma emission  with a temperature of several keV, and  were likely dominated by the emission from the gas heated up by the shock. The hardness ratio of the X-ray emission rapidly decreased with the time and the observed X-ray emission entered the SSS phase 
 about 45~days after the eruption. In SSS~phase, we fitted the data taken by $\verb|Swift|$ and $\verb|NICER|$ using the WD's atmosphere model and obtained the surface temperature of $k_BT\sim 50$~eV and the luminosity of $2\times 10^{37}~{\rm erg~s^{-1}}$, respectively. We cannot constrain the WD mass with the atmosphere emission model due to the quality of the data.  We reconfirmed the periodic oscillation with a period of 79.10$\pm$1.98~s in SSS phase, which is consistent with the finding reported by \cite{Dethero2023ATel16167}. Additionally, we demonstrated that the phase location of the pulse peak is not stable over time. This instability suggests that the hotter region on the white dwarf's surface shiftted over time.

 The gamma-ray and X-ray emission properties of V1716 Sco are similar to those of other classical novae. However, unlike other classical novae, the X-ray emission initially resolved by the \verb|Swift| occurred earlier, during a period when the gamma-ray emission was still at a detectable flux level for \verb|Fermi|-LAT observations. In particular, the \verb|Swift| observations indicated that the flux of the thermal 
 emission from the shock reached the peak value at  about 20~days after the eruption.  From the \verb|NuSTAR| and \verb|Swift| observations, 
 we interpreted that the ejector initially expanded with a constant speed of $v_s\sim 3\times 10^8~{\rm cm~s^{-1}}$ and then entered the deceleration phase at about 20~days after the eruptions.  Previous studies have argued that some classical novae produce a slow and less dense outflow, which  is eventually overtaken by the fast and dense outflow that causes the main outburst. The  hydrogen column density observed by the \verb|Nustar| and $\verb|Swift|$ suggested that the thermal X-ray emission from V1716~Sco was initially powered by the forward shock that propagates in the slow outflow.

\section*{Acknowledgements}
We acknowledge with thanks the variable star observations from the AAVSO International Database contributed by observers worldwide and used in this research. This work made use of data supplied by the UK Swift Science Data Centre at the University of Leicester.
J.T. is  supported by the National Key Research and Development Program of China (grant No. 2020YFC2201400) and the National Natural Science Foundation of
China (grant No. 12173014). L.C.-C.L. is supported by NSTC through grants  110-2112-M-006-006-MY3 and 112-2811-M-006-019. 

Facilities: Fermi, NuSTAR, Swift, NICER.

\bibliography{sample}


\appendix
\restartappendixnumbering
\section{The pulse profile from NICER data}
\label{appendix}
\begin{figure*}
    \centering
\includegraphics[scale=0.5,angle=270]{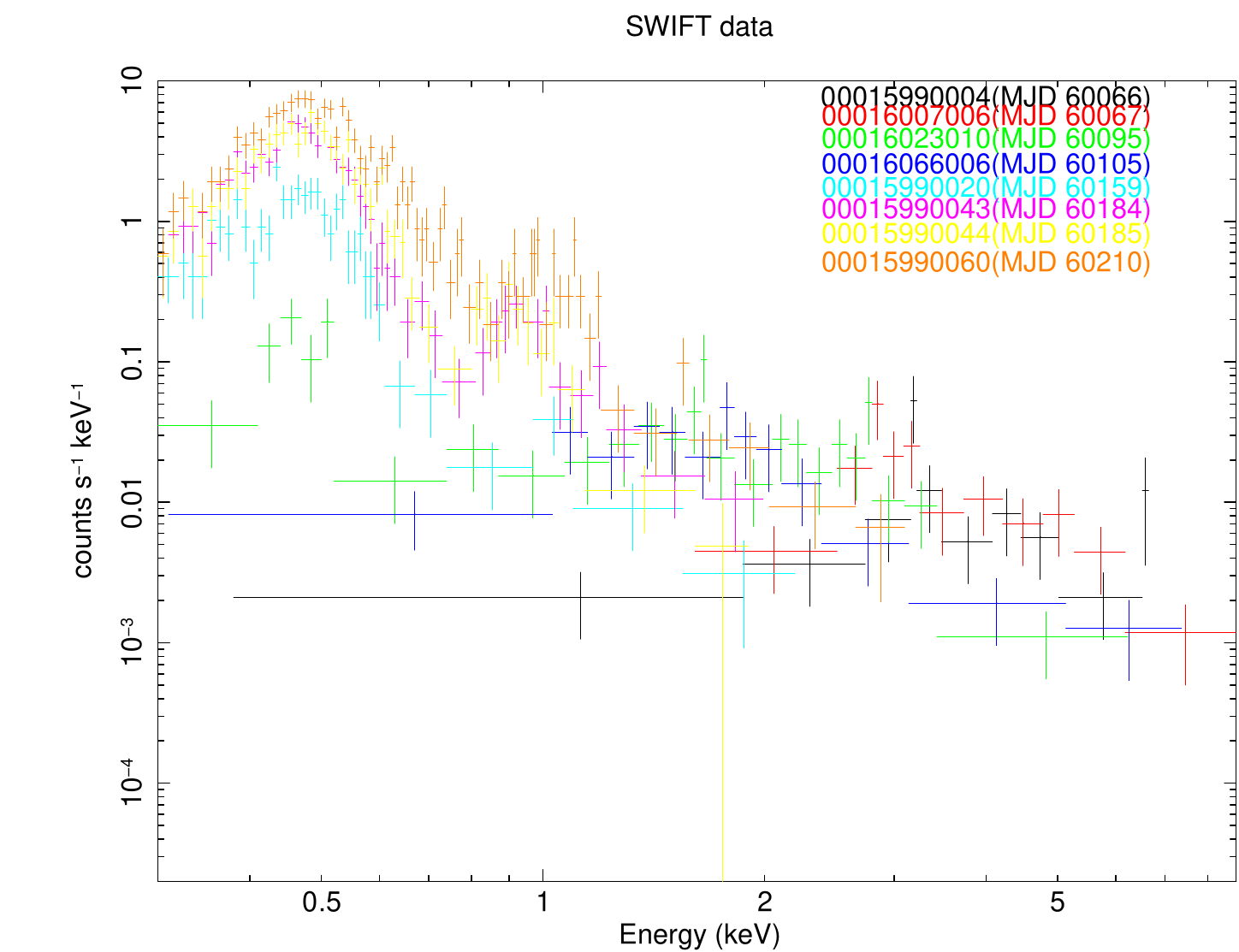}
\caption{The phase average spectra of each data sets from Swift-XRT.}
    \label{combin-swift-s}
\end{figure*}

Figure~\ref{combin-swift-s} presents spectra of the several data sets of \verb|Swift| observations. Result of the fitting of more  data sets are presented
in Table~\ref{table-model}  and Table~\ref{table-model-swift-wd-atom}.

Figure~\ref{nicer-fold-every}
shows  the light curves  folded  79.10~s using the \verb|NICER| data.
The the vertical dashed and solid lines in the figure  define the on-pulse  and off-pulse phases. Figure~\ref{nicer-resolve-every} presents the pulsed spectrum fitted with BB, Figure~\ref{nicer-resolve-every-wd-atom} shows the pulsed spectrum fitted with WD atmosphere, which
is created by subtracting the spectrum in the off-pulse phase from that in the on-pulse phase, with using the \verb|NICER| data. The results of the fitting using
the BB model and WD atmosphere model are summarized in  Tables~\ref{table-nicer-spec}
 and ~\ref{table-nicer-spec-wd-atom}, respectively.

\begin{figure*}
    \centering
    \includegraphics[scale=0.4]{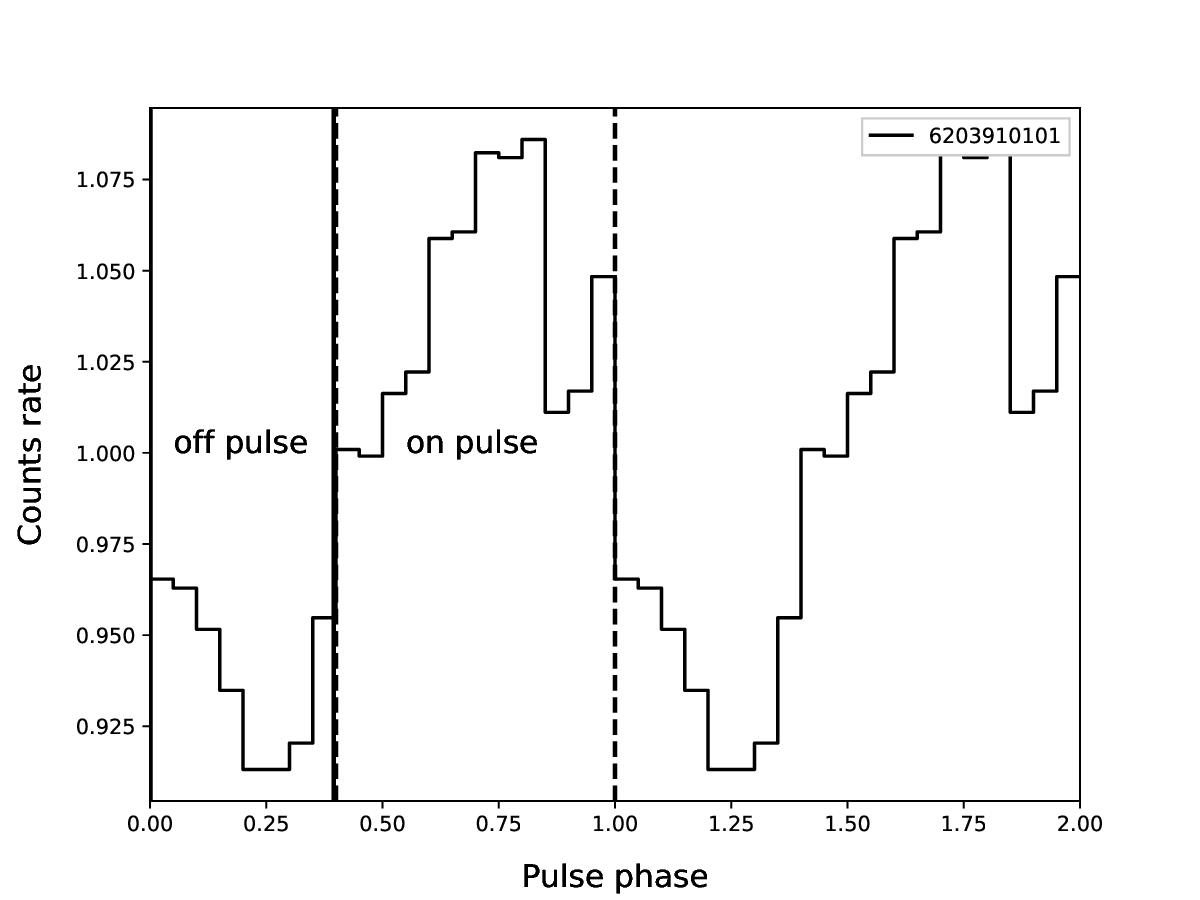}
    \includegraphics[scale=0.4]{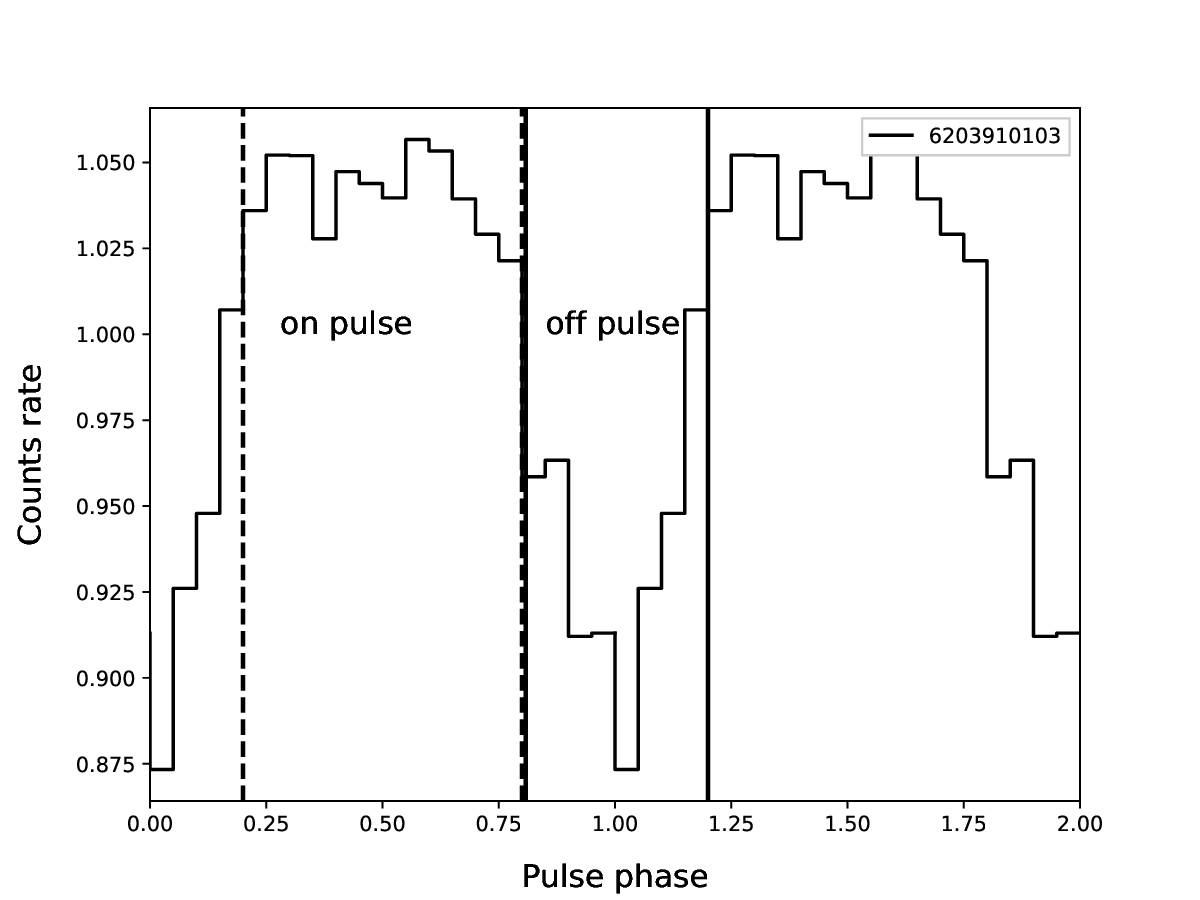}
    \includegraphics[scale=0.4]{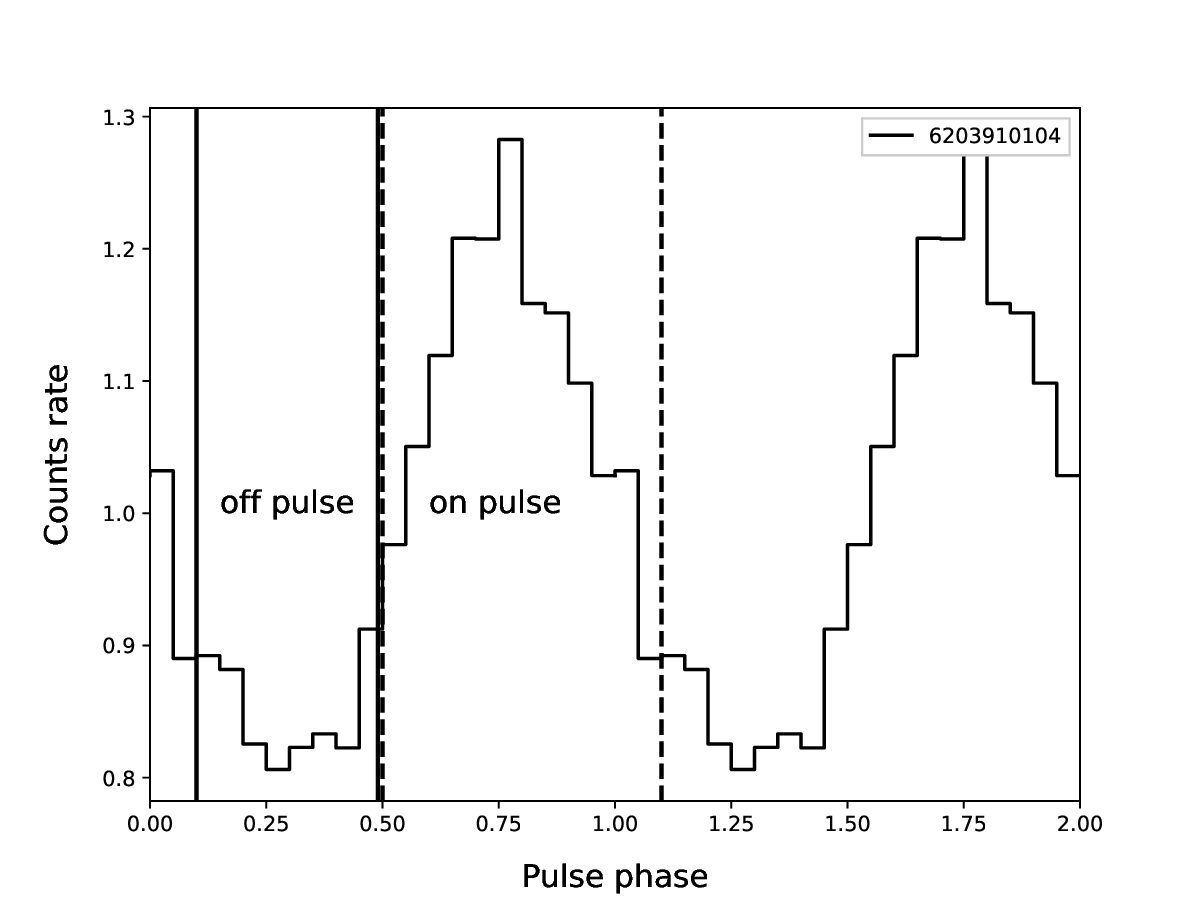}
    \includegraphics[scale=0.4]{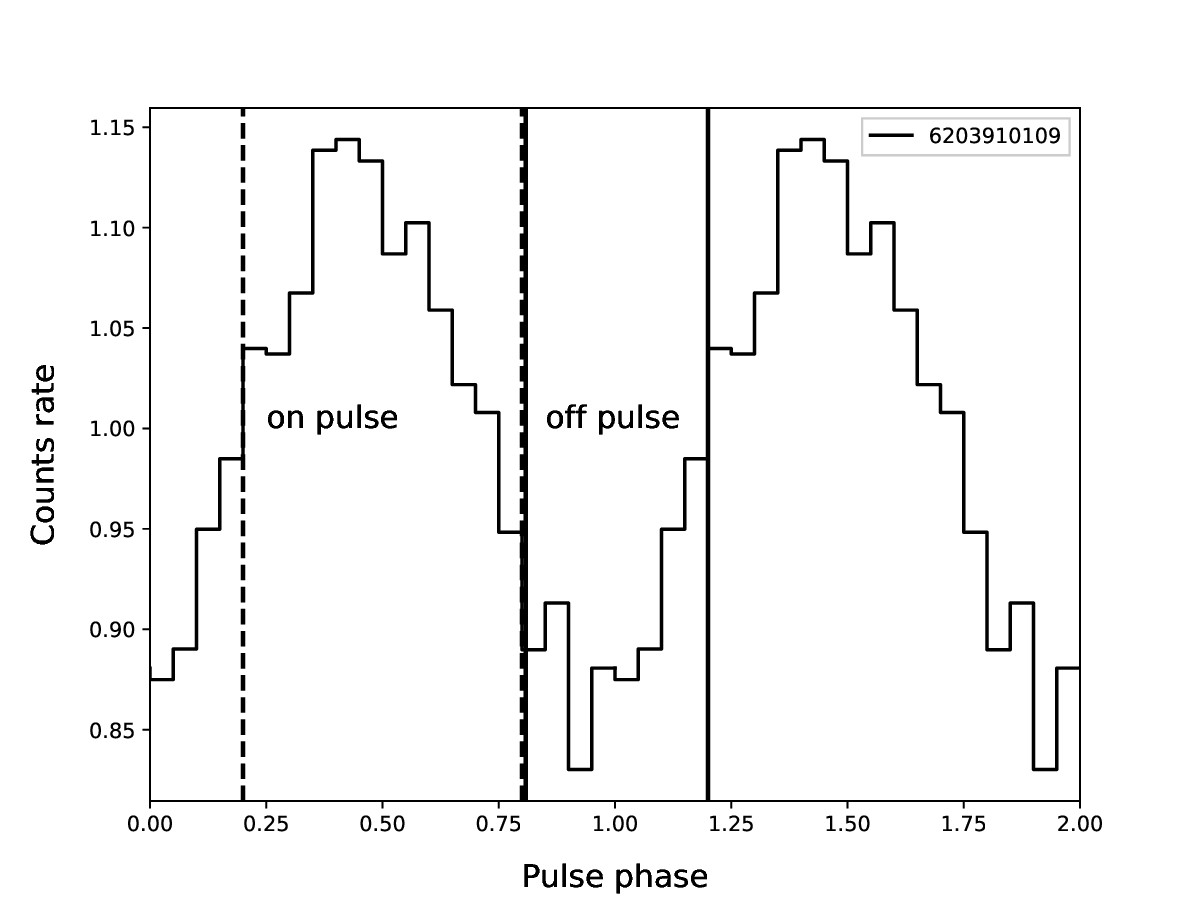}
    \includegraphics[scale=0.4]{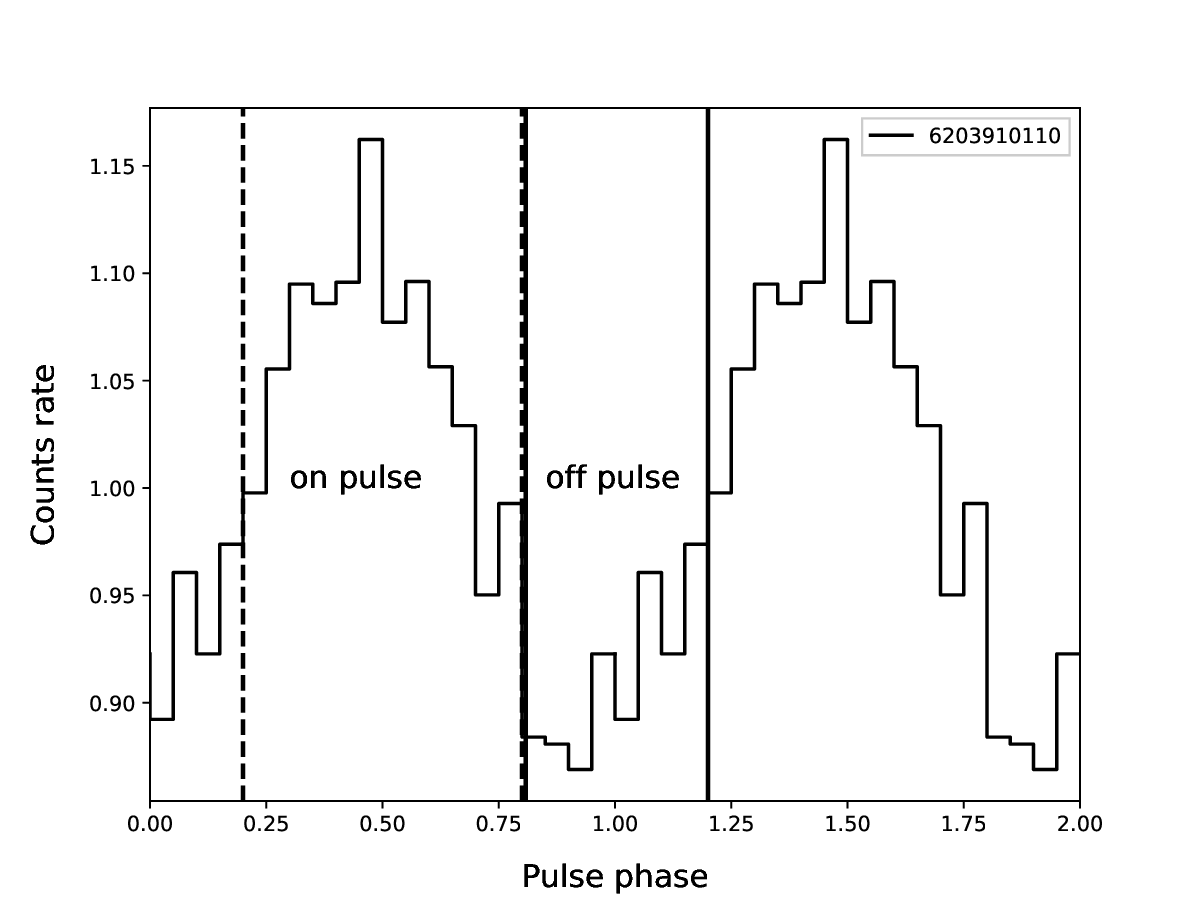}
    \includegraphics[scale=0.4]{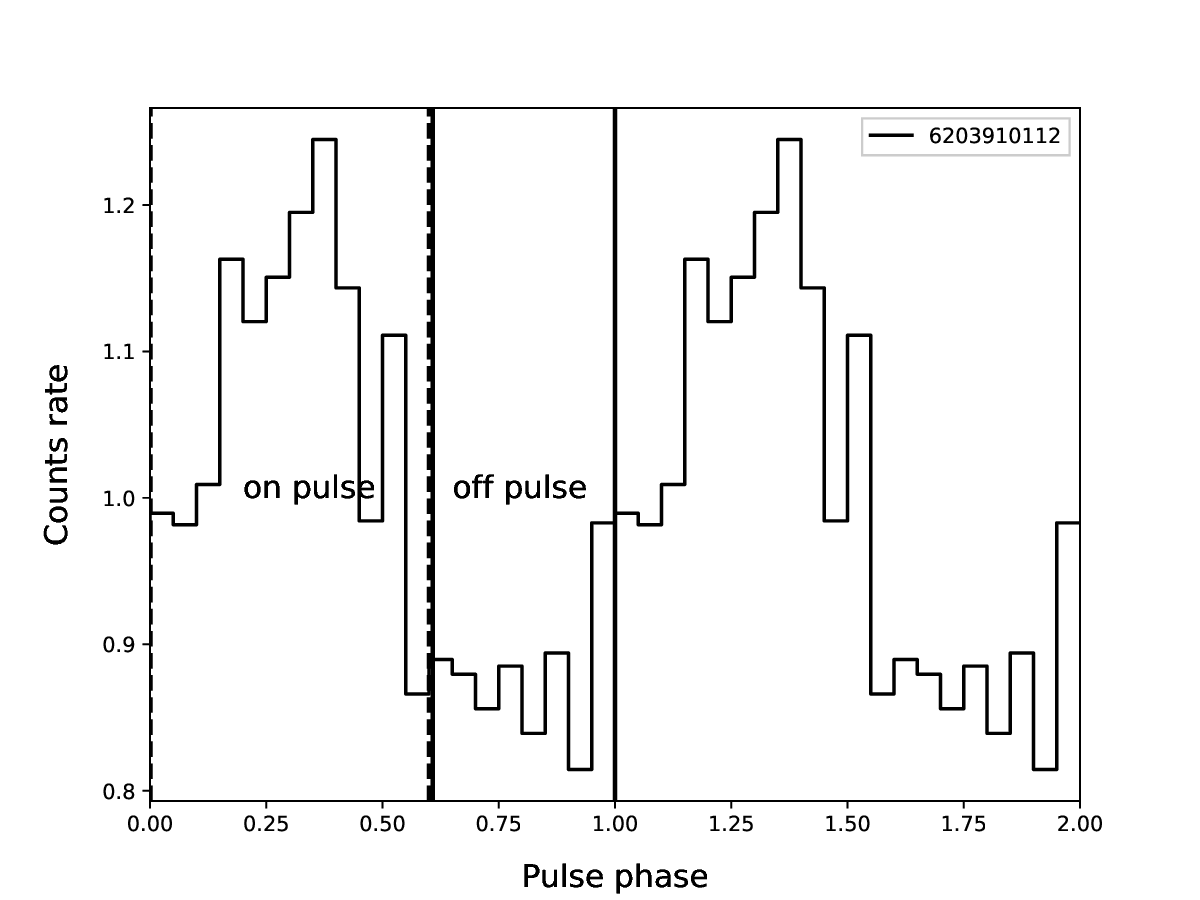}
    \includegraphics[scale=0.4]{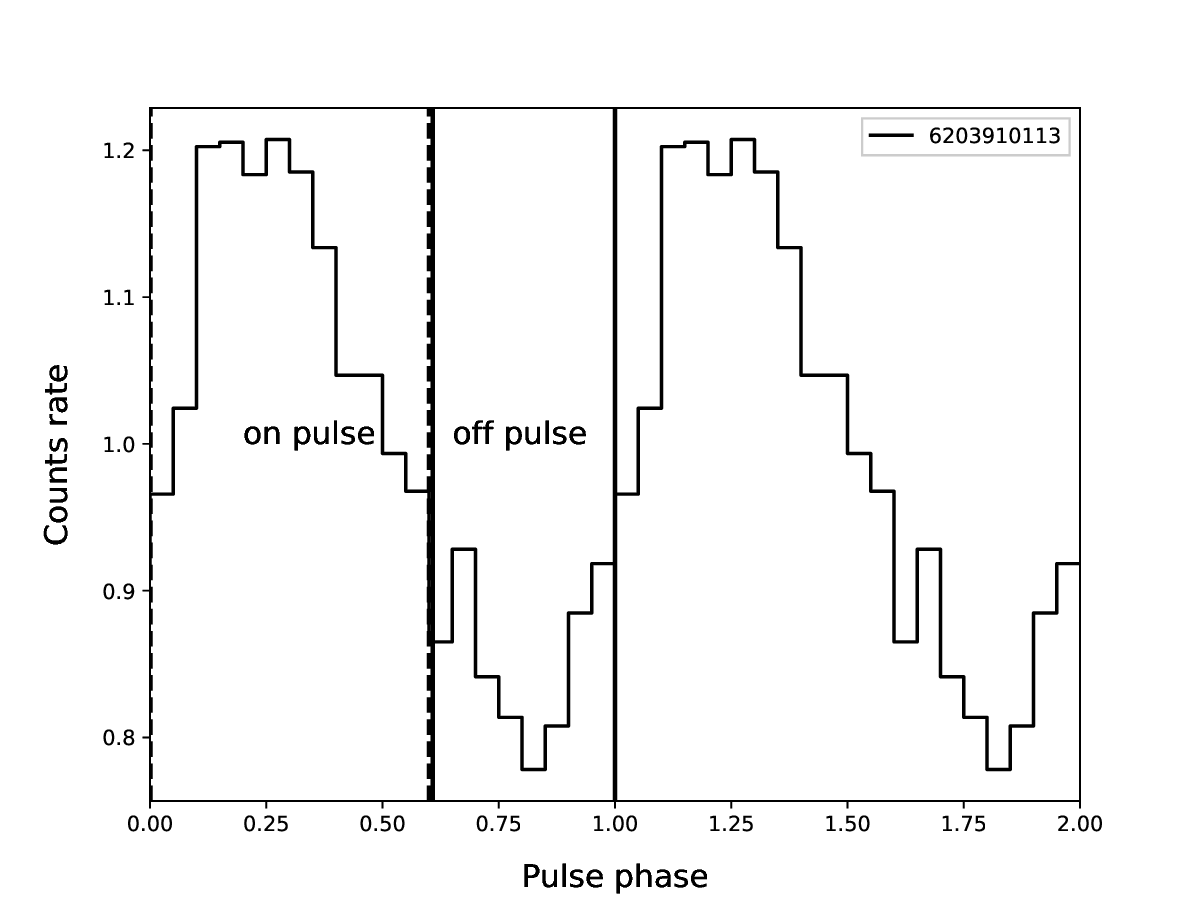}
    \includegraphics[scale=0.4]{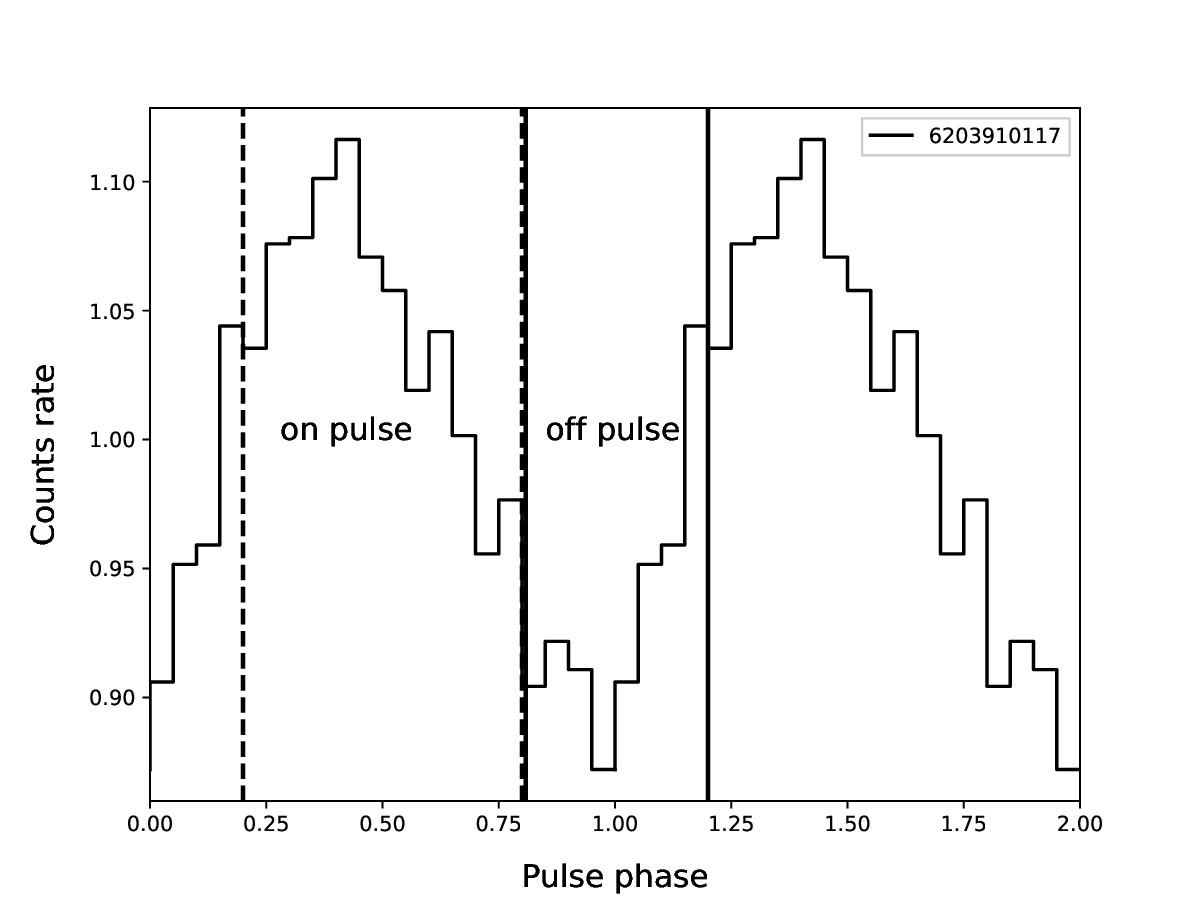}
\end{figure*}
\begin{figure*}
    \centering
    \includegraphics[scale=0.4]{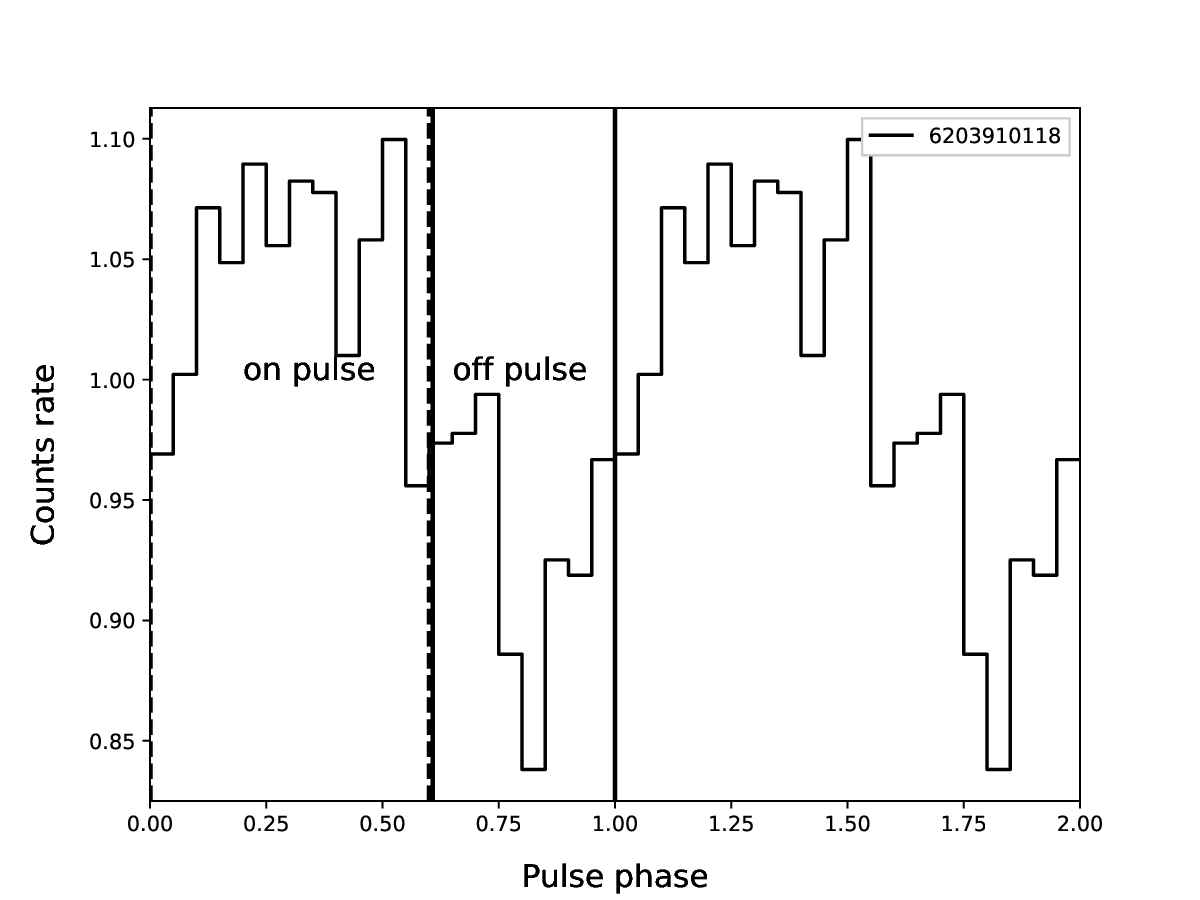}
    \includegraphics[scale=0.4]{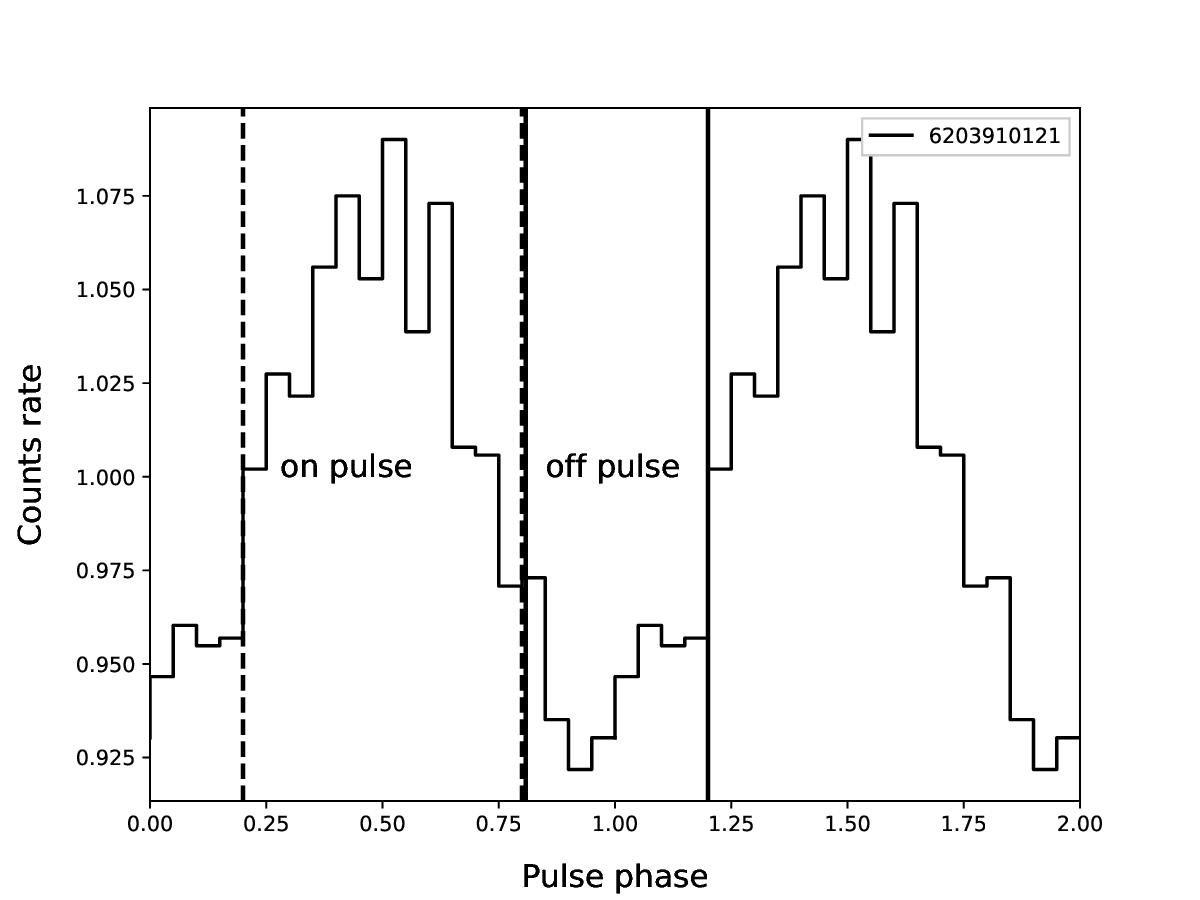}
    \includegraphics[scale=0.4]{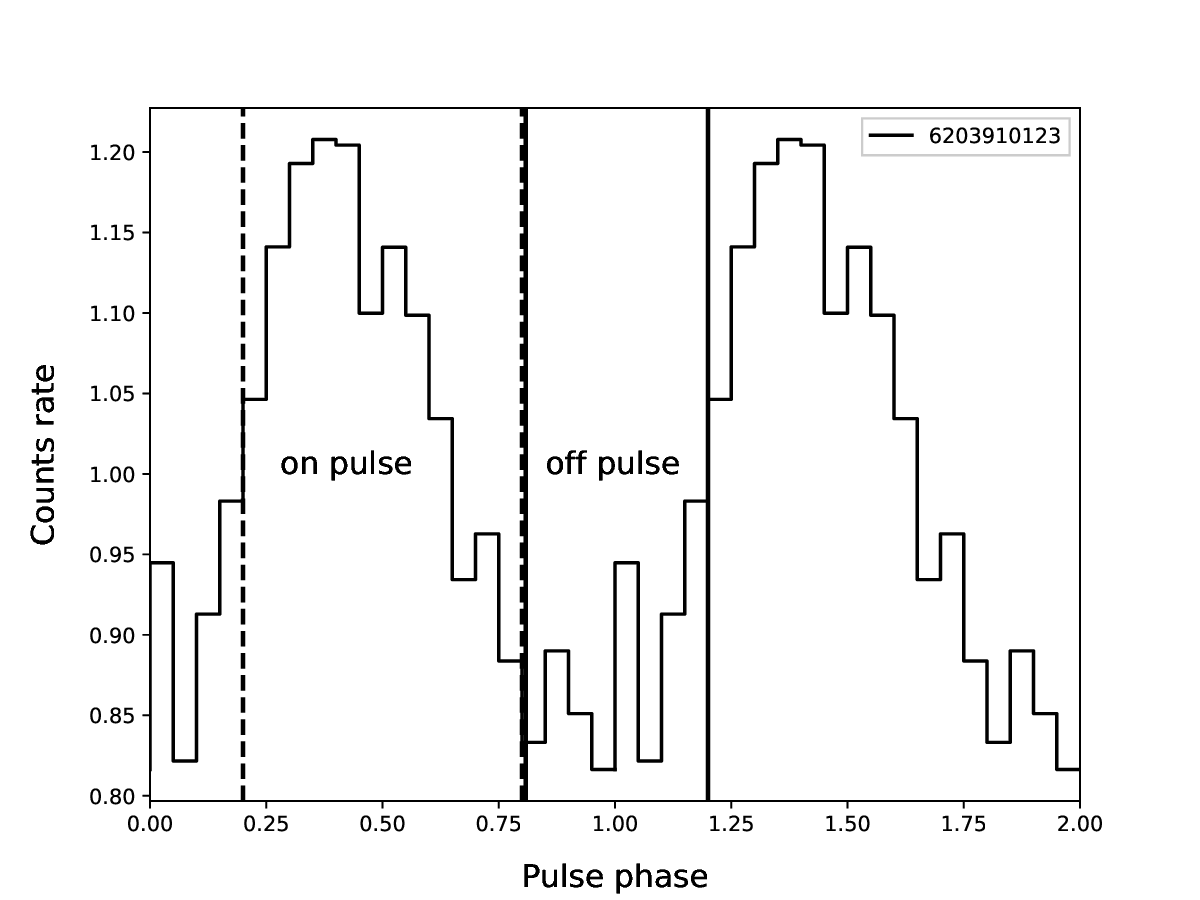}
    \includegraphics[scale=0.4]{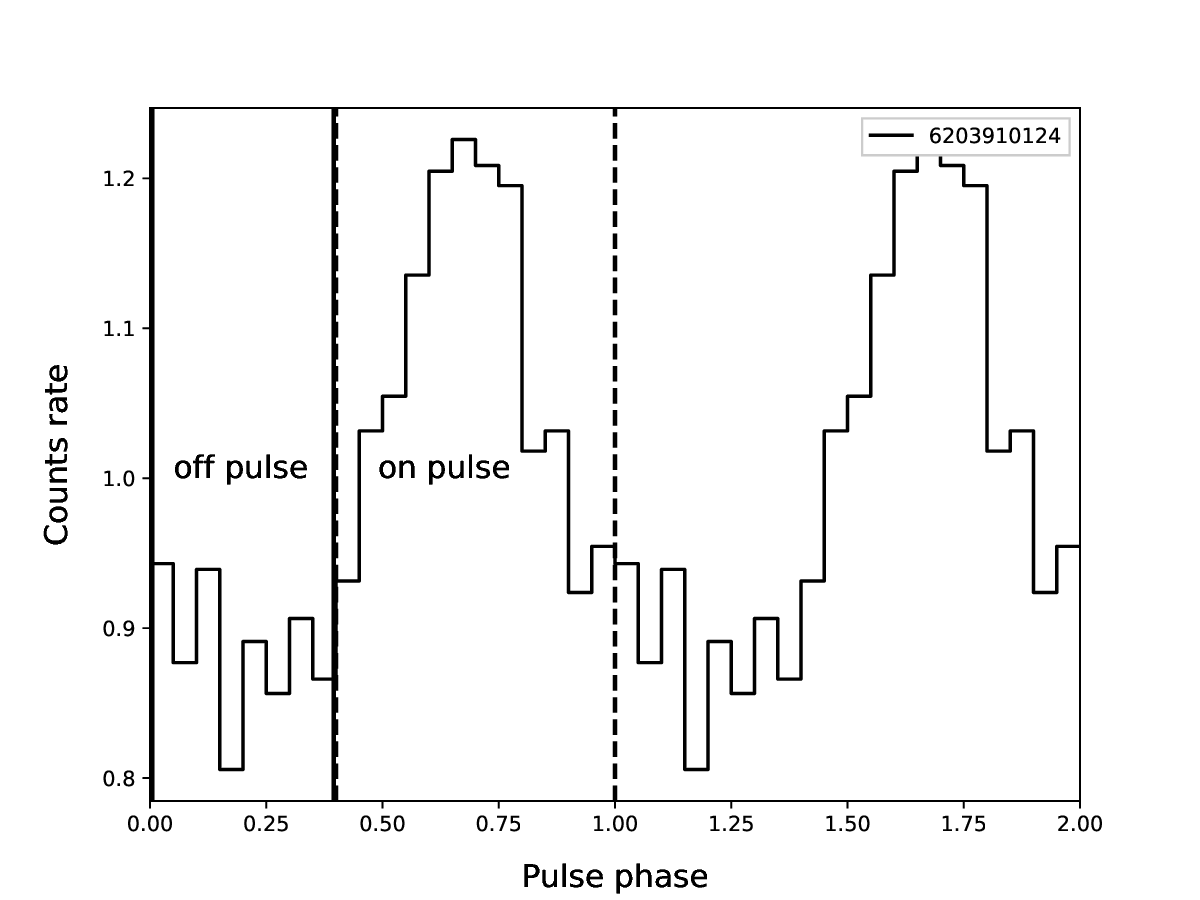}
    \includegraphics[scale=0.4]{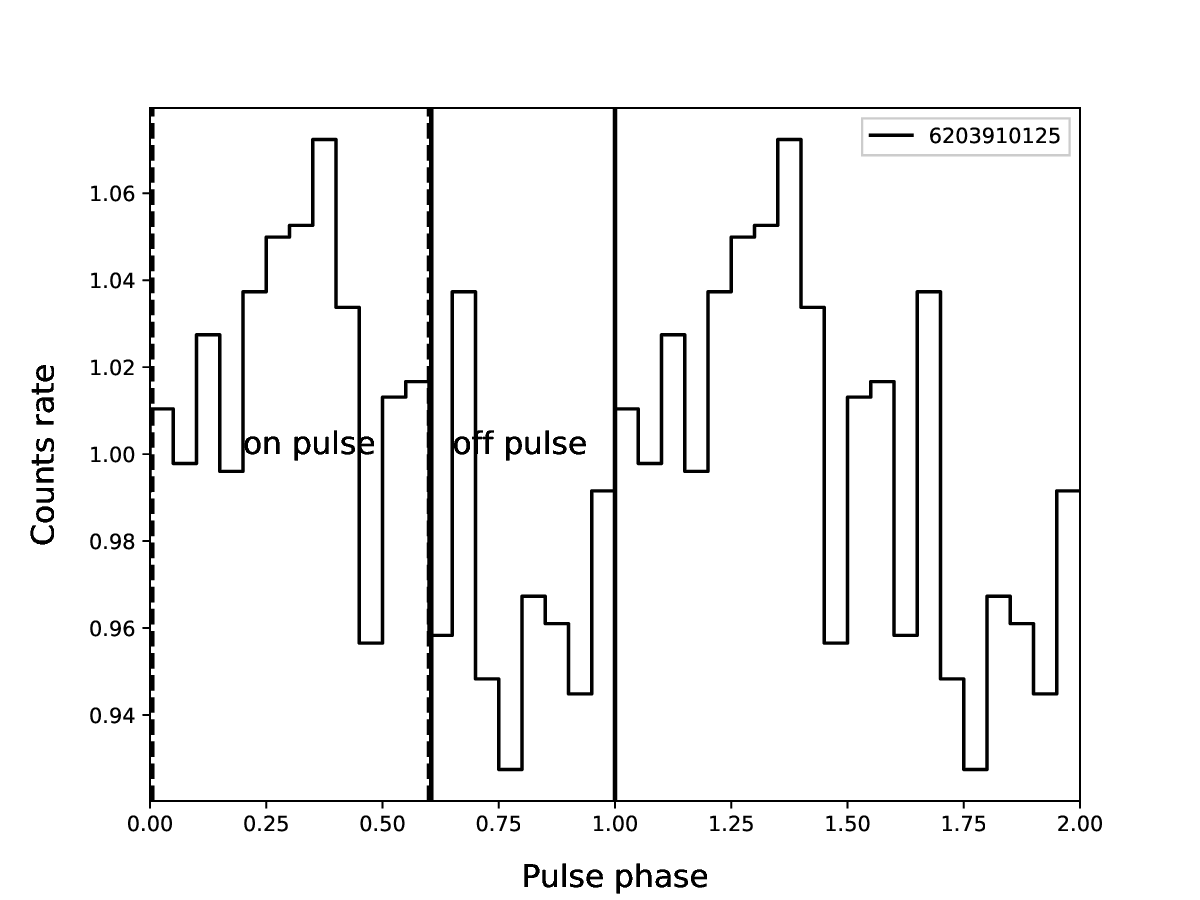}
    \includegraphics[scale=0.4]{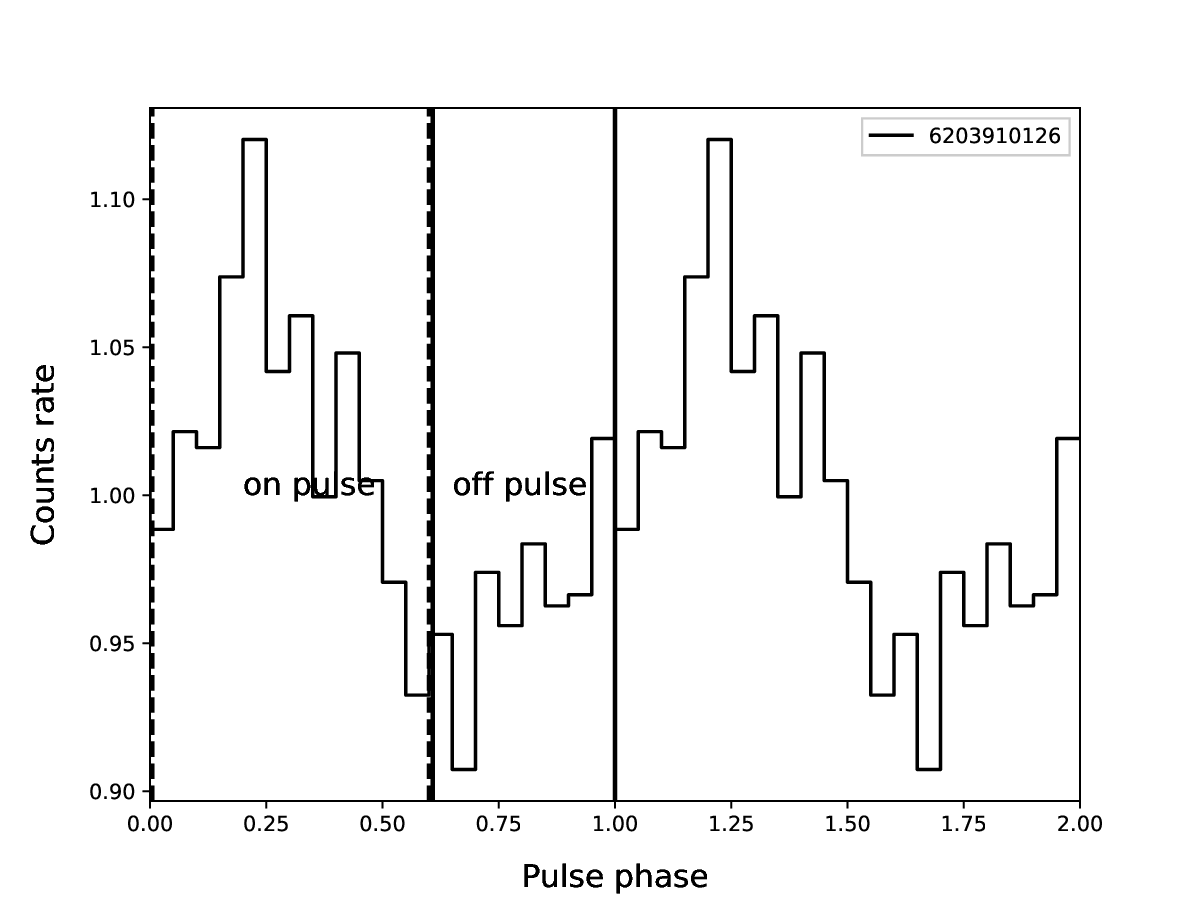}
    \caption{The fold pulse profile using the period get from all the NICER data sets list in Table~\ref{table-lsp}, the period from LS periodogram is 79.10 second. The vertical solid lines and  dashed lines define the on-pulse  and off-pulse phases, respectively.}
    \label{nicer-fold-every}
\end{figure*}

\begin{figure*}
\caption{The phase resolve spectra of each data sets from NICER fitted with BB model.}
\centering
\includegraphics[scale=0.3,angle=270]{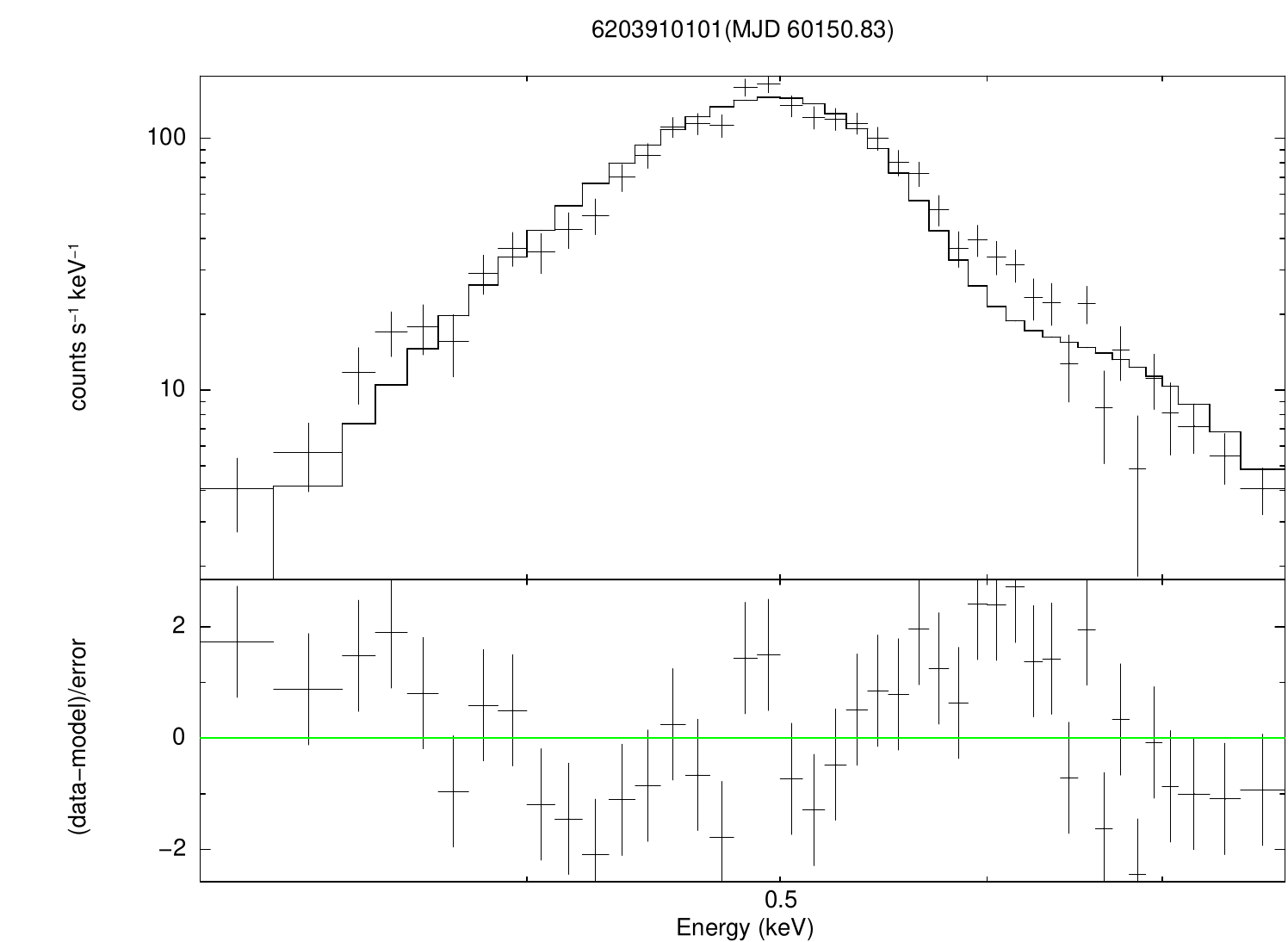}
\includegraphics[scale=0.3,angle=270]{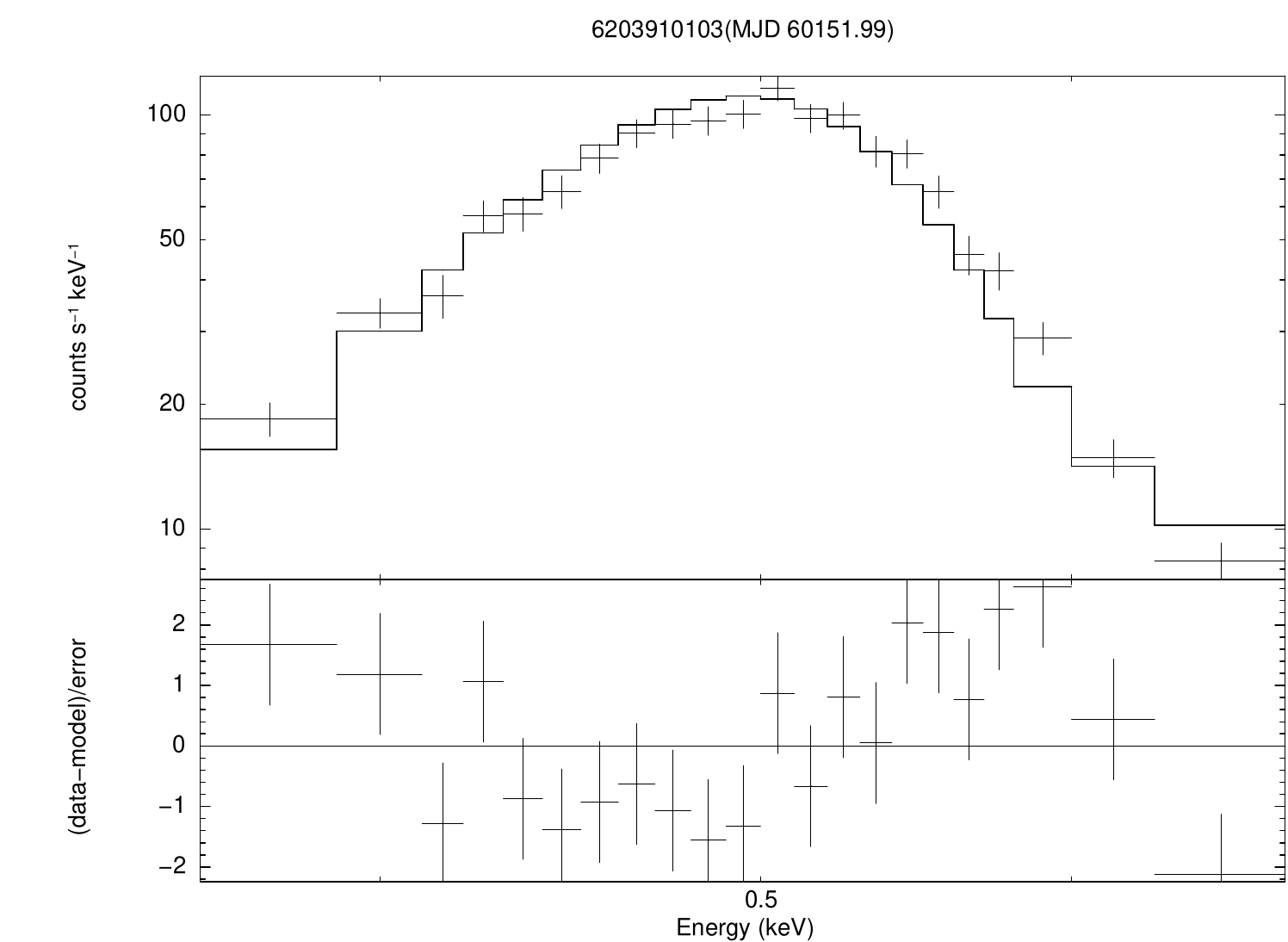}
\includegraphics[scale=0.3,angle=270]{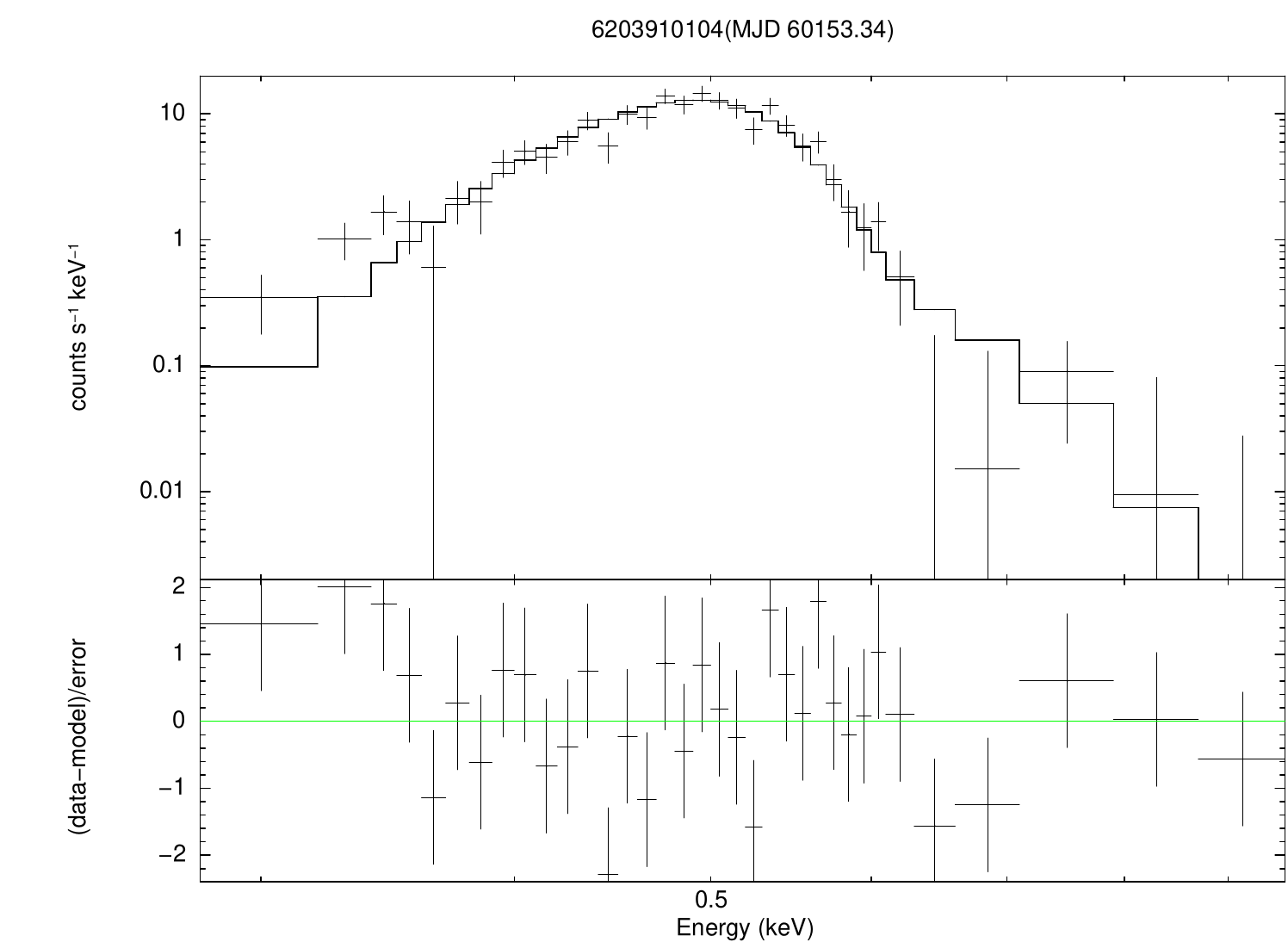}
\includegraphics[scale=0.3,angle=270]{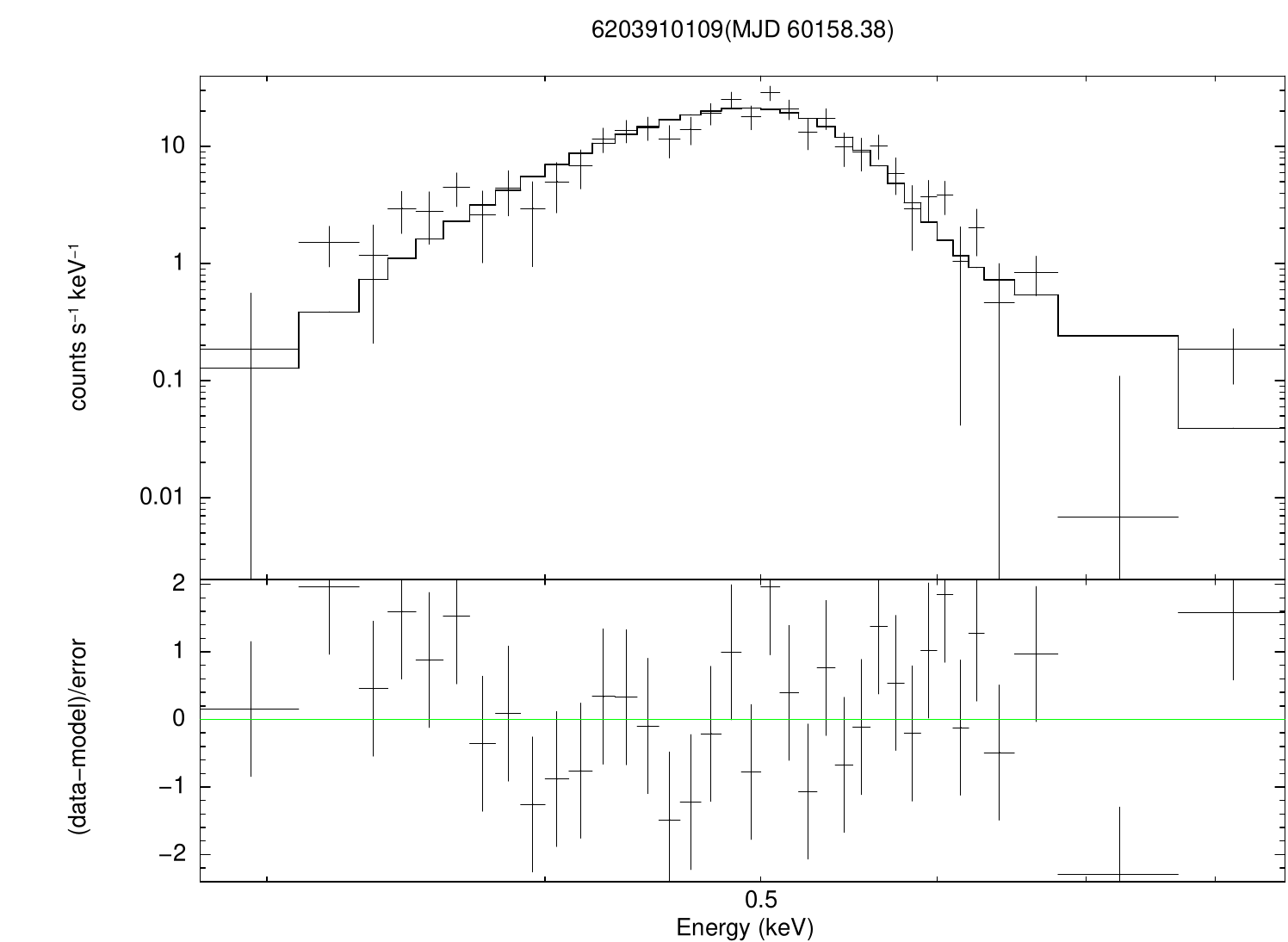}
\includegraphics[scale=0.3,angle=270]{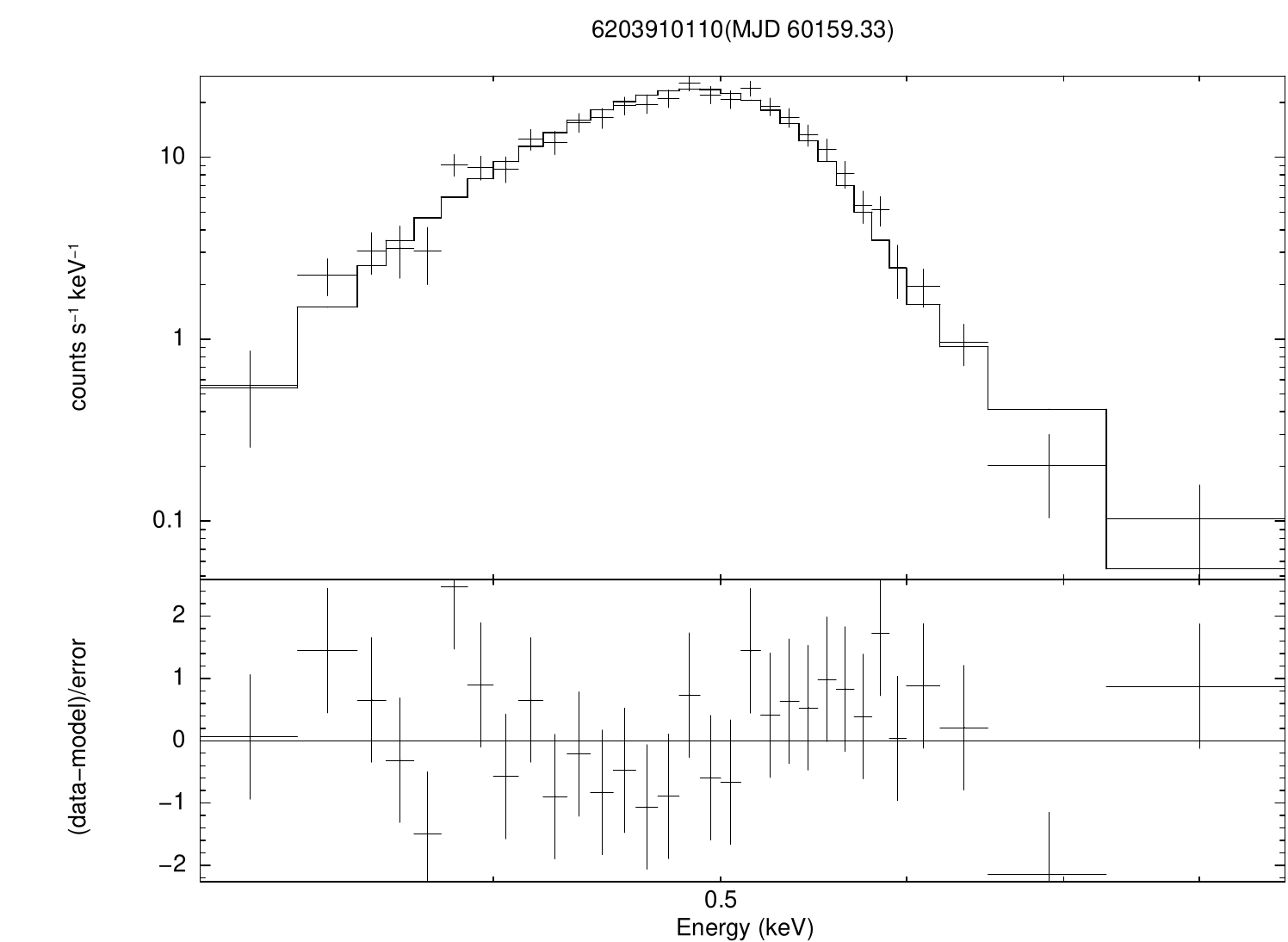}
\includegraphics[scale=0.3,angle=270]{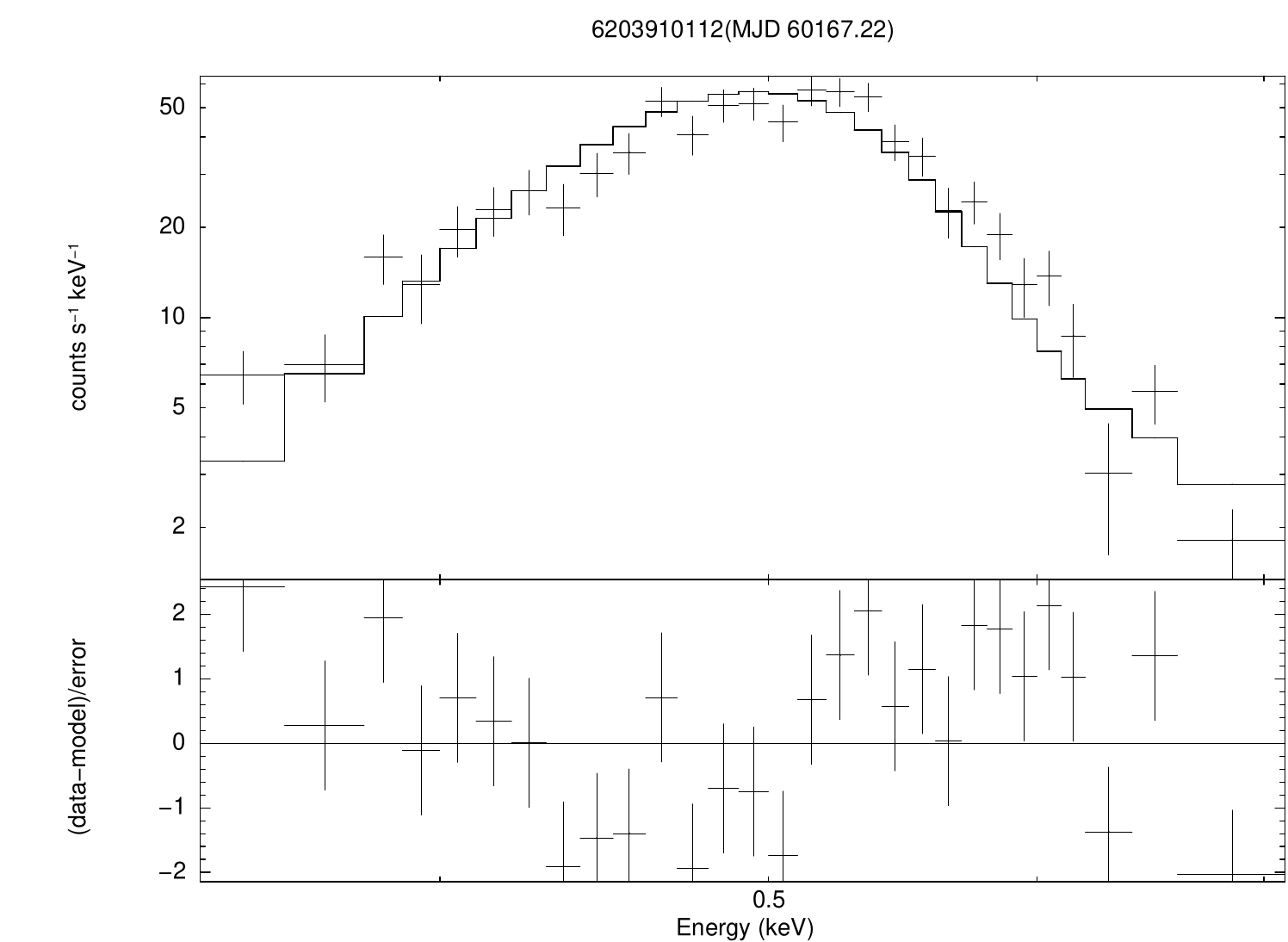}
\includegraphics[scale=0.3,angle=270]{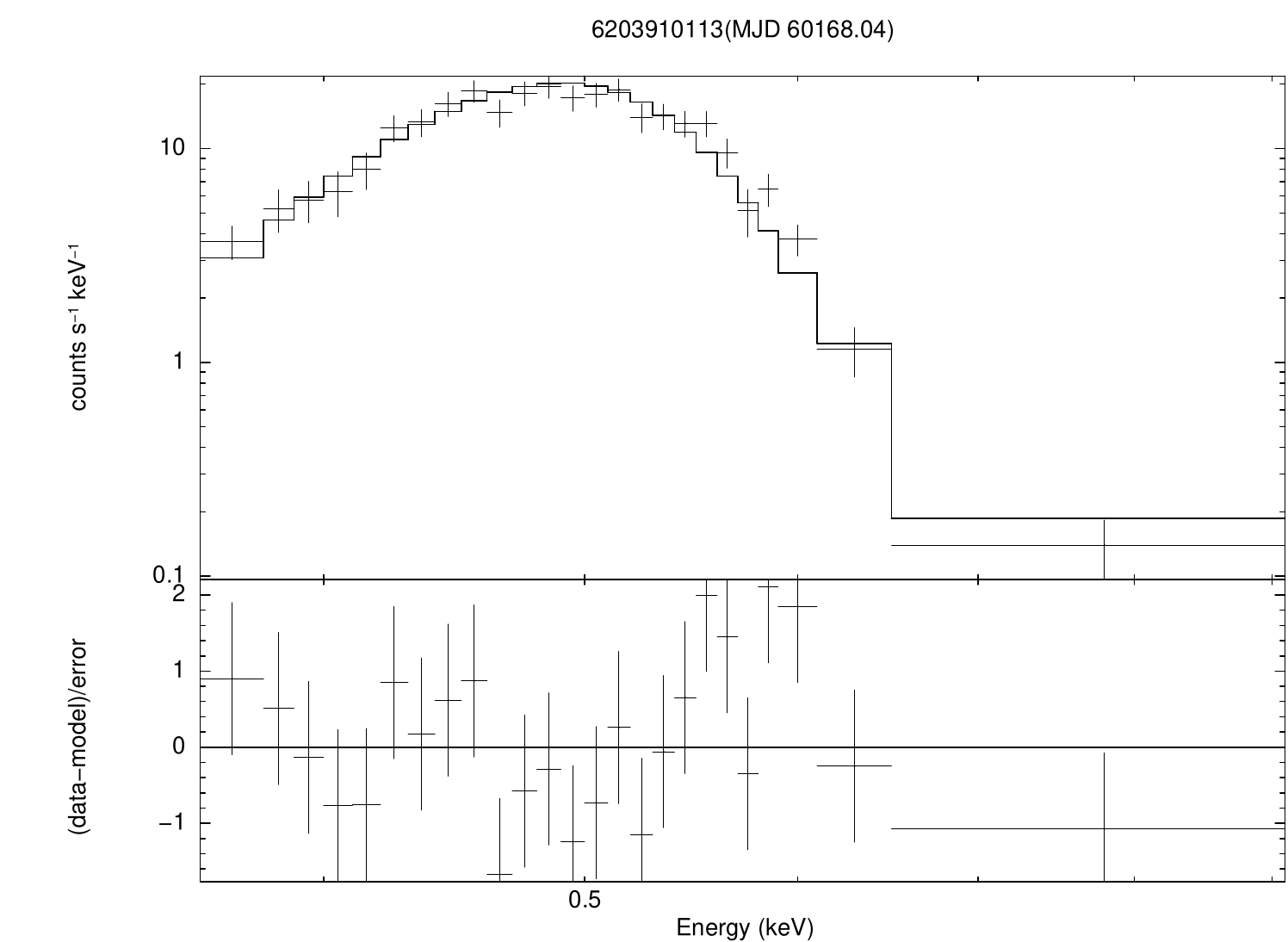}
\includegraphics[scale=0.3,angle=270]{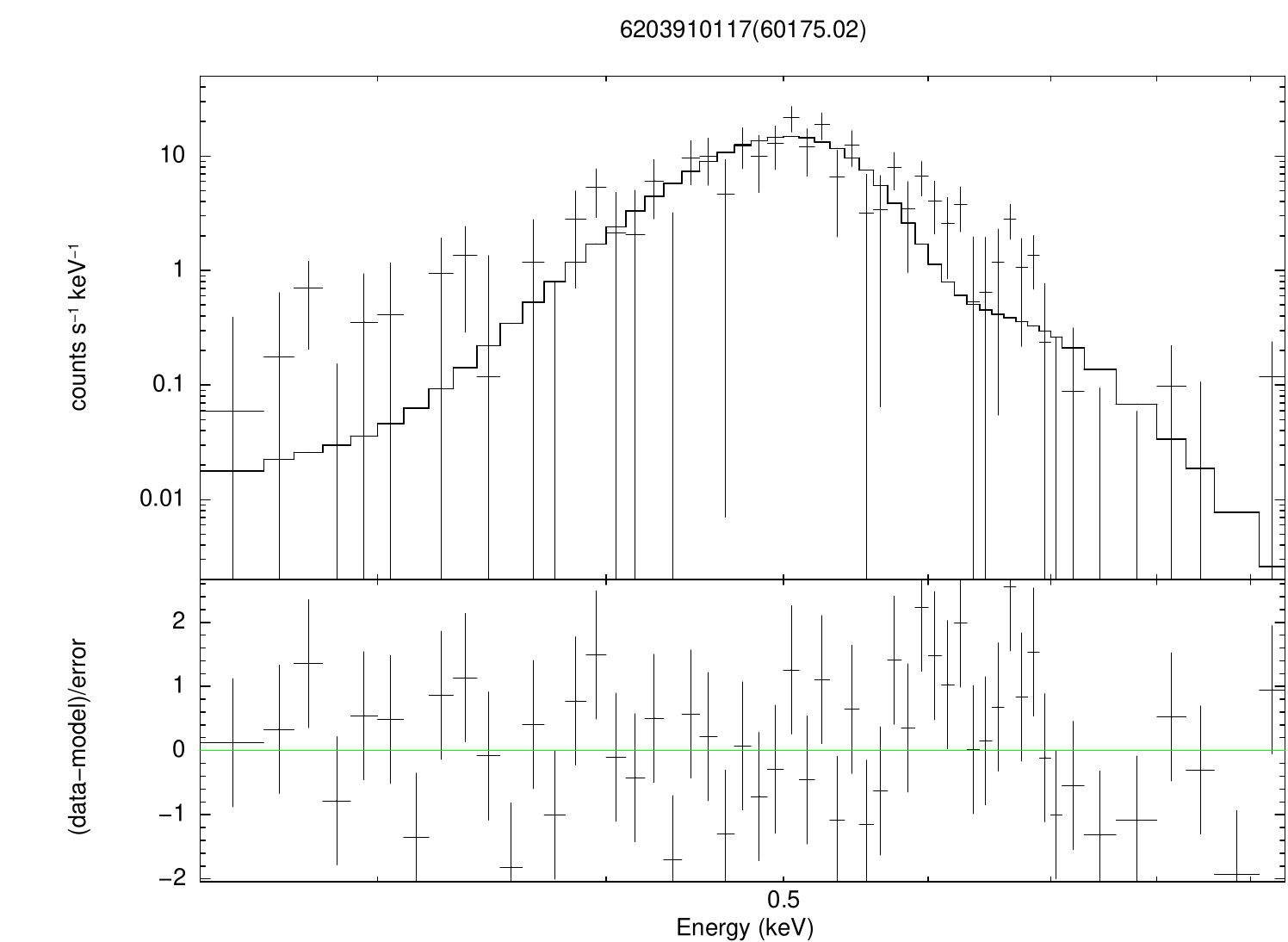}
\end{figure*}

\begin{figure*}
\centering
\includegraphics[scale=0.3,angle=270]{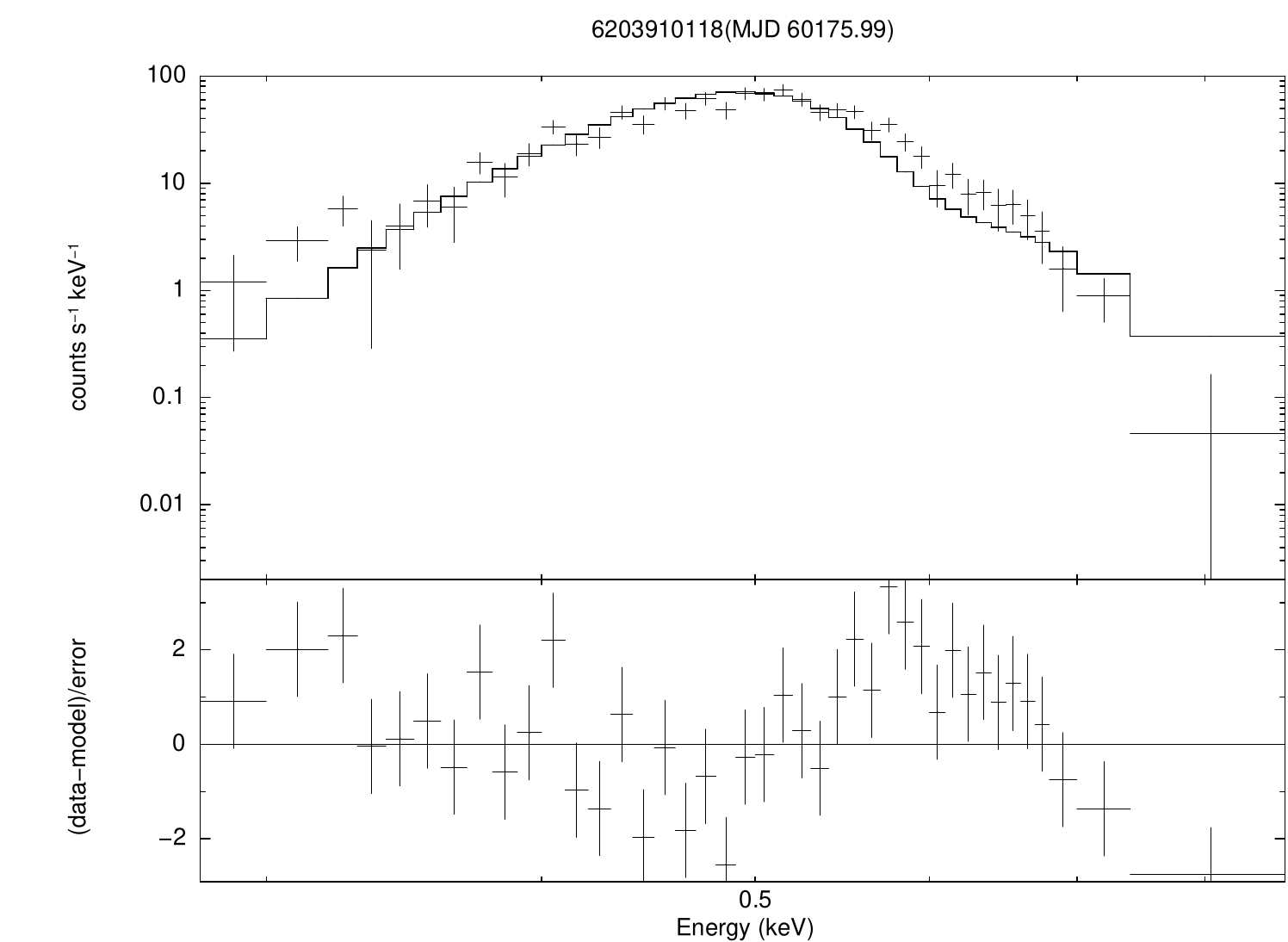}
\includegraphics[scale=0.3,angle=270]{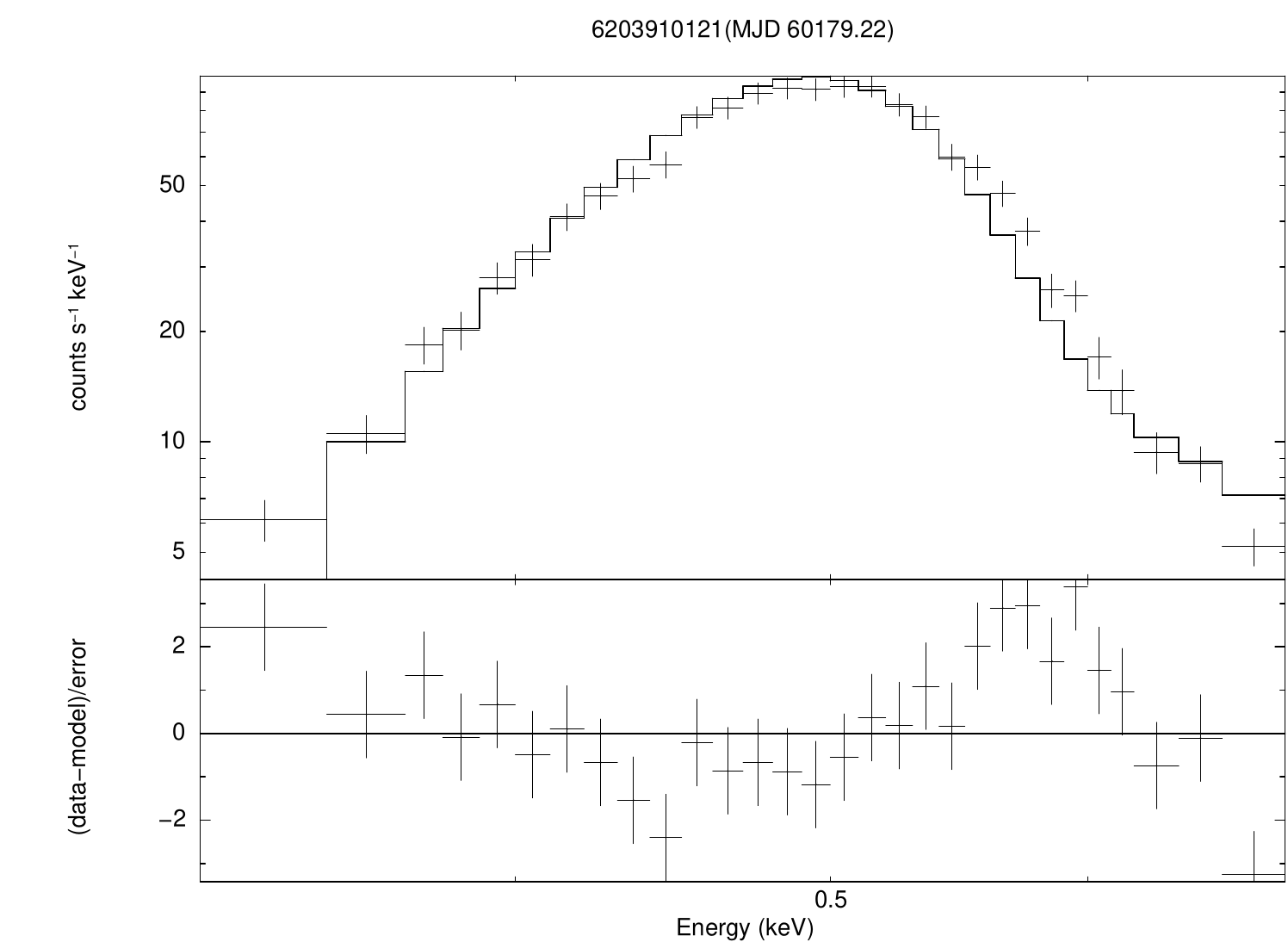}
\includegraphics[scale=0.3,angle=270]{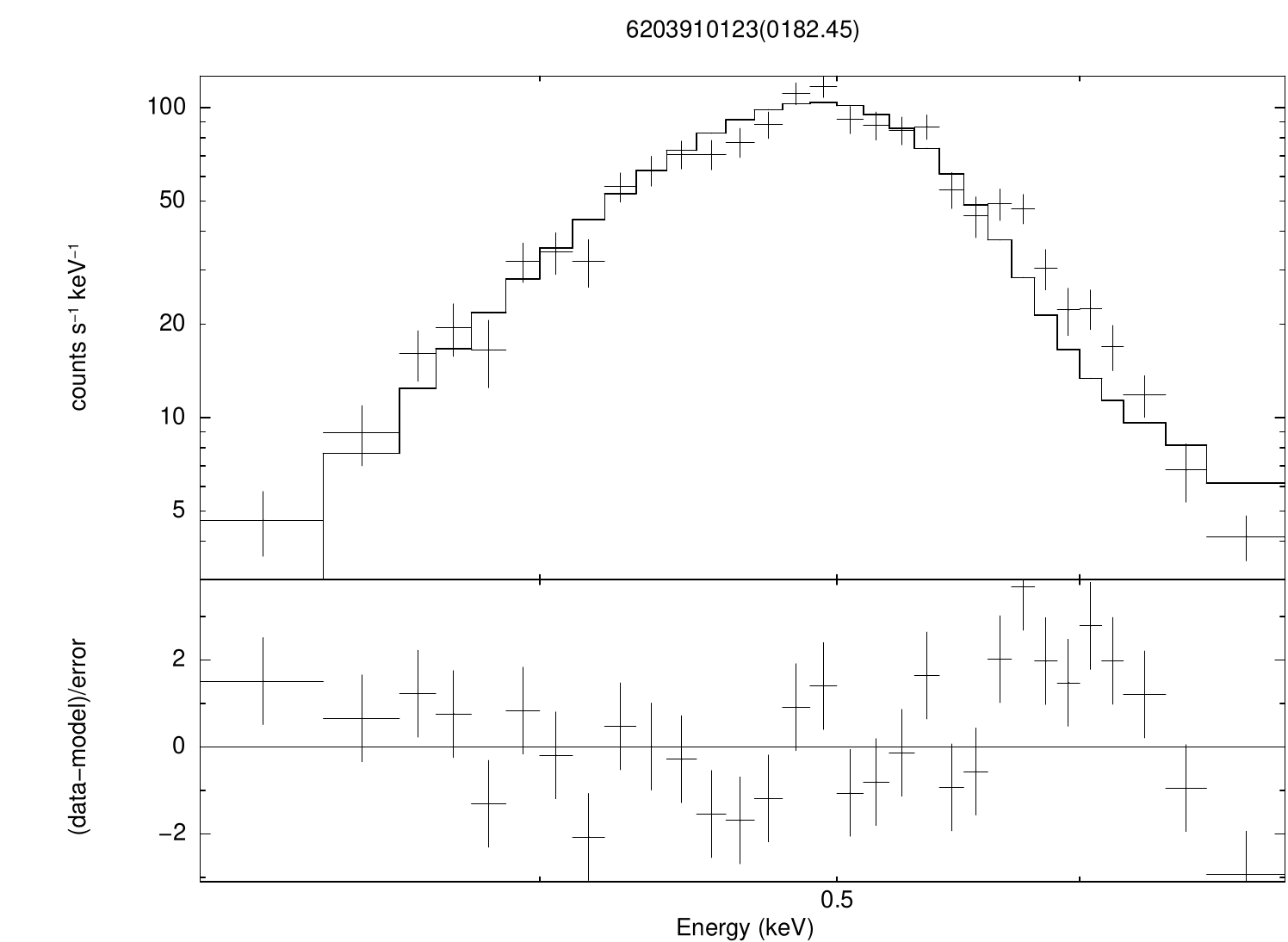}
\includegraphics[scale=0.3,angle=270]{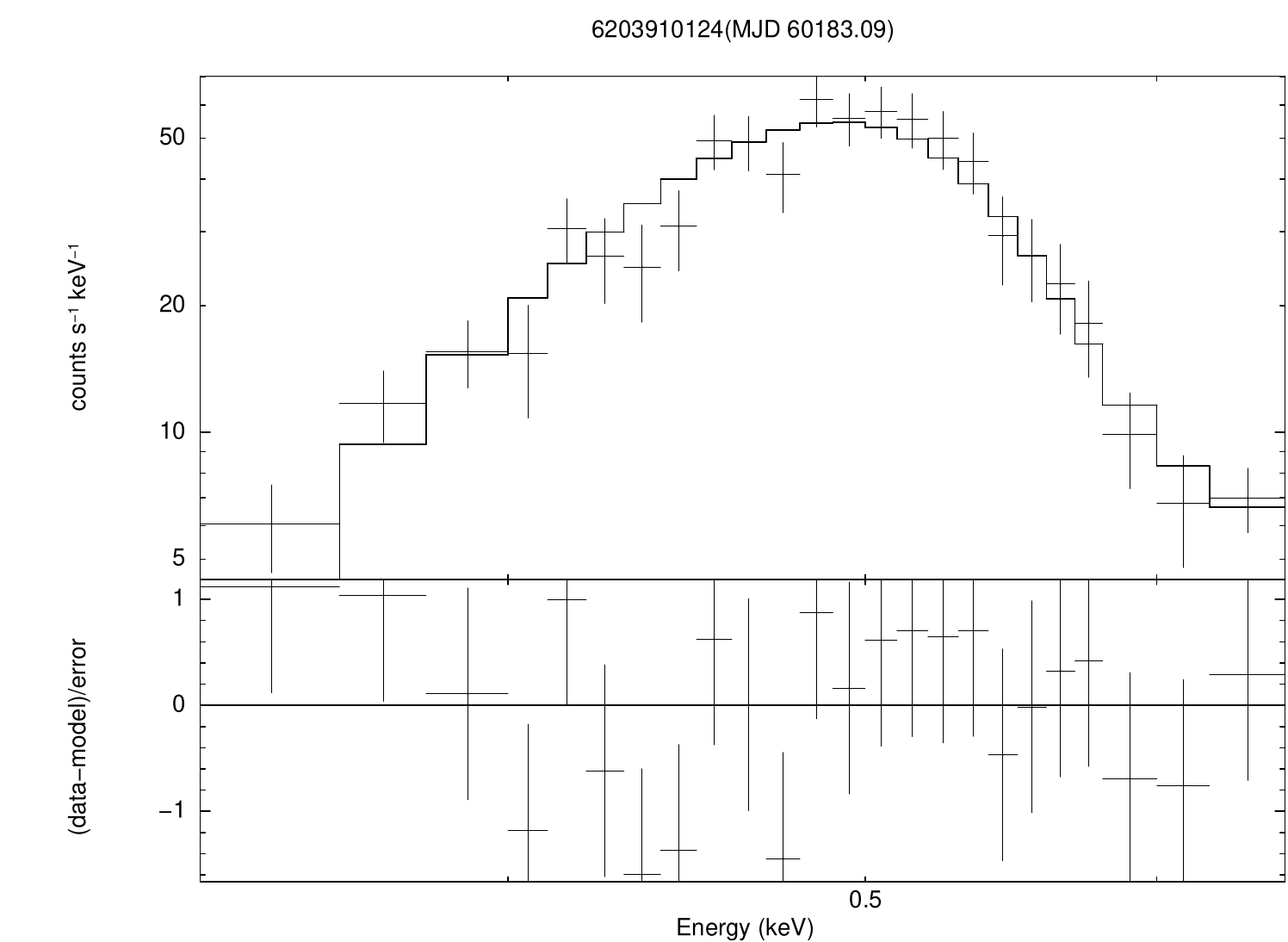}
\includegraphics[scale=0.3,angle=270]{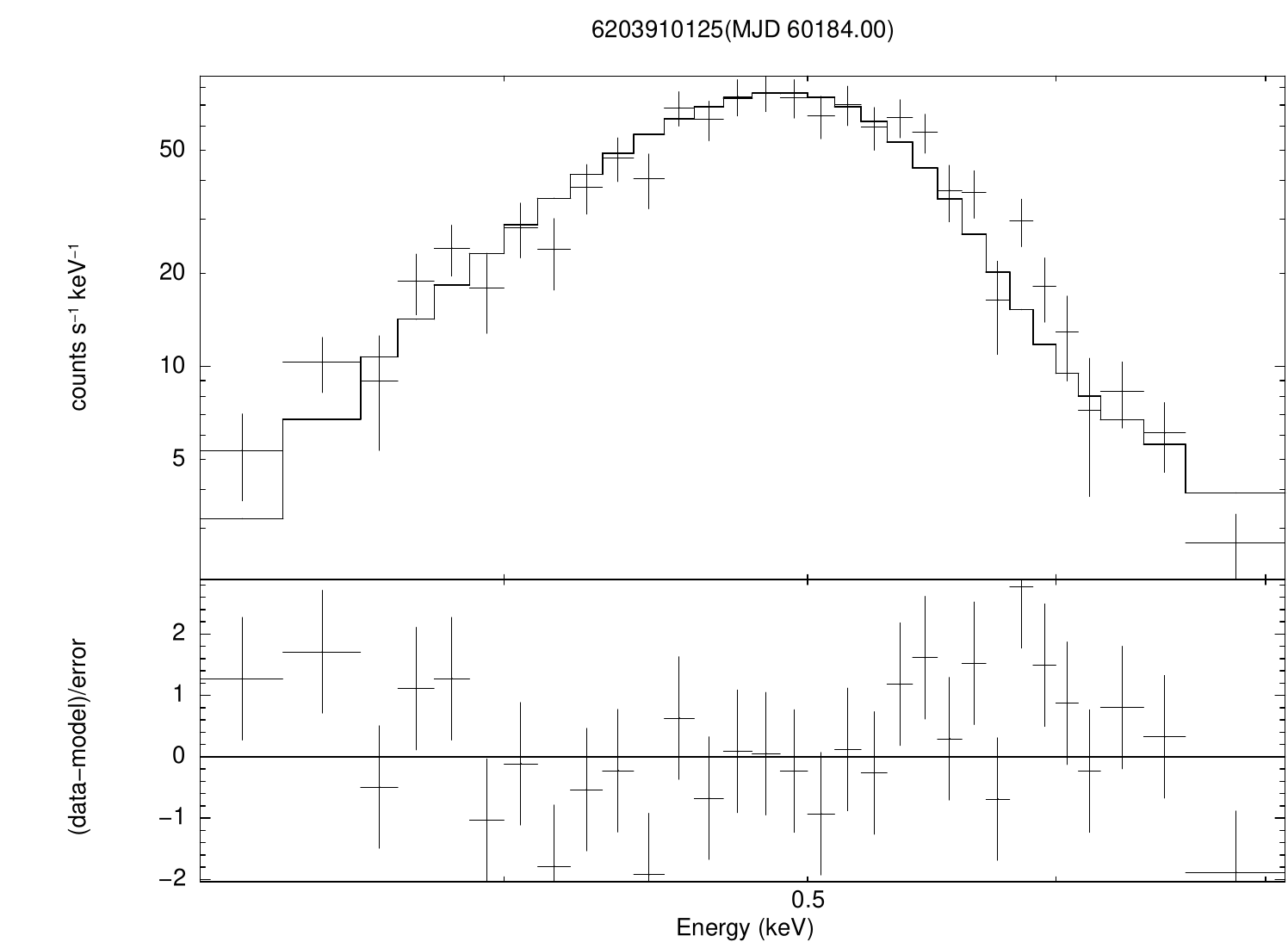}
\includegraphics[scale=0.3,angle=270]{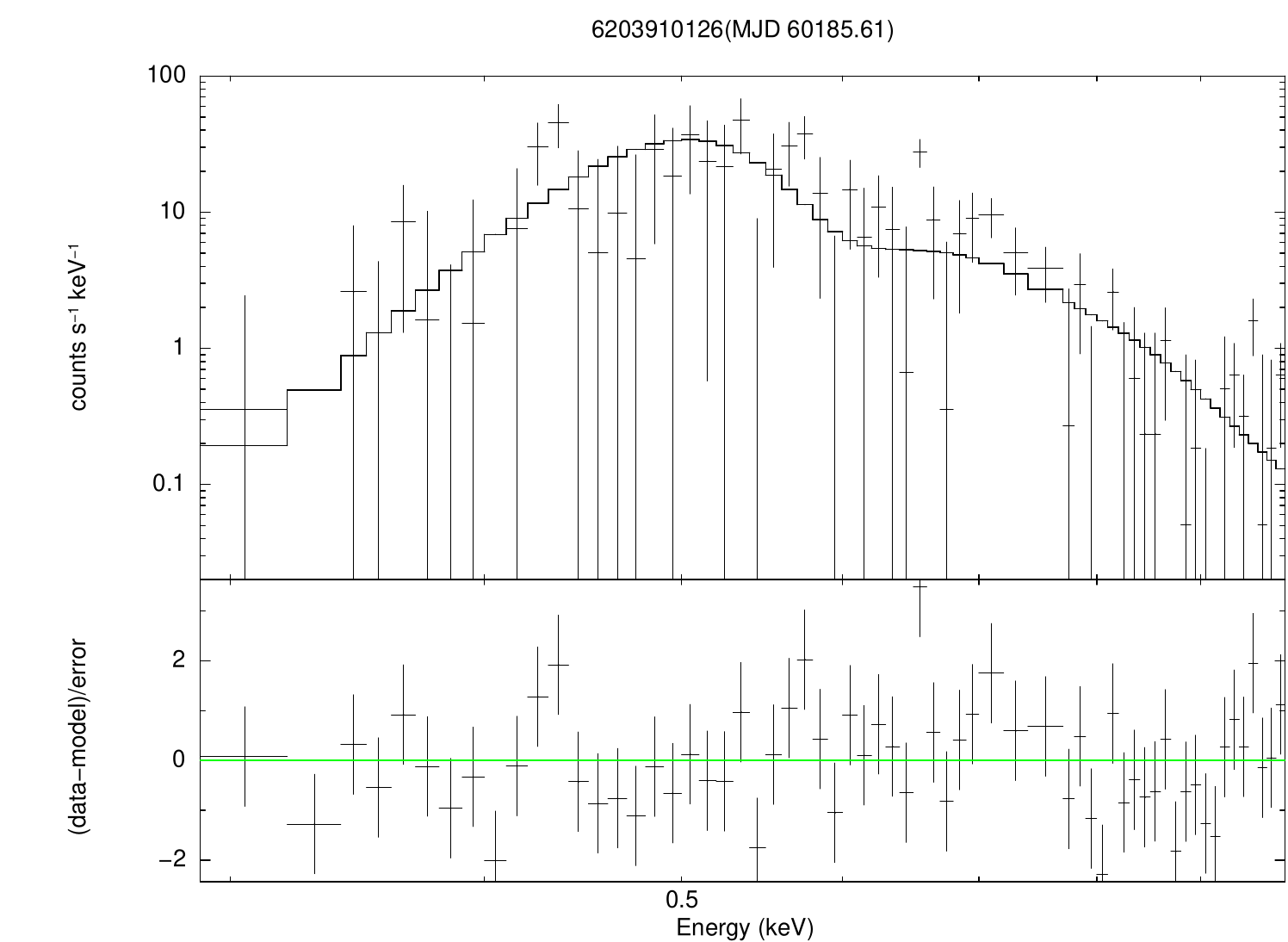}
    \label{nicer-resolve-every}
\end{figure*}

\begin{figure*}
\caption{The phase resolve spectra of each data sets from NICER fitted with WD atmosphere model.}
\centering
\includegraphics[scale=0.3,angle=270]{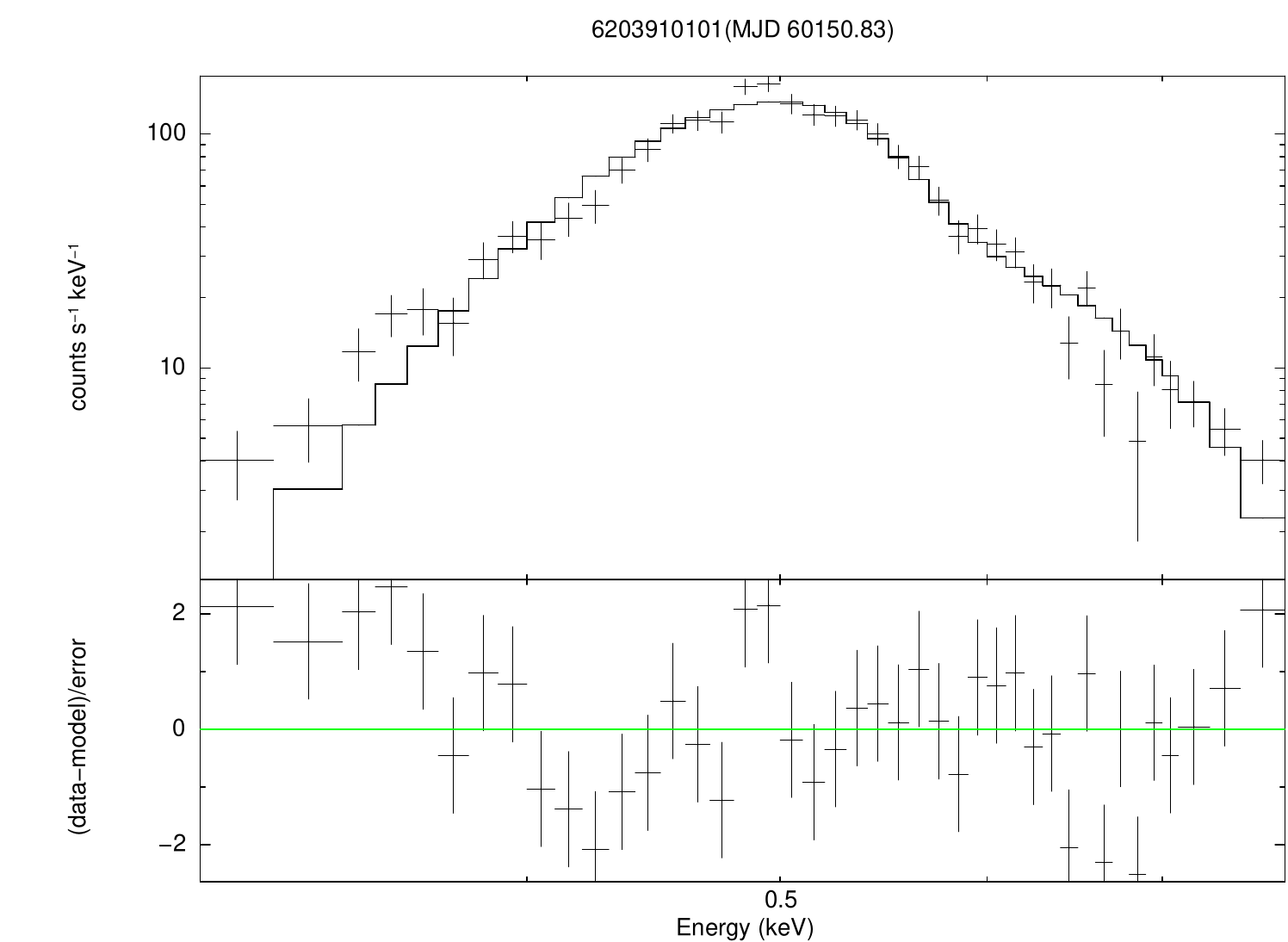}
\includegraphics[scale=0.3,angle=270]{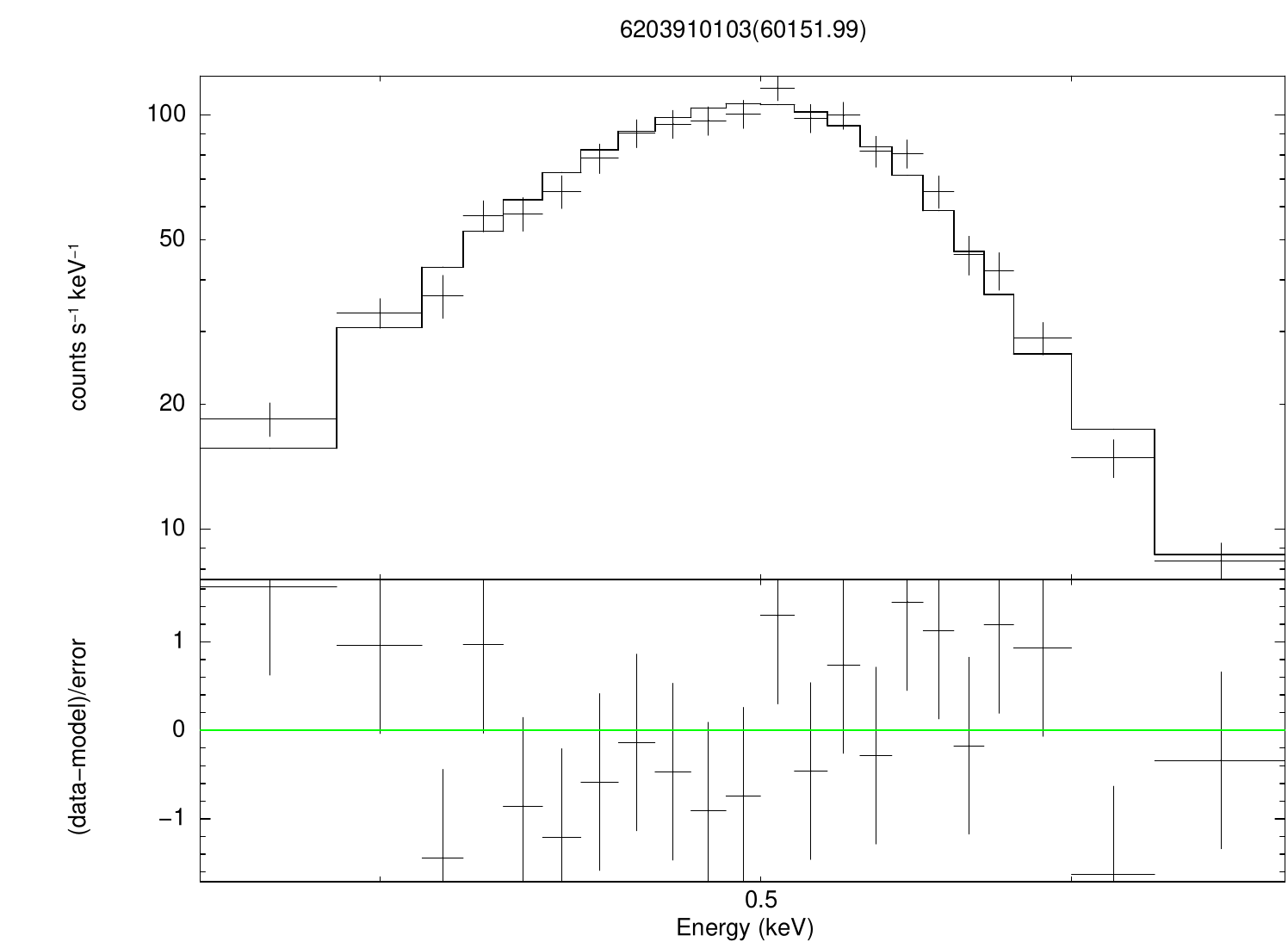}
\includegraphics[scale=0.3,angle=270]{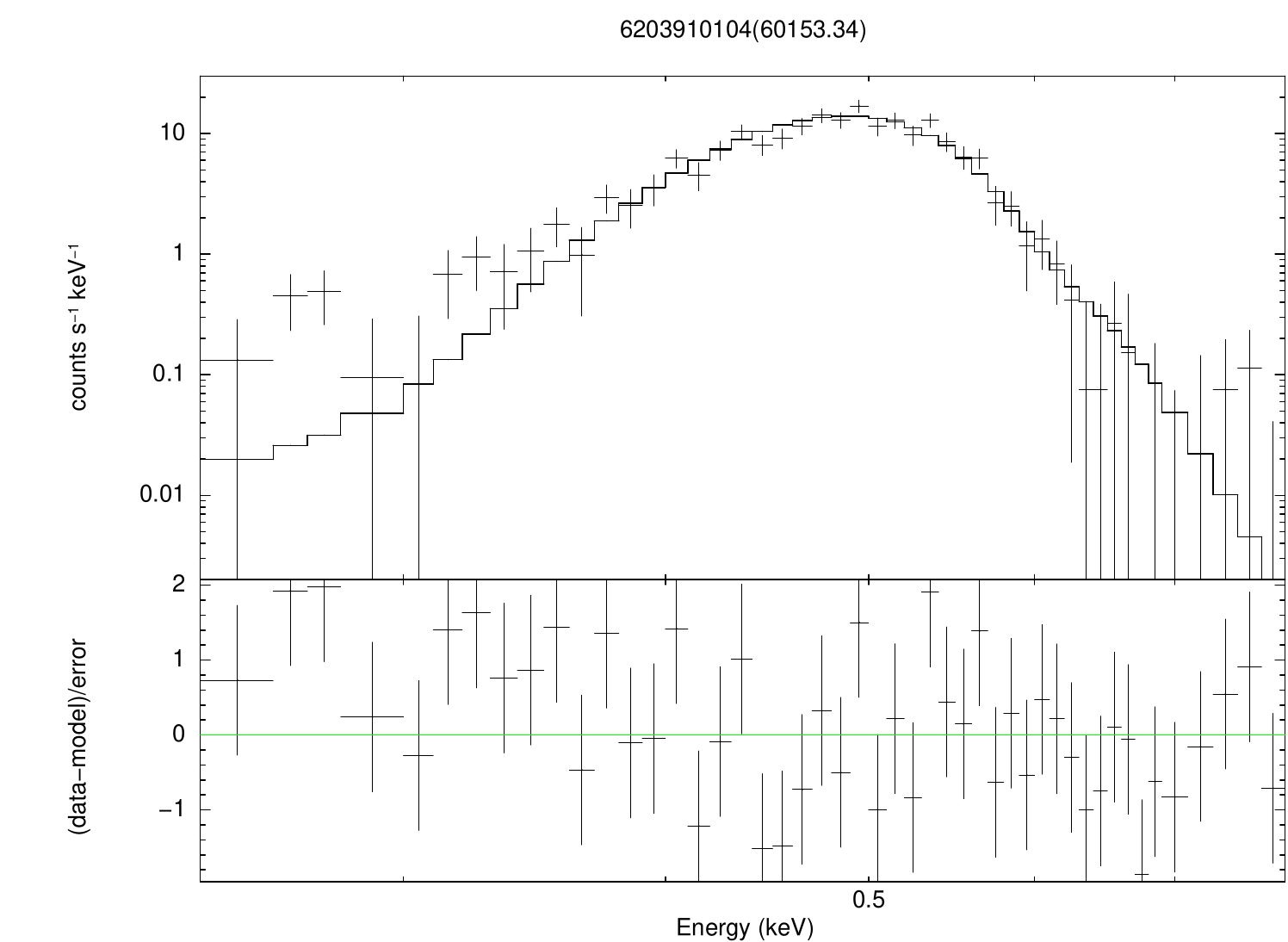}
\includegraphics[scale=0.3,angle=270]{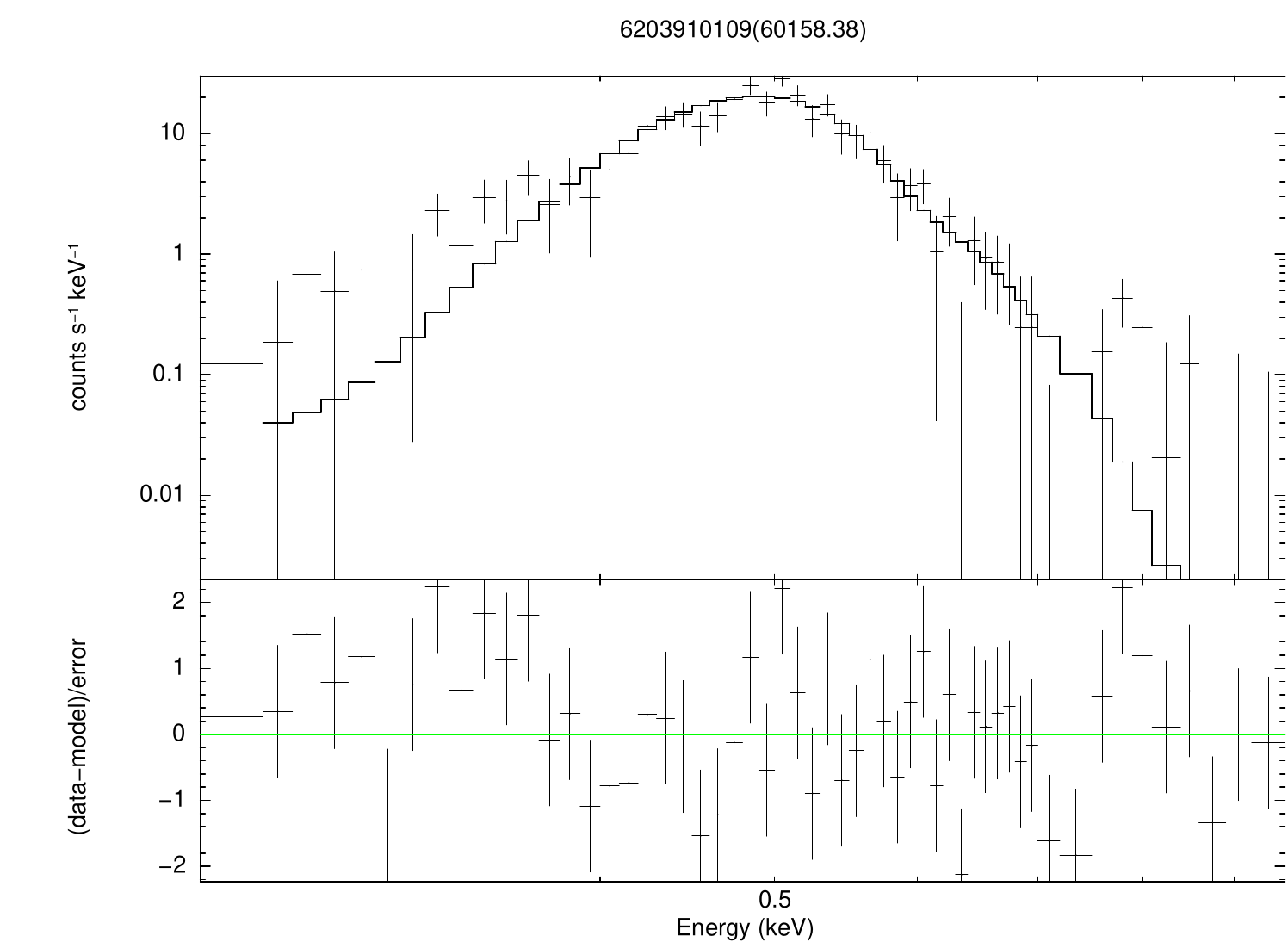}
\includegraphics[scale=0.3,angle=270]{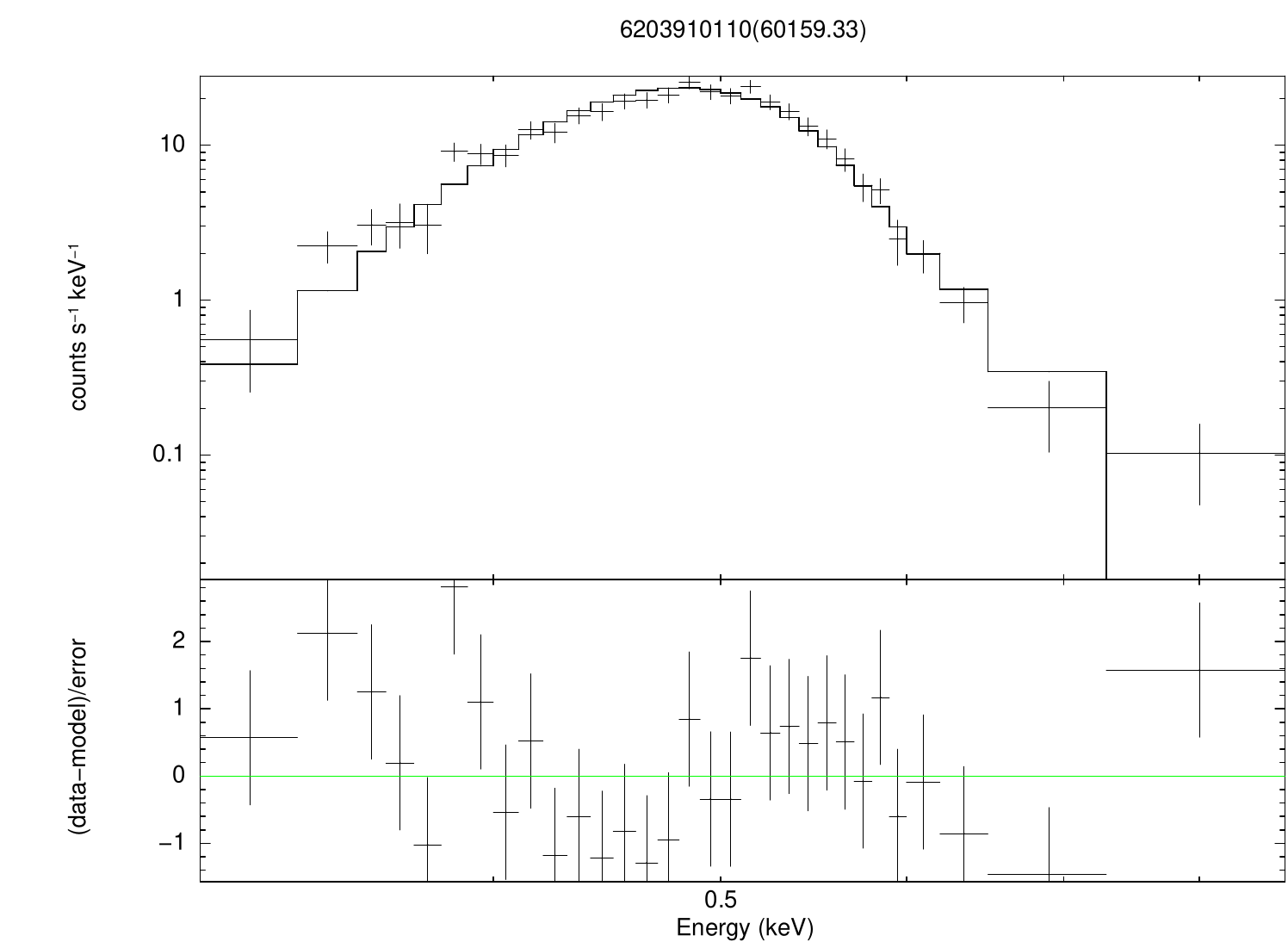}
\includegraphics[scale=0.3,angle=270]{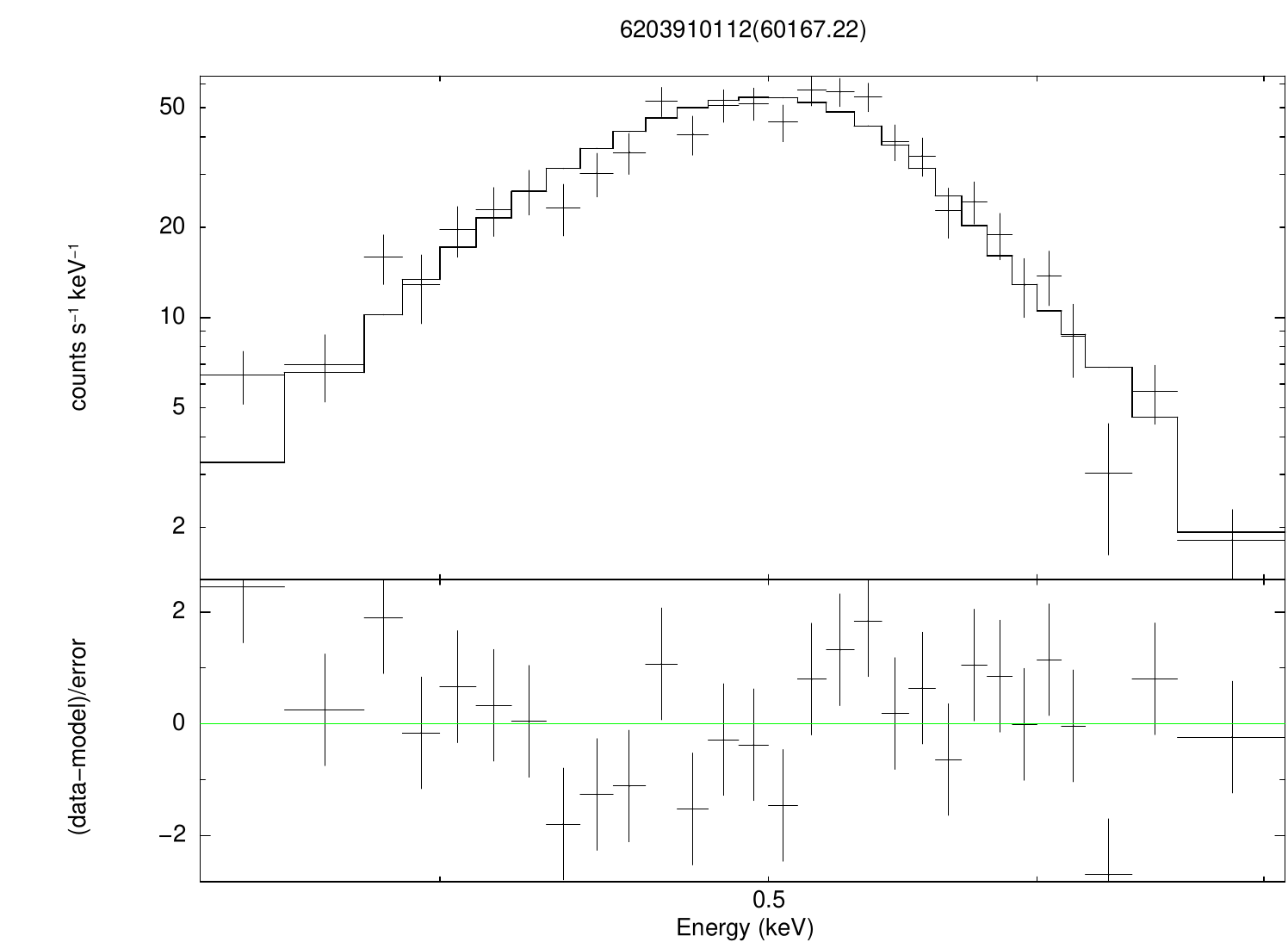}
\includegraphics[scale=0.3,angle=270]{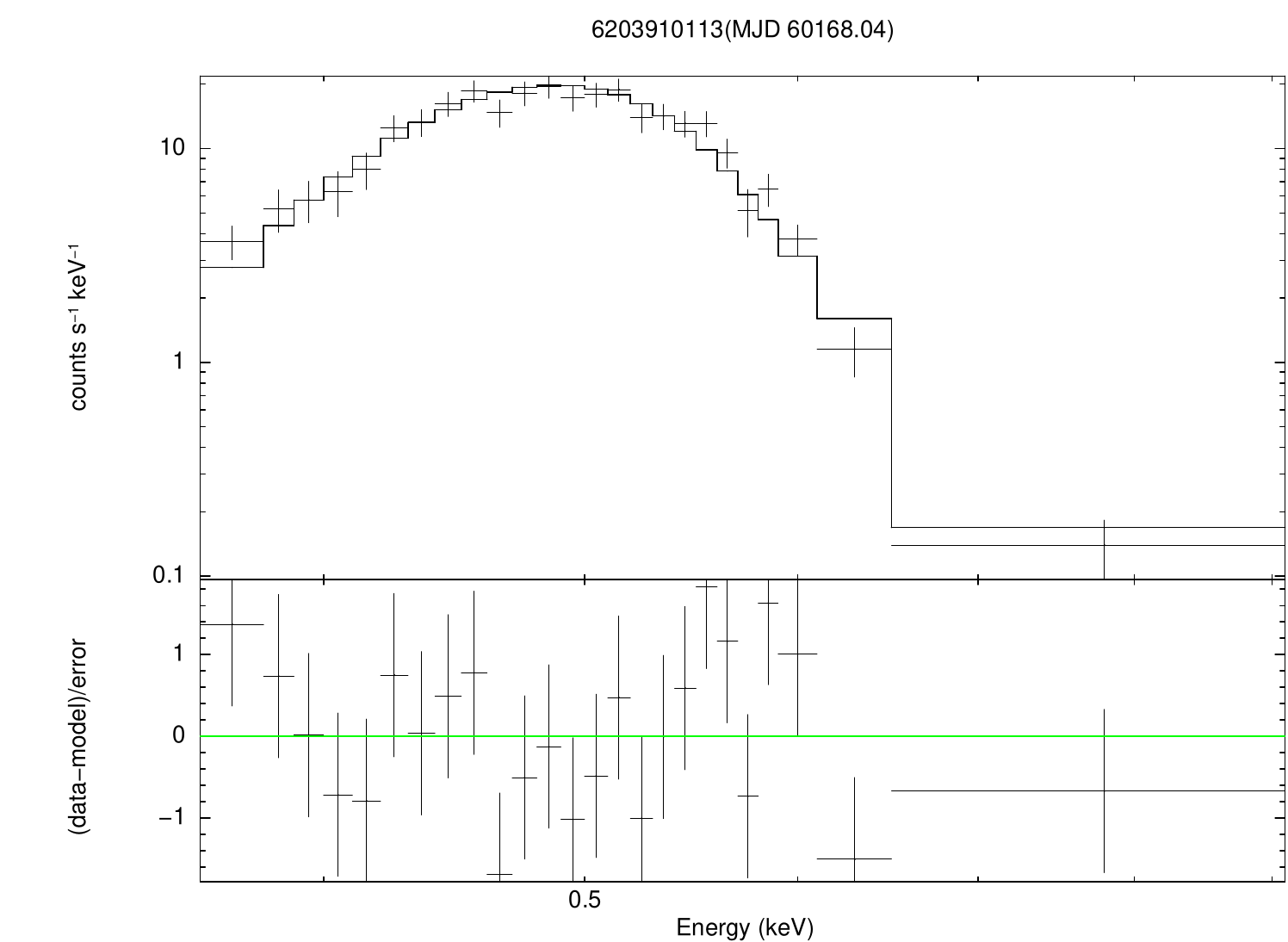}
\includegraphics[scale=0.3,angle=270]{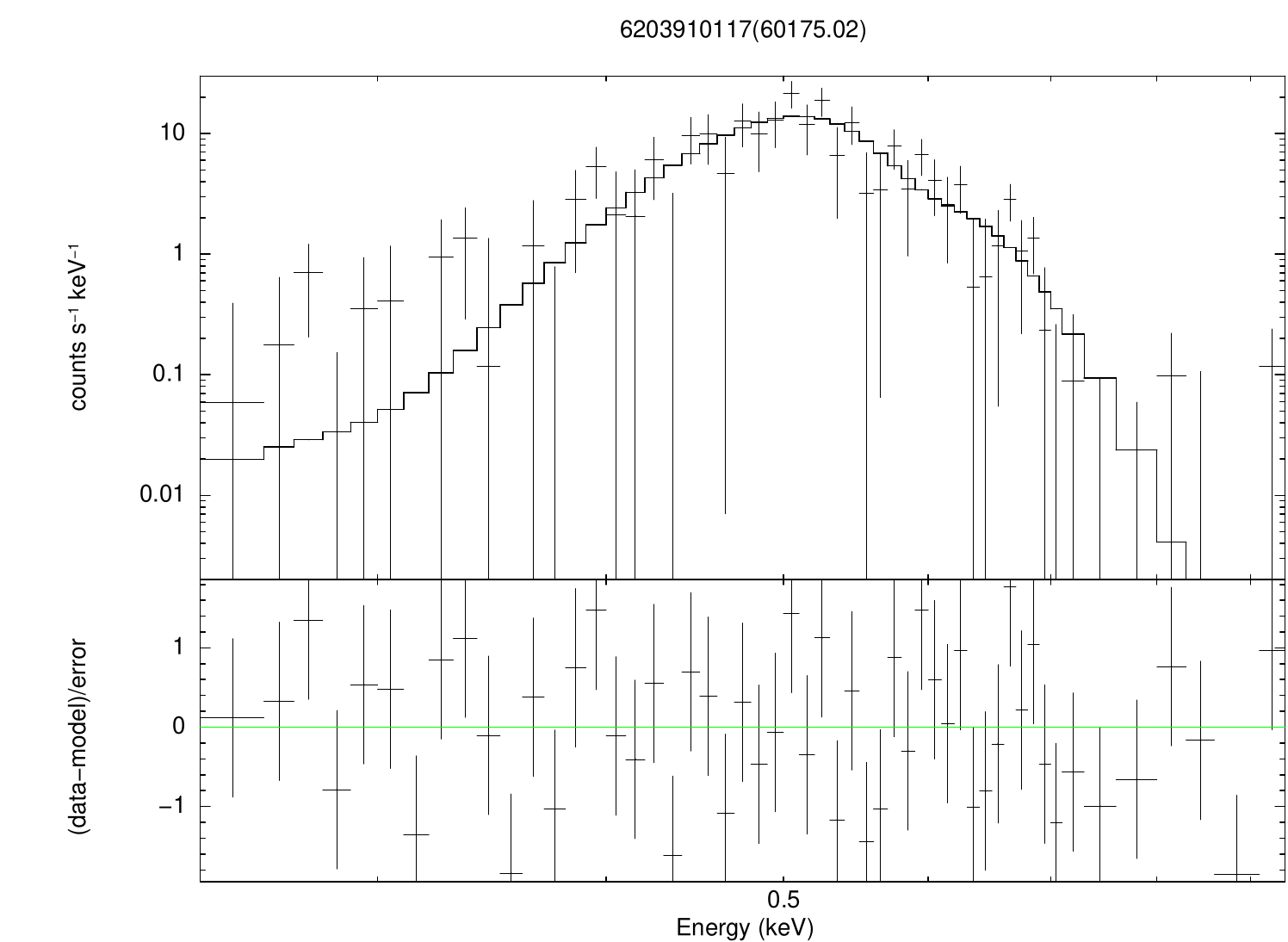}
\end{figure*}

\begin{figure*}
\centering
\includegraphics[scale=0.3,angle=270]{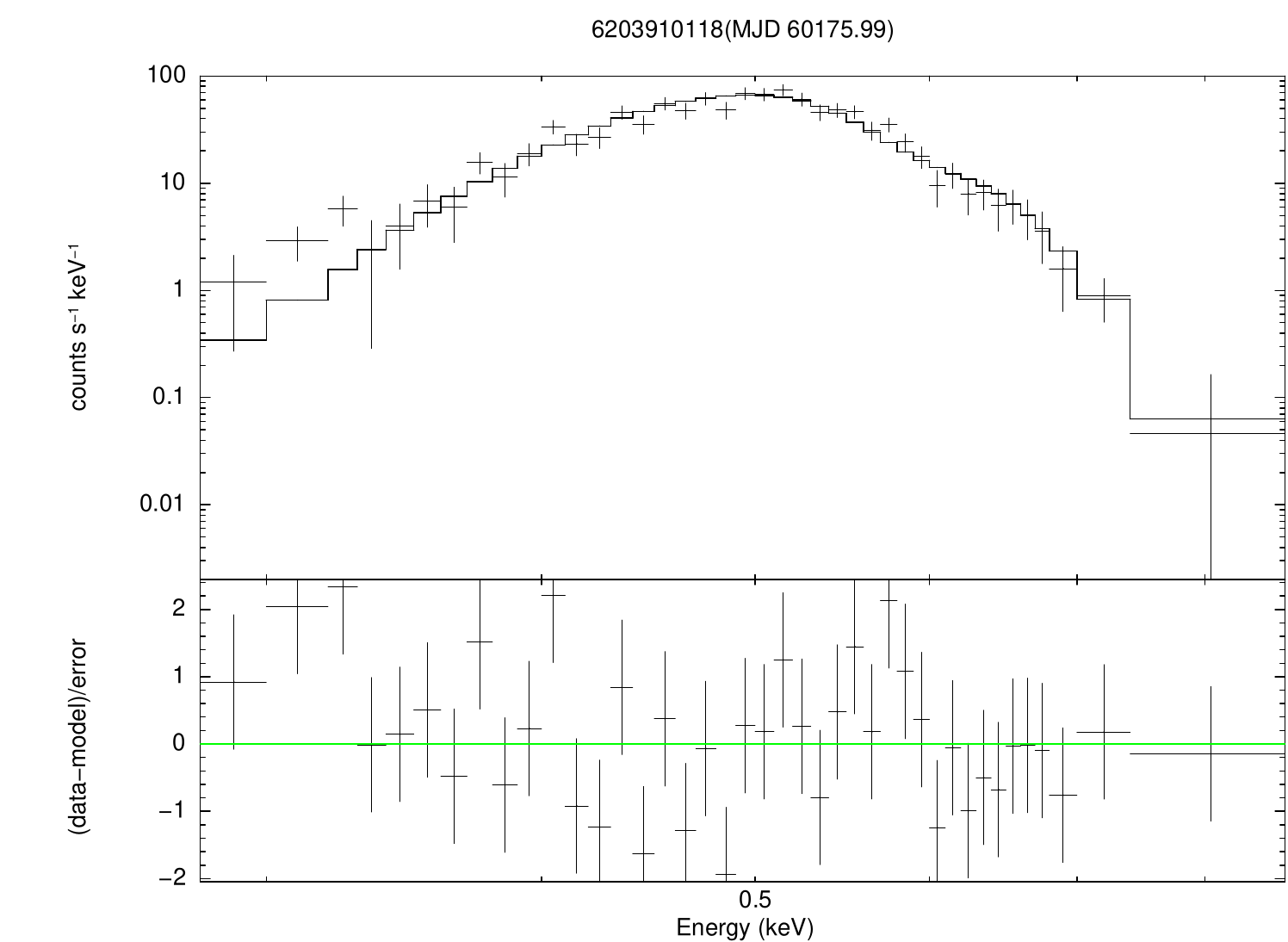}
\includegraphics[scale=0.3,angle=270]{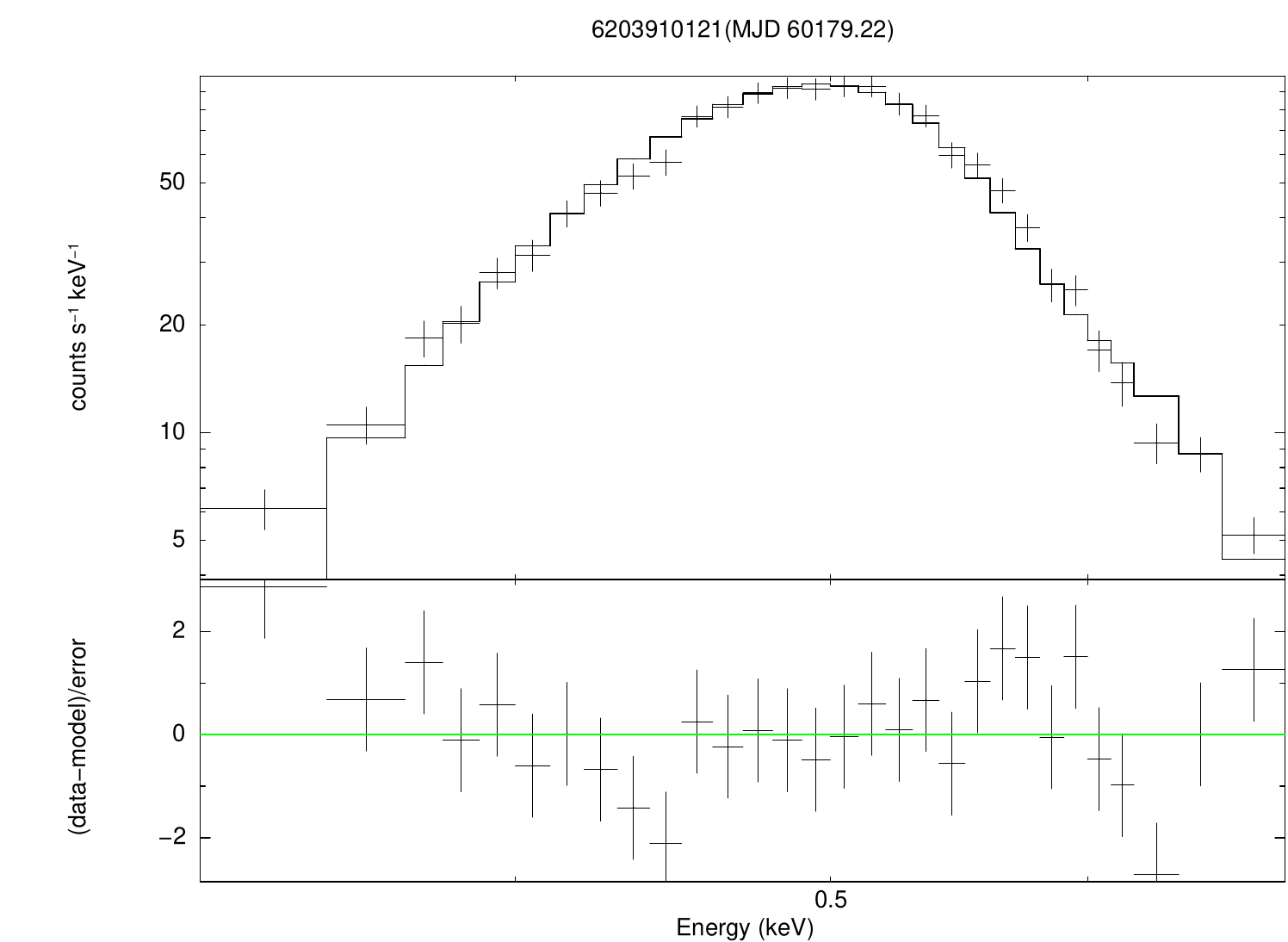}
\includegraphics[scale=0.3,angle=270]{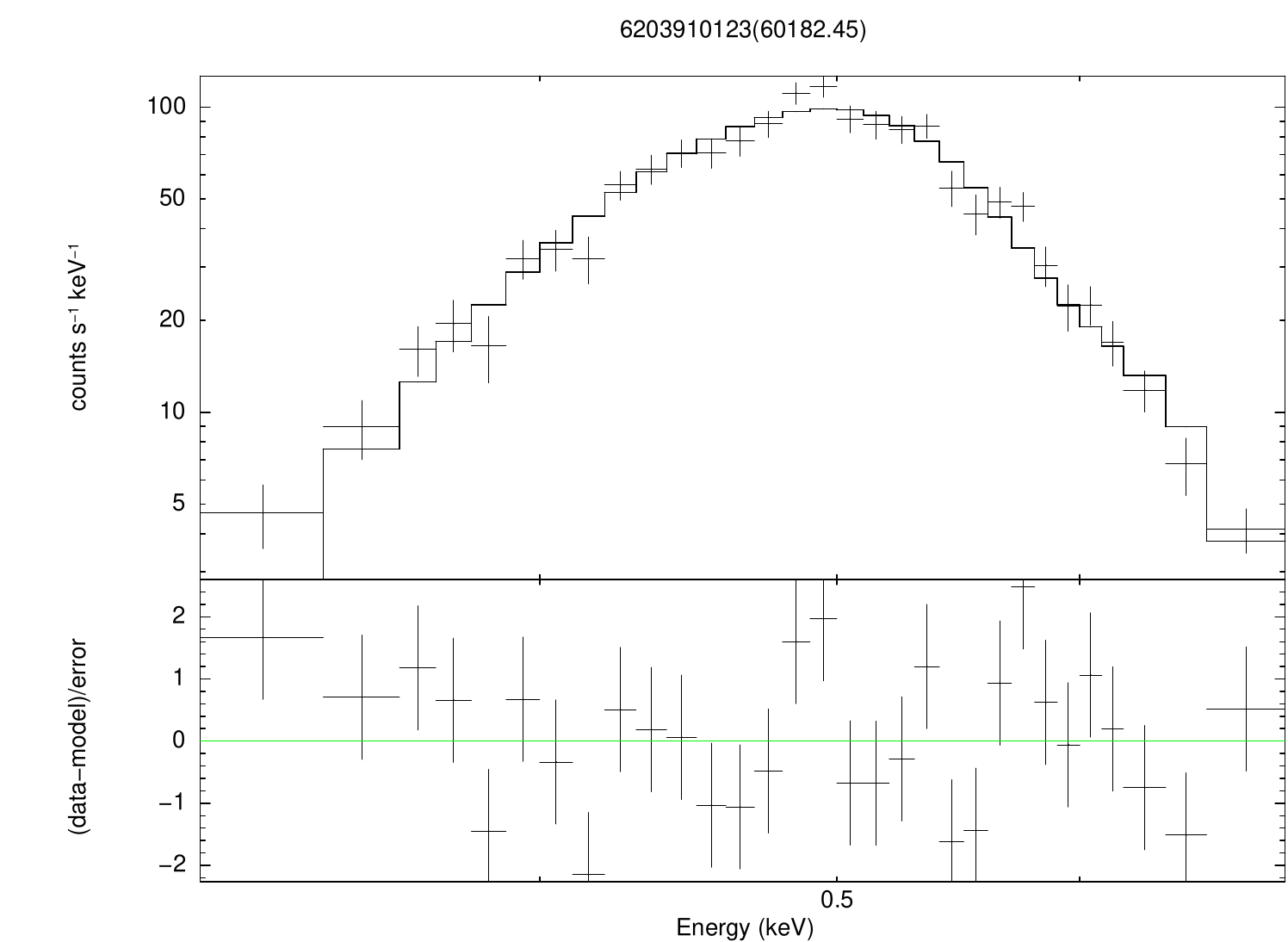}
\includegraphics[scale=0.3,angle=270]{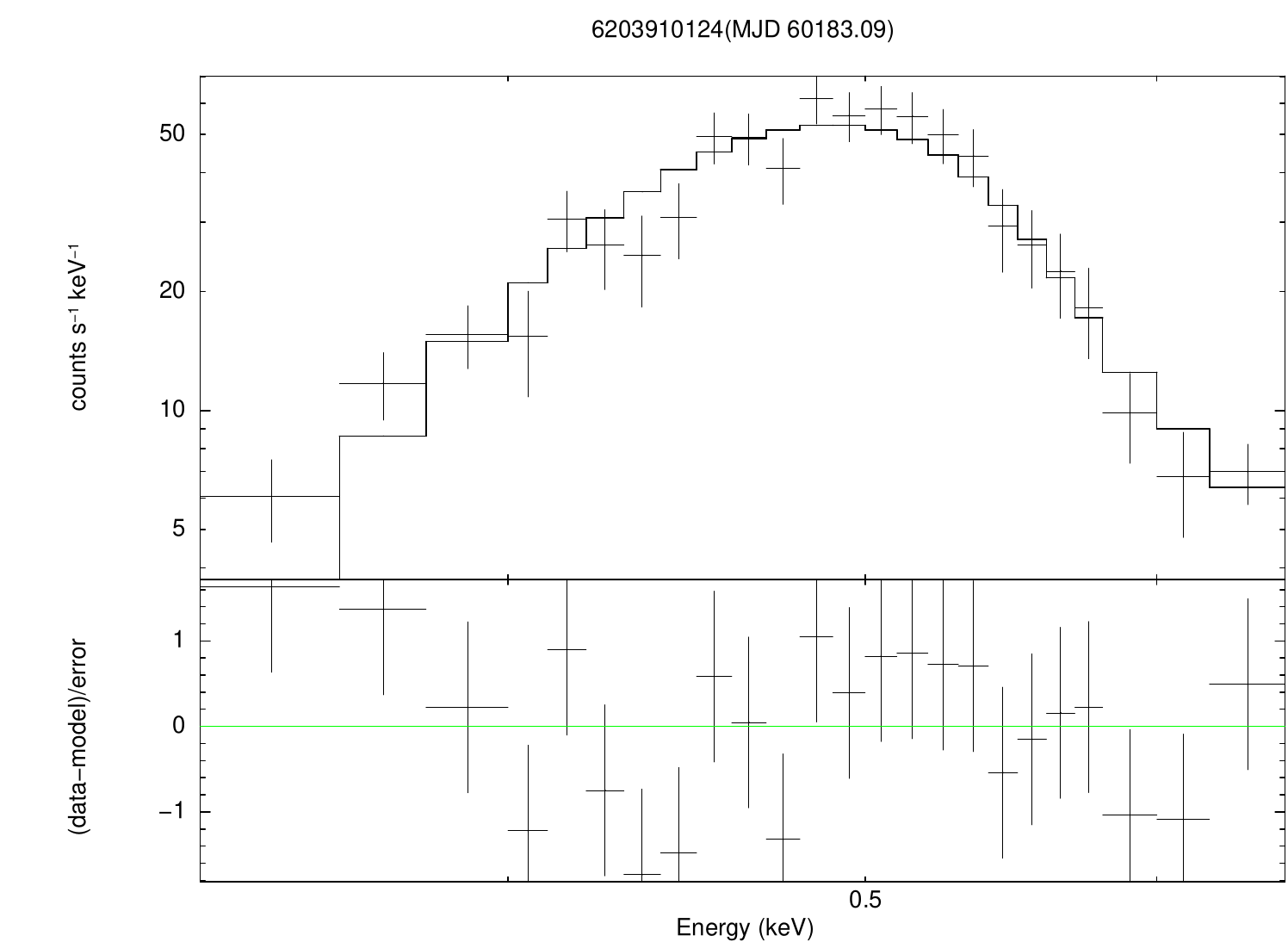}
\includegraphics[scale=0.3,angle=270]{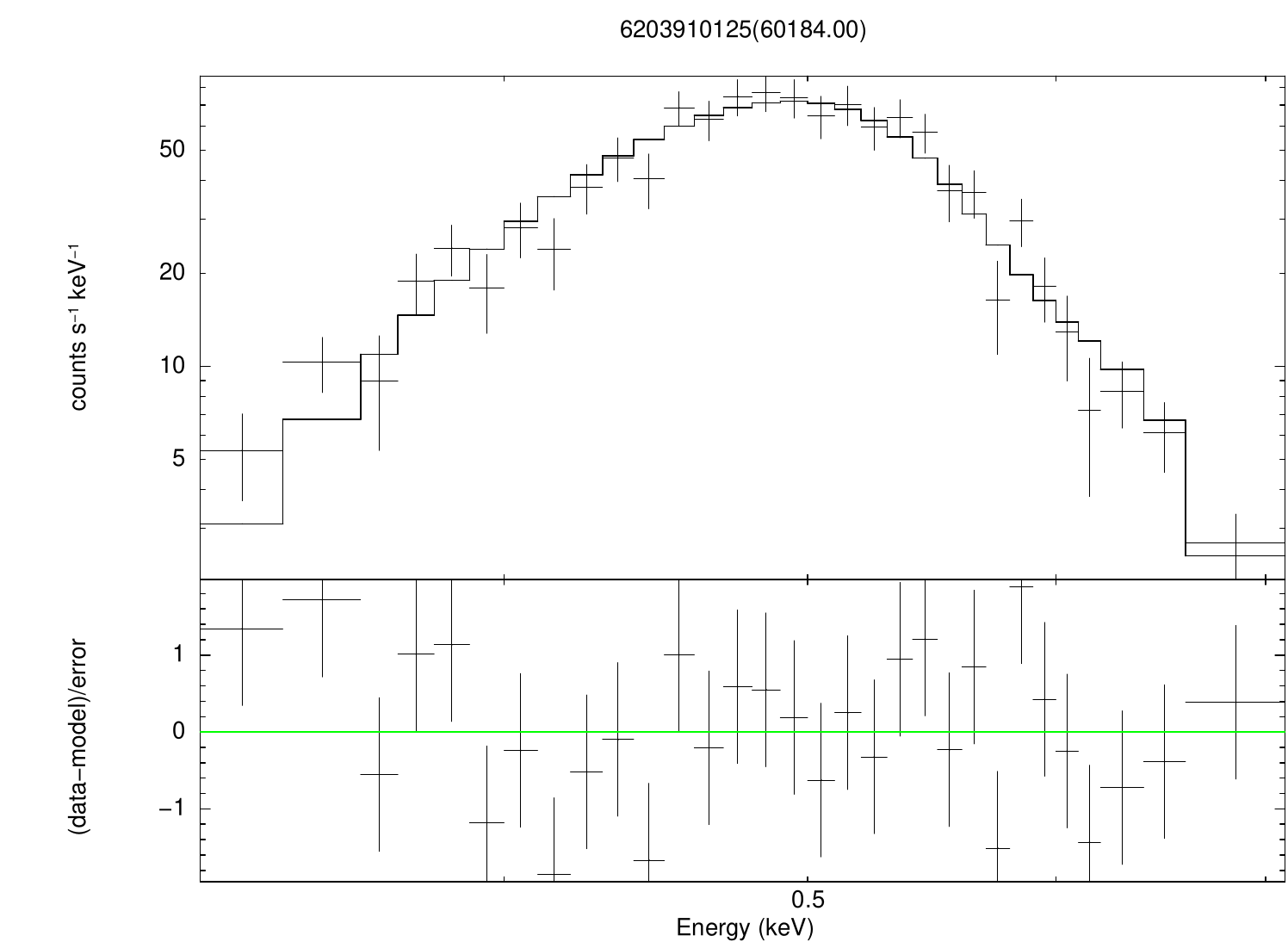}
\includegraphics[scale=0.3,angle=270]{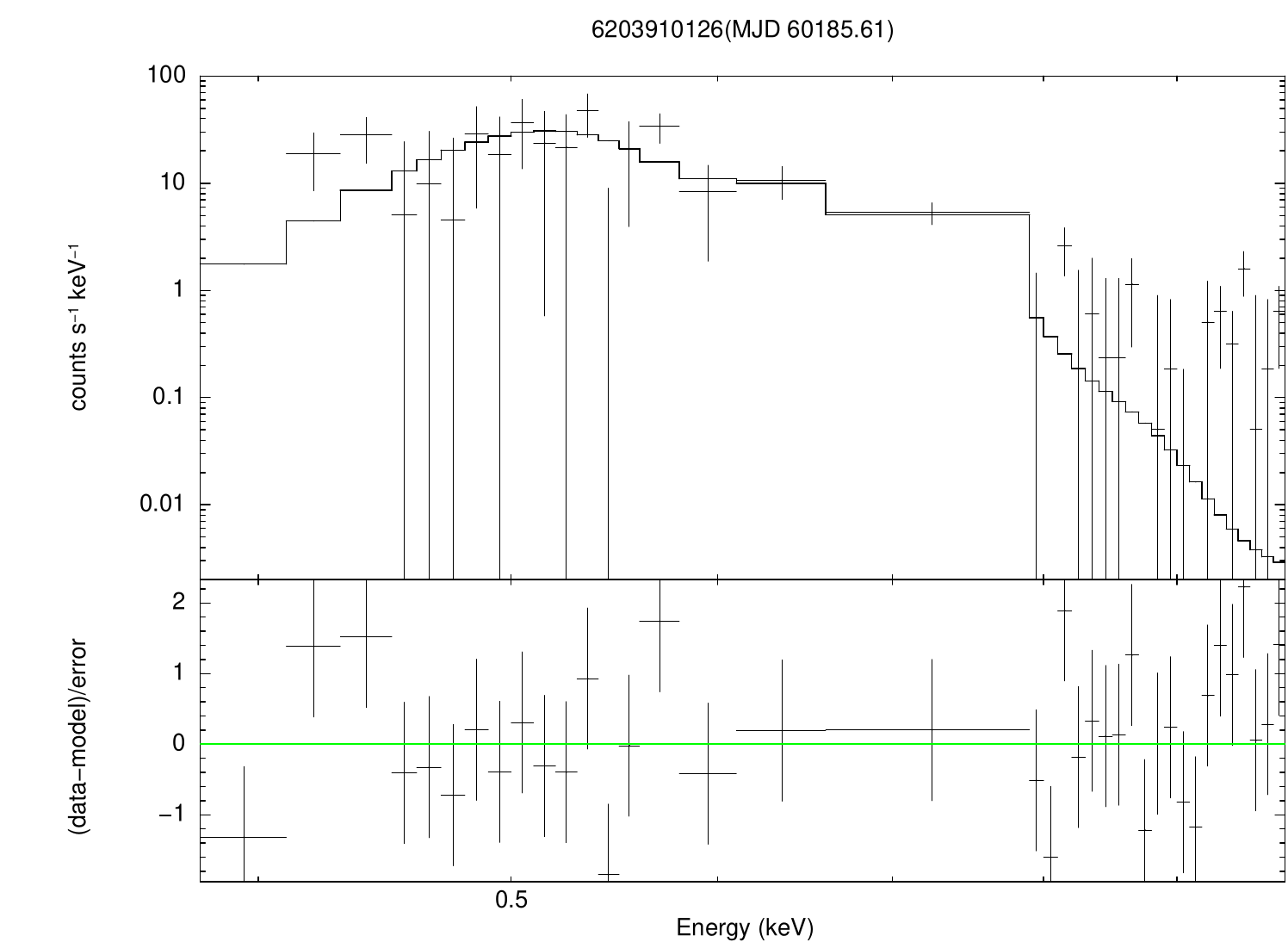}
    
    \label{nicer-resolve-every-wd-atom}
\end{figure*}

\begin{table*}
\centering
\caption{The parameters of X-ray data from NICER, using the model of BB.}
\begin{tabular}{|c|c|c|c|c|c|c|}
    \hline
ObsID&Start time(MJD) 
&N$_H$($10^{22}~\text{cm}^{-2}$)
&$k_{B}T$$_{bb}$~(eV)
&R$_{bb}(10^{3}~\text{km})$&L($10^{38}~\text{ergs}~\text{s}^{-1}$)
&C-Stat/dof  \\
\hline

6203910101&60150.83&0.66$\pm$0.024&37.77$\pm$0.71&70.36$\pm$15.29&0.41$\pm$0.085&79/42\\
6203910103&60151.99&0.65$\pm$0.024&37.37$\pm$0.91&63.57$\pm$15.73&2.31$\pm$0.47&43/22\\
6203910104&60153.34&0.91$\pm$0.074&25.89$\pm$0.88&464.82$\pm$168.15&13.10$\pm$0.74&36/35\\
6203910109&60158.38&0.83$\pm$0.064&29.75$\pm$1.83&390.19$\pm$102.34&20.06$\pm$5.01&41/36\\
6203910110&60159.33&0.71$\pm$0.033&28.31$\pm$1.41&308.40$\pm$121.28&10.16$\pm$0.85&32/31\\
6203910112&60167.22&0.75$\pm$0.043&33.17$\pm$1.27&173.97$\pm$76.65&9.93$\pm$0.86&55/29\\
6203910113&60168.04&0.76$\pm$0.046&28.84$\pm$1.29&337.78$\pm$155.41&15.39$\pm$2.55&26/25\\
6203910117&60175.02&0.80$\pm$0.024&28.26$\pm$2.79&501.07$\pm$118.58&29.65$\pm$9.80&60/53\\
6203910118&60175.99&0.72$\pm$0.039&31.58$\pm$1.06&226.03$\pm$90.79&11.41$\pm$1.88&88/41\\
6203910121&60179.22&0.62$\pm$0.017&37.03$\pm$0.68&59.37$\pm$10.93&2.55$\pm$0.52&72/30\\
6203910123&60182.45&0.64$\pm$0.025&35.68$\pm$0.91&80.09$\pm$21.56&3.56$\pm$0.99&73/31\\
6203910124&60183.09&0.54$\pm$0.042&40.12$\pm$1.46&18.91$\pm$9.60&0.45$\pm$0.28&16/24\\
6203910125&60184.00&0.61$\pm$0.035&35.35$\pm$1.33&66.13$\pm$26.19&2.26$\pm$1.32&39/31\\
6203910126&60185.61&0.72$\pm$0.040&37.32$\pm$0.92&78.58$\pm$24.58&0.85$\pm$0.36&42/38\\
  \hline
    \end{tabular}
    \label{table-nicer-spec}
\end{table*}

\begin{table*}
\centering
\caption{The parameters of X-ray data from NICER, using the model of WD atmosphere.}
\begin{tabular}{|c|c|c|c|c|c|c|}
    \hline
    \hline
ObsID&Start time(MJD) 
&N$_H$($10^{22}~\text{cm}^{-2}$)
&$k_{B}T$$_{atmo}$~(eV)&M$_{WD}$/M$_\sun$
&L($10^{38}~\text{ergs}~\text{s}^{-1}$)
&C-Stat/dof  \\
         \hline
6203910101&60150.83&0.63$\pm$0.015&50.84$\pm$1.25&1.35$\pm$0.32&0.19$\pm$0.63&79/42 \\
6203910103&60151.99&0.64$\pm$0.014&50.92$\pm$1.18&1.53$\pm$0.37&0.42$\pm$0.11&28/22\\
6203910104&60153.34&0.85$\pm$0.092&50.37$\pm$1.95&1.17$\pm$0.50&26.15$\pm$11.44&46/44\\
6203910109&60158.38&0.77$\pm$0.13&49.26$\pm$4.91&8.86$\pm$6.01&154.01$\pm$104.17&41/36\\
6203910110&60159.33&0.76$\pm$0.010&50.99$\pm$1.41&1.47$\pm$0.35&0.11$\pm$0.076&38/31\\
6203910112&60167.22&0.68$\pm$0.017&50.41$\pm$1.20&1.47$\pm$0.36&0.34$\pm$0.16&42/29\\
6203910113&60168.04&0.75$\pm$0.020&51.37$\pm$1.11&2.45$\pm$0.54&0.21$\pm$0.11&22/25\\
6203910117&60175.02&0.33$\pm$0.031&60.21$\pm$3.16&2.99$\pm$1.63&5.21$\pm$2.94&46/53\\
6203910118&60175.99&0.69$\pm$0.014&50.45$\pm$0.76&1.44$\pm$0.21&0.15$\pm$0.067&44/41\\
6203910121&60179.22&0.64$\pm$0.073&50.56$\pm$0.54&1.45$\pm$0.15&0.25$\pm$0.052&35/30\\
6203910123&60182.45&0.64$\pm$0.013&50.55$\pm$1.18&1.48$\pm$0.38&0.24$\pm$0.071&40/31\\
6203910124&60183.09&0.69$\pm$0.071&50.04$\pm$1.89&1.60$\pm$0.84&0.11$\pm$0.054&21/24\\
6203910125&60184.00&0.64$\pm$0.059&48.97$\pm$0.97&1.22$\pm$0.23&0.13$\pm$0.052&29/31\\
6203910126&60185.61&0.86$\pm$0.52&57.63$\pm$10.24&2.34$\pm$3.88&0.37$\pm$0.36&36/38\\
  \hline
    \end{tabular}
    \label{table-nicer-spec-wd-atom}
\end{table*}

\end{document}